\documentclass[12pt,letterpaper]{article}

\usepackage[margin=1in]{geometry} 
\usepackage{setspace}
\usepackage{authblk}
\usepackage{amsmath,amsfonts,amssymb,amsthm}
\usepackage{graphicx}
\usepackage{booktabs}
\usepackage{multirow}
\usepackage{enumitem}
\usepackage{xcolor}
\usepackage{hyperref}
\usepackage{natbib}
\usepackage{caption}

\newcommand{\pr}{\text{pr}}
\newcommand{\reals}{\mathbb{R}}

\newcommand{\E}{E}

\newcommand{\vu}{u}

\newcommand{\vy}{y}

\newcommand{\vz}{z}
\newcommand{\ind}{\textit{I}}

\newcommand{\vtheta}{\theta}

\newcommand{\R}{\mathbb{R}}

\newtheorem{theorem}{Theorem}%
\newtheorem{corollary}{Corollary}
\newtheorem{proposition}[theorem]{Proposition}%
\newtheorem{example}{Example}%
\newtheorem{lemma}{Lemma}%
\newtheorem*{remark}{Remark}

\newtheorem{definition}{Definition}
\newtheorem{assumption}{Assumption}

\newcounter{stepcounter} 

\makeatletter 
\newenvironment{step}[1][] 
  {
   \par\addvspace{\medskipamount}
   \refstepcounter{stepcounter}
   \noindent 
   \textit{
     Step \thestepcounter
     \def\@optarg{#1}
     \ifx\@optarg\@empty 
     \else
        \space(#1)
     \fi 
     .
   }
   \quad 
  }%
  {
   \par\addvspace{\medskipamount}
  }
\makeatother 

\title{Semiparametric Spatial Point Processes}

\author[1]{Xindi Lin}
\author[2]{Bumjun Park}
\author[3]{Christopher Zahasky}
\author[1]{Hyunseung Kang}

\affil[1]{Department of Statistics, University of Wisconsin-Madison}
\affil[2]{Department of Biostatistics, University of Washington}
\affil[3]{Department of Geoscience, University of Wisconsin-Madison}

\begin{document}
\maketitle

\abstract{We introduce a broad class of models called semiparametric spatial point process for making inference between spatial point patterns and spatial covariates. These models feature an intensity function with both parametric and nonparametric components. For the parametric component, we derive the semiparametric efficiency lower bound under Poisson point patterns
and propose a point process double machine learning estimator that can achieve this lower bound. The proposed estimator for the parametric component is also shown to be consistent and asymptotically normal for non-Poisson point patterns. For the nonparametric component, we propose a kernel-based estimator and characterize its rates of convergence.  Computationally, we introduce a fast, numerical approximation that transforms the proposed estimator into an estimator derived from weighted generalized partial linear models. We conclude with a simulation study and two real data analyses from ecology and hydrogeology.}


\section{Introduction}

\subsection{Background and motivation}
Spatial point patterns are a type of datasets that describe the spatial locations of events and are widely studied in criminology (e.g., \citet{mohler2011self,ang2012geometrically,reinhart2018review,hessellund2022semiparametric}), spatial transcriptomics (e.g., \citet{sun2020statistical,zhu2021spark,yan2025categorization}), spatial causal inference (e.g., \citet{papadogeorgou2022causal,zhou2024estimating,mukaigawara2025spatiotemporal}), 
and spatial epidemiology \citep{wang2024review}; see \citet{baddeley2015spatial} for textbook examples. A major goal in the study of spatial point patterns is to estimate and infer the relationship between the spatial point patterns and a specific subset of ``target'' spatial covariates while adjusting for ``nuisance'' covariates; we refer to the effect of the target covariates on the spatial point patterns as the \emph{target parameter} or the \emph{target effect} and the effect of the nuisance covariates on the point patterns as the \emph{nuisance parameter} or the \emph{nuisance effect}. For example, \cite{park2024statistical} studied the effect of land use on the spatial distribution of per-and polyfluoroalkyl substances (PFAS), which are synthetic and harmful chemicals that are increasingly being detected in groundwater, while adjusting for key environmental and spatial socioeconomic covariates. In causal inference with spatiotemporal data  \citet{papadogeorgou2022causal}, \citet{zhou2024estimating}, and \citet{mukaigawara2025spatiotemporal} studied the  effect of a spatial exposure while adjusting for relevant confounders. More generally, \citet{hessellund2022semiparametric}, \citet{chu2022quasi}, and \citet{cheng2023semi} proposed methods to estimate the effect of spatial covariates on spatial point patterns in the presence of a latent background process and applied their methods to study spatial point patterns from criminology and forestry.

A popular and well-studied approach to studying the relationship between spatial point patterns and spatial covariates is to use parametric spatial point processes (e.g., \citet[Chapter 9, Chapter 13]{baddeley2015spatial}, \citet[Chapter 8]{cressie2015statistics}). \citet{rathbun1994asymptotic} propose a maximum likelihood estimator for parametric Poisson spatial point processes. Later, \citet{schoenberg2005consistent}, \citet{guan2007thinned}, and \citet{waagepetersen2009two} 
showed that the estimator in \cite{rathbun1994asymptotic} 
remained consistent and asymptotically normal even for non-Poisson spatial point processes. To address computational challenges of solving the maximum likelihood estimator, \cite{berman1992approximating}, \cite{baddeley2000practical}, and \cite{baddeley2014logistic} proposed numerical approximation methods, which transformed the original optimization problem into estimating parameters of a particular type of generalized linear models. 
These methods for spatial point processes are widely accessible through the R package \texttt{spatstat} \citep{baddeley2005spatstat}. 


Unfortunately, a major limitation of the current literature is the restrictive assumptions on the nuisance effect. Some (e.g., \citet{fithian2015bias,papadogeorgou2022causal,zhou2024estimating,park2024statistical,mukaigawara2025spatiotemporal}) model the nuisance effect as a parametric effect and as we show in Section \ref{subsection:sim_estimator_compare}, mis-specifying  the nuisance effect can lead to biased inference of the target effect. Others model the nuisance effect non-parametrically, but the nuisance covariates are restricted to only spatial coordinates. Also, estimation of such nuisance effects depend on the existence of extra ``control'' point patterns \citep{rodrigues2010semiparametric,zhuang2019semiparametric,xu2019stochastic,hessellund2022semiparametric} or strong modeling assumptions, such as Gaussianity \citep{li2014functional} and piecewise linearity \citep{cheng2023semi}. 
Finally, to the best of our knowledge, there is no work on statistical optimality of complex spatial point processes with respect to efficiency lower bounds. Specifically, while efficiency lower bounds are well understood for simple parametric Poisson processes \citep{rathbun1994asymptotic}, efficiency lower bounds for more advanced models—particularly those with nonparametric components—remains an open question.


\subsection{Our Contributions}
This paper introduces \textit{semiparametric spatial point processes}, a more general class of models for spatial point patterns where the target effect is modeled parametrically and the nuisance effect is modeled non-parametrically. These models encompass a variety of existing models that have been used in practice
and are designed to unify the theoretical study of related spatial point processes in the literature; see Section \ref{sec:semi_ppp} for examples.


Our first contribution is to establish the semiparametric efficiency lower bound of estimating the target parameter in a semiparamtric spatial point process. Our result extends both the work of parametric Cramer-Rao bound for spatial point processes in \cite{rathbun1994asymptotic} and the classic work in semiparametric statistics with i.i.d. data (e.g., \citet{newey1990semiparametric,severini1992profile,hardle2000partially}) to spatial point patterns. Our second contribution is to propose a double machine learning framework \citep{chernozhukov2018double} for spatial point processes to estimate the target parameter with well-established asymptotic properties. 
Third, 
we propose a fast, numerical approximation that transforms the optimization into solving weighted generalized partial linear models in i.i.d. settings. 

The rest of the paper is organized as follows. Section \ref{sec:semi_ppp} introduces semiparametric spatial point processes. Section \ref{sec:semi_eff} derives the semiparametric efficiency lower bound. Section \ref{sec:dml} presents the double machine learning estimator. Section \ref{section:theory} discusses the asymptotic properties and Section \ref{sec:kernel_estimation} proposes the kernel-based nuisance estimator. Section \ref{sec:computation} presents the numerical approximation algorithm. Section \ref{section:data_analysis} presents the simulation and real data analysis. The implementation code is available on the author's GitHub page at \url{https://github.com/XindiLIN}. The technical derivations and additional simulation results are deferred to the Supplementary Material.

\section{Semiparametric Spatial Point Processes}\label{sec:semi_ppp}
Let $X$ denote a spatial point process in $\R^2$ and $\mathrm{d}\vu$ denote the Lebesgue measure of an infinitesimal neighborhood around a spatial location $\vu\in \R^2$.
The intensity function of $X$ is the function 
$\lambda: \mathbb{R}^2 \to [0, \infty)$ such that for any bounded subset $A\subset\R^2$, we have
$$\E\left\{\sum_{\vu\in X}\ind(\vu\in A)\right\}=\int_A \lambda(\vu)\mathrm{d}\vu,$$
\noindent where $\ind(\cdot)$ is the indicator function. 
The second-order intensity function of the spatial point process $X$ is the function $\lambda_2: \R^2\otimes \R^2 \to [0,\infty)$ such that for any two bounded subsets $A,B\subset \R^2$, we have $$\E\left\{\sum_{\vu\in X,v\in X}^{\vu\neq v}\ind(\vu\in A,v\in B)\right\} = \int_A\int_B \lambda_2(\vu,v)\mathrm{d}\vu \mathrm{d}v.$$ 
The pair correlation function of $X$ is the normalized second-order intensity function, i.e., $g(\vu,v) := \lambda_2(\vu,v)/\lambda(\vu)\lambda(v).$ 

We assume the intensity of $X$ depends on two sets of bounded, spatial covariates, $\vy_{\vu} \in\mathcal{Y}\subset \R^p$ and $\vz_{\vu} \in\mathcal{Z}\subset \R^q$ taking the following form: 
\begin{equation}\label{eq:semi_pp}
    \lambda(\vu;\vtheta,\eta) = \Psi \left\{\tau_{\vtheta}(\vy_{\vu}),\eta(\vz_{\vu})\right\},  \quad{} \vtheta\in \Theta,  \eta\in\mathcal{H}.
\end{equation}
The term $\tau_{\vtheta}(\cdot)$ is an user-specified function of $\vy_\vu$ and is parameterized by an unknown, finite-dimensional parameter $\vtheta$ in a Euclidean space $\Theta$, for instance $\tau_{\vtheta}(\vy_{\vu}) = \vtheta^\top \vy_{\vu}$ and $\vtheta$ is a vector of real numbers. The term $\eta(\cdot)$ is an unknown, infinite-dimensional function of $\vz_{\vu}$ in a linear space $\mathcal{H}$, for instance, the space of smooth functions in a Hilbert space $L^2(\mathcal{Z})$. The function $\Psi(\cdot)$ is a user-specified link function; a popular choice is the exponential function. We let $\vtheta^*$, $\eta^*$ and $g^*$ denote the true values of the target parameter $\vtheta$, nuisance parameter $\eta$, and the pair correlation function $g$ respectively. Our goal is to estimate and conduct inference on the true target parameter $\vtheta^*$.
We refer to a spatial point processes with an intensity function in equation \eqref{eq:semi_pp} as \emph{a semiparametric spatial point process} and some examples of such processes are listed below.

\begin{example}[Parametric Nuisance Effect] Spatial point processes with the intensity function 
    $\lambda(\vu) = \exp(\vtheta^\top\vy_\vu+\eta^\top \vz_\vu)$. 
    The parameters $\vtheta$ and $\eta$ are finite-dimensional and characterize the effect of target covariate $\vy_\vu$ and nuisance covariates $\vz_\vu$, respectively (e.g. \citet{fithian2015bias,park2024statistical}).
\end{example}

\begin{example}[Nonparametric Baseline Process] Spatial point processes with intensity function
    $\lambda(\vu) = \lambda_0(\vu) \exp\left(\vtheta^\top \vy_\vu\right) $
    where $\lambda_0(\vu)$ is a nonparametric, ``baseline'' intensity function, and $\vtheta$ characterizes the parametric effect of covariates $\vy_\vu$. (e.g. \citet{diggle1990point,chu2022quasi,cheng2023semi}).
\end{example}

\begin{example}[Spatial Causal Inference] 
Let $\vy_\vu$ be a spatial exposure/treatment and let $\vz_\vu$ be observed, spatial confounders. Under model \eqref{eq:semi_pp}, the confounder's effects are adjusted nonparametrically. This is in contrast with the current spatiotemporal causal inference literature (e.g., \citet{papadogeorgou2022causal,zhou2024estimating,mukaigawara2025spatiotemporal}), which typically models the propensity as Poisson point process with parametric intensity function $\lambda(\vu;\eta) = \exp(\eta^\top \vz_\vu).$ 
    
\end{example}

\begin{example}[Log-Linear \textit{Semiparametric Spatial Point Processes}] Consider a spatial point process with a log-linear intensity function (e.g., \citet[chap. 9.2.3]{baddeley2015spatial};  \citet{hessellund2022semiparametric})
\begin{equation}\label{eq:semi_pp_log_linear}
\lambda(\vu;\vtheta,\eta)=\exp\left\{\vtheta^\top\vy_{\vu}+\eta(\vz_{\vu})\right\}.
\end{equation}
Log-linear intensity functions are common due to their theoretical and practical conveniences (e.g., \citet[chap. 9.2.3]{baddeley2015spatial};  \citet{hessellund2022semiparametric}). As we will demonstrate, they also possess some unique properties for our method.

    
\end{example}



We make a couple of remarks about model \eqref{eq:semi_pp}. First, we use $\lambda$ and $\Psi$ 
to distinguish two parametrizations of the intensity function: one that is a function of the spatial location $\vu$ (i.e., $\lambda$) and another that is a function of the spatial covariates $\vy_\vu$ and $\vz_\vu$ at the spatial location $\vu$ (i.e., $\Psi$). While cumbersome from a modeling perspective, the latter parametrization helps to distinguish the theoretically distinct roles that the two types of spatial covariates plays in inference. Second, for an observational window $A\subset \R^2$, we define the pseudo-log-likelihood function of a semiparametric spatial point process $X$ to be
\begin{equation}\label{eq:pseudo-log-likelihood}
    \ell(\vtheta,\eta; X) = \sum_{\vu\in X\cap A}\log\lambda(\vu;\vtheta,\eta)-\int_A\lambda(\vu;\vtheta,\eta)\mathrm{d}\vu.
\end{equation}

\section{Semiparametric Efficiency Lower Bound for Poisson Spatial Point Process}\label{sec:semi_eff}


To derive the semiparametric efficiency lower bound, we introduce several new notations and definitions. To the best of our knowledge, this is the first work discussing the semiparametric efficiency in spatial point process models. Notably, unlike similar derivations for i.i.d. data, an important technical issue that needs to be resolved is the inherent difference between the spatial domain and the covariate domain. The pseudo-log-likelihood in equation \eqref{eq:pseudo-log-likelihood} is defined over points in the spatial observation window $A \subset \mathbb{R}^2$. In contrast, the spatial covariates themselves are defined over the support of the covariates, i.e., $\mathcal{Y} \subset \mathbb{R}^p$ and $\mathcal{Z} \subset \mathbb{R}^q$. To bridge this gap, we define the ``covariate-conditioned expectation'' over a push-forward measure. 


\begin{definition}[Covariate-Conditioned Expectation]\label{def:conditioning}  
Suppose the mapping from the spatial domain to the covariate domain,  $\vu\mapsto (\vy_\vu,\vz_\vu),\vu\in A\subset \R^2$, induces a push-forward measure with Radon-Nikodym derivative, $f_A(\vy,\vz)$, such that for any function $\phi(\vy,\vz)$ defined on the covariate domain, we have
$$\int_A\phi(\vy_\vu,\vz_\vu)\mathrm{d}\vu=\int_{\mathcal{Y}\times\mathcal{Z}}\phi(\vy,\vz)f_A(\vy,\vz)\mathrm{d}\vy\mathrm{d}\vz.$$
    Then, for any given nuisance covariate value $\vz\in\mathcal{Z}$, we define the ``covariate-conditioned expectation'' of the pseudo-log-likelihood in \eqref{eq:pseudo-log-likelihood} as:
\begin{equation}\label{eq:conditioning_covariates}
\E[\ell(\vtheta,\eta;X)|\vz]  := \int_{\mathcal{Y}} \left(\log\left\{\Psi [\tau_{\vtheta}(\vy),\eta(\vz)]\right\}\Psi[\tau_{\vtheta^*}(\vy),\eta^*(\vz)] -\Psi[\tau_{\vtheta}(\vy),\eta(\vz)]\right)f_A(\vy,\vz)\mathrm{d}\vy.
\end{equation}

\end{definition}

    


Given Definition \ref{def:conditioning}, deriving the semiparametric efficiency lower bound involves two  steps: (a) constructing a class of parametric submodels for the processes, and
(b) determining the supremum of the Cramer-Rao bound over these submodels, which yields the semiparametric efficiency lower bound. 
Specifically, for every 
$\theta\in\Theta$, we define $\eta_{\vtheta} \in \mathcal{H}$ as the nuisance function that maximizes the expectation of the pseudo-log-likelihood in \eqref{eq:pseudo-log-likelihood}, i.e., 
\begin{equation} \label{eq:eta_theta}
\eta_{\vtheta} = \arg\max_{\eta\in \mathcal{H}}\E\left\{\ell(\vtheta,\eta; X)\right\}.
\end{equation}
We also define $\nu^*$ as the derivative of this map at $\vtheta^*$:
\begin{equation}\label{eq:nuisance_derivative}
  \nu^*(\vz) = \frac{\partial}{\partial\vtheta}\eta_\vtheta(\vz)\big|_{\vtheta = \vtheta^*}.
\end{equation}
Consider a parametric submodel of $X$ with intensity function $\lambda(\vu;\theta,\bar\eta_{\vtheta})$ where $ \bar\eta_{\vtheta}$ is a known, twice continuously differentiable map from $\Theta$ to $\mathcal{H}$. We use the notation $\bar\eta_{\vtheta}$ to distinguish this map from the other map $\eta_{\vtheta}$ defined in \eqref{eq:eta_theta}. 
By the chain rule of the Gâteaux derivative (see Section 5.5 of the Supplementary Material), the Cramer-Rao lower bound for estimating $\vtheta$ within this submodel is the inverse of the sensitivity matrix 
$S^{-1}(\vtheta^*,\eta^*,\nu)$:
\begin{equation}\label{eq:sensitivity_submodel}
    S(\vtheta^*,\eta^*,\nu) = \int_A \lambda(\vu;\vtheta^*,\eta^*)\left[\frac{\partial}{\partial\vtheta}\log\lambda(\vu;\vtheta^*,\eta^*)+\frac{\partial}{\partial\eta}\log\lambda(\vu;\vtheta^*,\eta^*)
\left\{ \nu(\vz_{\vu})\right\}\right]^{\otimes 2}\mathrm{d}\vu.
 \end{equation}
An important observation from \eqref{eq:sensitivity_submodel} is that the Cramer-Rao lower bound depends on 
the map $\bar\eta_{\vtheta}$ only through $\nu$, i.e., the derivative of $\bar\eta_{\vtheta}$ with respect to $\vtheta$ at the true parameter $\vtheta^*$. In other words, finding the worst-case Cramer-Rao lower bound among all parametric submodels is equivalent to finding $v$
that maximizes the Cramer-Rao lower bound and Theorem \ref{thm:least_favorable_direction} formally derives the supremum of the Cramer-Rao bounds over all parametric submodels.
 \begin{theorem}[Semiparametric Efficiency Lower Bound for Poisson Processes]\label{thm:least_favorable_direction} 
The supremum of the 
\noindent Cramer-Rao lower bound (i.e., the minimum of \eqref{eq:sensitivity_submodel}) is attained at $\nu^*$ defined in \eqref{eq:nuisance_derivative} and is equal to
\begin{equation}\label{eq:least_favorable_curve}
    \nu^*(\vz) = -\left\{\frac{\partial^2}{\partial\eta^2}\E\left\{\ell(\vtheta^*,\eta;X)\mid \vz\right\}\bigg|_{\eta=\eta^*}\right\}^{-1}\frac{\partial^2}{\partial\vtheta\partial\eta}\E\left\{\ell(\vtheta^*,\eta;X)\mid \vz\right\}\bigg|_{\eta=\eta^*}.
\end{equation}
\end{theorem} 
\noindent

\noindent We remark that the term $\nu^*$ in Theorem \ref{thm:least_favorable_direction} that attains the efficiency bound is expressed using the ``covariate-conditioned expectation'' defined in \eqref{eq:conditioning_covariates}. 

\section{Double Machine Learning for Spatial Point Processes}\label{sec:dml}

Double machine learning \citep{chernozhukov2018double} is a popular framework for estimating the parametric component in semiparametric models using i.i.d. data. 
Broadly speaking, double machine learning employs cross-fitting, a form of sample splitting, to prevent overfitting bias from modern machine learning methods and enable root-n consistent estimation and valid inference for a target parameter. 
In this section, we propose a double machine learning framework for semiparametric spatial point processes. 

 Developing this framework requires overcoming several fundamental challenges not found in prior works. First, a naive, uniform splitting of the observed points fails to produce independent subsets of points, which is crucial for double machine learning.
Second, there is no existing literature for estimating the nonparametric component in \eqref{eq:semi_pp}. 
Third, even if the nonparametric component is estimated parametrically, we are unaware of works that characterize the asymptotic properties with sample splitting in point processes, especially for asymptotic normality and efficiency. The following sections details how we address each of these challenges.

\subsection{V-fold random thinning}\label{sec:thinning}

Historically, thinning has been used in point processes or Bayesian inference to efficiently draw samples from complex, often analytically intractable processes \citep{moller2010thinning,chiu2013stochastic}. More recently, \cite{cronie2024cross} developed a cross-validation for point processes by using thinning to split a point pattern into pairs of training and validation sets. In our paper, we re-purpose thinning as a general form of data splitting to construct a ``cross-fitting'' estimator for spatial point processes. We call this implementation of thinning as \emph{V-fold random thinning} and it is formally defined below.



\begin{definition}[V-fold Random Thinning]\label{def:thinning}
 Suppose $X$ is a spatial point process and $V \geq 2$ is an integer. For each point in $X$, we uniformly sample a number $v$ from $\{1,\ldots, V\}$ and assign the point to the sub-process $X_v$.   
 \end{definition}
For each sub-process $X_v$, we estimate the nuisance components in \eqref{eq:semi_pp} using data from the other folds, and use these out-of-sample nuisance estimates to estimate the target parameter. Proposition \ref{proposition:thinning} establishes three useful properties of the sub-process $X_v$ constructed from V-fold random thinning. Notably, all these properties allow investigators to use modern machine learning methods for estimating the nuisance parameter $\eta$ without introducing overfitting bias.
  
 \begin{proposition}[Properties of V-fold Random Thinning]\label{proposition:thinning} 
  For any set $I\subset [V]=\{1,2,...,V\}$, let $|I|$ be the cardinality of $I$, and $X^{(I)}:=\bigcup_{j\in I}X_{j}$. Then, $X^{(I)}$ satisfies the following properties: (i) the intensity function of $X^{(I)}$ is $V^{-1}|I|\lambda(\vu)$; (ii) the pseudo-log-likelihood function of $X^{(I)}$ has the same first-order properties as $X$. 
    (iii) if $X$ is a Poisson process, then $X^{(I)}$ is independent with its complement $X\backslash X^{(I)}$.
  
\end{proposition}
\noindent The first-order preservation property in (ii) from Proposition \ref{proposition:thinning} refers to the property that for any set $I$,  $\E\left\{\ell(\vtheta,\eta;X^{(I)})\right\}$ is maximized by the same true values $\vtheta^*,\eta^*$ and for any fixed $\vtheta$, $\E\left\{\ell(\vtheta,\eta;X^{(I)})\right\}$ is maximized by the same $\eta_{\vtheta}$ defined in \eqref{eq:nuisance_derivative}.

 \noindent   


 

\subsection{Double Machine Learning Framework} \label{sec:spatial_cross_fitting}
Our double machine learning framework for estimating $\vtheta^*$ is stated below and is broken into four steps.

\begin{step}[V-fold random thinning]  Use V-fold random thinning in Definition \ref{def:thinning} to partition the point process $X$ into $V$ sub-processes $X_1,\ldots,X_V$.
\end{step}
\begin{step}[Estimate $\eta_\theta$ with sub-processes excluding $X_v$]\label{step1}
 For every $v \in \{1,\ldots,V\}$, let $X_v^c := \bigcup_{j\neq v} X_j$ be the sub-processes that exclude $X_v$. Estimate $\eta_{\theta}$ in equation \eqref{eq:eta_theta} using the observations in $X_v^c$, i.e., $\hat\eta_{\vtheta}^{(v)} = \arg\max_{\eta\in\mathcal{H}}\ell(\theta,\eta;X_v^c).$ 
\end{step}
\begin{step}[Estimate $\vtheta^*$ with sub-process $X_v$] \label{step2}
Estimate the target parameter $\vtheta^*$ by plugging in the estimate $\hat\eta_{\vtheta}^{(v)}$ from step \ref{step1} into the pseudo-log-likelihood and
maximizing it over the observations in $X_v$, i.e., 
$\hat\vtheta^{(v)} =\arg\max_{\vtheta\in \Theta}\ell(\vtheta,\hat{\eta}^{(v)}_{\vtheta};X_v).$
\end{step}
\begin{step}[Aggregation] \label{step4} Repeat steps \ref{step1} and \ref{step2}
for each $v$ and return the aggregated estimate of $\vtheta^*$, i.e.,
$\hat\vtheta = V^{-1} \sum_{v=1}^V\hat\vtheta^{(v)}$.
\end{step}

Roughly speaking, steps \ref{step1} and \ref{step2} resemble a profile-likelihood estimator where we maximize with respect to the nuisance parameter (i.e., step \ref{step1}), and then maximize with respect to the target parameter (i.e., step \ref{step2}). The important difference between the usual profile-likelihood estimator (e.g., \citet{severini1992profile}) and our method is that we use different observations at each maximization step. As mentioned above, this is critical to prevent over-fitting of the nuisance parameter; see Remark \ref{remark:ml} and Section \ref{section:theory} for further discussions.


\begin{remark}[Step \ref{step1} and using nonparametric estimators] \label{remark:ml}
In step \ref{step1}, we do not specify a particular nonparametric estimator of $\eta_\theta$, or more specifically the function class $\mathcal{H}$, to estimate $\eta_{\vtheta}$. 
In Section \ref{section:theory}, we show that as long as $\hat\eta_{\vtheta}^{(v)}$ is consistent at the desired rate, the estimator $\hat\vtheta$ is $\sqrt{n}$-consistent. Importantly, these rate conditions are ``high-level'' conditions that are agnostic to the function class of $\mathcal{H}$ or how investigators fine-tuned their nonparametric estimators.
For more discussions on this agnostic framework for estimating infinite-dimensional nuisance parameters in semiparametric models, see \citet{van2011targeted}, \citet{chernozhukov2018double} and \citet{athey2019machine}. 

\end{remark}

\begin{remark}[Step \ref{step4} and semiparametric efficiency] 
In step 4, any one of $\hat{\vtheta}^{(1)},\ldots,\hat{\vtheta}^{(V)}$ will be consistent and asymptotically normal under the assumptions laid out in Sections \ref{sec:consistency} and \ref{sec:normality}. But, as we show in Section \ref{sec:semi_eff}, aggregating $\hat{\vtheta}^{(v)}$ in step \ref{step4} and consequently, using the sub-processes multiple times are critical to regain the full semiparametric efficiency of the estimator; see \citet{schick1986asymptotically}  for a similar phenomena in i.i.d. settings.
\end{remark}


\begin{remark}[Choosing the number of folds $V$ for random thinning]
The choice of $V$ has no asymptotic consequences for our estimator as long as $V$ is bounded.
Practically speaking, while $V=2$ was more than sufficient in our numerical experiments, we generally recommend choosing $V$ based on both data and computational availability. 
\end{remark}

\section{Asymptotic Framework}\label{section:theory}

\subsection{Overview and definitions}
We develop asymptotic theory—including consistency, asymptotic normality, and semiparametric efficiency—for our double machine learning estimator in Section \ref{sec:dml}. We also propose a consistent estimator of the asymptotic variance to enable statistical inference on the target parameter and due to space constraints, the details are in Section 1 of the Supplementary Material. 
Throughout the sections, we consider the asymptotic regime where there is a sequence of expanding observation windows $A_n$ in $\R^2$ where $A_1\subset A_2\cdots \subset A_n$ and the area of $A_n$, denoted as $|A_n|$, goes to infinity.  Notationally, we use the subscript $n$ to denote the same quantities defined in the previous section. 
For example,  $\ell_n(\vtheta,\eta; X)$ denotes the pseudo-log-likelihood in \eqref{eq:pseudo-log-likelihood} by setting $A=A_n$, and $\hat\vtheta_n,\hat\vtheta^{(v)}_n,\hat\eta_{\vtheta,n}^{(v)}$ denote the sequences of estimators when $A=A_n$. For more details on expanding window asymptotics, see \citet{rathbun1994asymptotic}. 


\subsection{Consistency}\label{sec:consistency} To establish consistency of our estimator, we make the following assumption. The expression $a_n = \Theta(b_n)$ denotes that $a_n$ and $b_n$ increase at the same rate.
 \begin{assumption}[Regularity conditions for consistency] \label{assumption:regularity} \hfill 
 
      \begin{enumerate}[label=(A\arabic*)]
        \item \label{condition:smoothness}(\textit{Smoothness of $\lambda$})  The intensity function $\lambda(\vu; \vtheta,\eta)$ 
        is twice continuously differentiable with respect to $\vtheta$ and $\eta$.
         \item \label{condition:Sufficient Separation}\textit{(Sufficient separation)} There exists positive constants $c_0$ and $c_1$ such that the set $C(\vtheta,\eta) = \{\vu:|\log\lambda(\vu;\vtheta,\eta)-\log\lambda(\vu;\vtheta^*,\eta^*)|\geq \min(c_0,c_1|\vtheta-\vtheta^*|)\}$ satisfies $|C(\vtheta,\eta) \cap A_n| = \Theta(|A_n|)$ uniformly over $\vtheta$ and $\eta$. 
        \item \label{condition:boundedness}(\textit{Boundedness of $\lambda$}) 
        There exists a positive constant $c_2$ such that the size of the set $\{\vu:\inf_{\vtheta,\eta}\lambda(\vu;\vtheta,\eta)<c_2\}\cap A_n$ is bounded for all $n$.
        \item \label{condition:pair_correlation}(\textit{Weak pairwise dependence}) The pair correlation function $g(\cdot,\cdot)$ in Section \ref{sec:semi_ppp} satisfies $\int_{\R^2}|g(0,\vu)-1|\mathrm{d}\vu < \infty$.
    \end{enumerate}
 \end{assumption}

Conditions in Assumption \ref{assumption:regularity} are semiparametric extensions of regularity conditions widely used in parametric spatial point processes. 
Specifically, condition \ref{condition:smoothness} corresponds to
condition (C1) in \citet{rathbun1994asymptotic}.  
Condition \ref{condition:Sufficient Separation} corresponds to condition (A3) in \cite{schoenberg2005consistent}.
Condition \ref{condition:boundedness} 
ensures that enough points are observed in window $A_n$. Condition \ref{condition:pair_correlation} is a mild constraint on the pairwise dependence of spatial point process and is common in the literature (e.g.,(C3) in \cite{hessellund2022semiparametric}).

We also make the assumption about the estimation of the nuisance parameter.
\begin{assumption}[Consistency of estimated nuisance parameter] \label{assumption:np_condition_uniform} For every $v\in[V]$ and $i=0,1,2$, the estimated nuisance parameter $\hat\eta_{\vtheta,n}^{(v)}$ satisfies:
$$\sup_{\vtheta\in\Theta,\vz\in\mathcal{Z}}\left|\frac{\partial^i}{\partial\vtheta^i}\hat\eta_{\vtheta,n}^{(v)}(\vz)-\frac{\partial^i}{\partial\vtheta^i}\eta_{\vtheta,n}^{*}(\vz)\right|=o_p(1). $$
\end{assumption}
\noindent 

\begin{theorem}[Consistency of $\hat\vtheta_n$]\label{thm:consistency}
Suppose Assumption \ref{assumption:regularity} and \ref{assumption:np_condition_uniform} hold. Then, $\hat \vtheta_n$ is consistent for $\vtheta^*$, i.e., $\hat\vtheta_n-\vtheta^*\rightarrow_p 0$. 
\end{theorem}


\subsection{Asymptotic normality and efficiency}\label{sec:normality}
Let $S(\vtheta^*,\eta^*,\nu^*)$ and $\Sigma(\vtheta^*,\eta^*,\nu^*,g^*)$ be the sensitivity matrix and the covariance matrix, respectively, of the semiparametric spatial point process, i.e.,
\begin{align*}
S(\vtheta^*,\eta^*,\nu^*) &:= \E\left\{-\frac{\partial^2}{\partial\vtheta^\top\partial\vtheta}\ell(\vtheta,\eta_{\vtheta};X)\bigg|_{\vtheta=\vtheta^*}\right\}, \\
\Sigma(\vtheta^*,\eta^*,\nu^*,g^*) &:={\rm var}\left\{ \frac{\partial}{\partial\vtheta}\ell(\vtheta,\eta_{\vtheta};X)\bigg|_{\vtheta=\vtheta^*}\right\}.
\end{align*}

\noindent 
Note that the sensitivity matrix depends on $\nu^*$ defined in \eqref{eq:nuisance_derivative}, and the covariance matrix depends on $\nu^*$ as well as the pair correlation function $g^*$. For an observation window $A_n$, let $S_n$ and $\Sigma_n$ denote the corresponding sensitivity matrix and covariance matrix. We let $\lambda_{\text{min}}(M)$ denote the smallest eigenvalue of a matrix $M$. 

\begin{assumption}[Regularity conditions for asymptotic normality of $\hat{\vtheta}_n$.]\label{assumption:normality} To establish asymptotic normality, we make the following assumption.
    \begin{enumerate}[label=(B\arabic*)]
    \item \label{condition:nonsigularity}\textit{(Nonsingular sensitivity matrix)}
        $\liminf_n \lambda_{\min}\left\{|A_n|^{-1}{S}_n(\vtheta^*,\eta^*,\nu^*) \right\}>0.$
        \item \label{condition:nonsingular_covariance}\textit{(Nonsingular covariance matrix)} $\liminf_n \lambda_{\min}\left\{ |A_n|^{-1}{\Sigma}_n(\vtheta^*,\eta^*,\nu^*,g^*) \right\}>0.$
       \item \label{condition:alpha-mixing}(\textit{$\alpha$-mixing})  The $\alpha$-mixing coefficient  satisfies $\alpha_{2,\infty}^X(r) = O(r^{-(2+\epsilon)})$ for some $\epsilon > 0.$
    \end{enumerate}      
\end{assumption}
\noindent Condition \ref{condition:nonsigularity} and \ref{condition:nonsingular_covariance} are common in spatial point processes literature (e.g., conditions (C4) and (N3) in \citet{hessellund2022semiparametric}).  
The $\alpha$-mixing rate in condition \ref{condition:alpha-mixing} enables a version of the central limit theorem for random fields and is widely used in the study of spatial point processes (e.g., condition (v) in Theorem 1 of \citet{waagepetersen2009two}). 
A formal definition of the $\alpha$-mixing condition is in Section 5.3 of the Supplementary Material. 

In addition to the above assumption, the estimated nuisance parameter needs to be consistent at the rate $o_p\left(|A_n|^{-{1}/{4}}\right)$ 
in order for the proposed estimator to be asymptotically normal; see Assumption \ref{assumption:np_condition_pointwise}. This convergence rate mirrors the rate $o_p\left(n^{-{1}/{4}}\right)$ in i.i.d settings where $n$ is the sample size. (e.g., Assumption 3.2 in \citet{chernozhukov2018double}).

\begin{assumption}[Rates of convergence of estimated nuisance parameter] \label{assumption:np_condition_pointwise} For every $v\in[V]$ and $i=0,1$, the estimated nuisance parameter
 $\hat\eta_{\vtheta,n}^{(v)}$ in step \ref{step1} satisfies:
$$\sup_{\vz\in\mathcal{Z}}\left|\frac{\partial^i}{\partial\vtheta^i}\hat\eta_{\vtheta^*,n}^{(v)}(\vz)-\frac{\partial^i}{\partial\vtheta^i}\eta_{\vtheta^*,n}(\vz)\right| = o_p\left(|A_n|^{-\frac{1}{4}}\right)$$
\end{assumption}

\begin{theorem}[Asymptotic Normality of $\hat\vtheta_n$]\label{thm:normality} 
Suppose Assumptions \ref{assumption:regularity} - 
\ref{assumption:np_condition_pointwise} hold, the intensity function is log-linear and $\hat\eta_{\vtheta,n}^{(v)}$ is estimated by \eqref{eq:spatial_kernel_regression}. Then, $\hat{\vtheta}_n$ is asymptotically normal, i.e., 
$${S} _n(\vtheta^*,\eta^*,\nu^*){\Sigma}_n^{-\frac{1}{2}}(\vtheta^*,\eta^*,\nu^*,g^*)(\hat\vtheta_n - \vtheta^*)\rightarrow_d N(0,I).$$ 
\end{theorem}
\begin{corollary}[Asymptotic Normality and Efficiency of $\hat\theta_n$ for Poisson Spatial Point Processes] \label{corallary:Efficiency}
    Suppose conditions \ref{condition:smoothness} - \ref{condition:boundedness}, \ref{condition:nonsigularity} and Assumption \ref{assumption:np_condition_pointwise} hold, and $X$ is Poisson. Then,  $\hat{\vtheta}_n$ achieves the semiparametric efficiency bound in Theorem\ref{thm:least_favorable_direction}, i.e., 
$${S}^{\frac{1}{2}}_n(\vtheta^*,\eta^*,\nu^*)(\hat\vtheta_n - \vtheta^*)\rightarrow_d N(0,I).$$
\end{corollary}

\noindent Unlike the results for parametric spatial point processes (e.g., Theorem 1 in \citet{waagepetersen2009two}), Theorem \ref{thm:normality} requires that the process $X$ has a log-linear intensity function. The theoretical justification for achieving asymptotic normality in a non-Poisson process using the kernel regression estimator in \eqref{eq:spatial_kernel_regression} is an interesting result, which we discuss further in Section 3.8 of the Supplementary Material. 





Theorem \ref{thm:normality} implies that ${S}_n^{-1}(\vtheta^*,\eta^*,\nu^*){\Sigma}_n(\vtheta^*,\eta^*,\nu^*,g^*){S}_n^{-1}(\vtheta^*,\eta^*,\nu^*)$ is the asymptotic covariance matrix of $\hat\theta$. To obtain a consistent estimator of the asymptotic covariance matrix, we use quadrature method to approximate $S_n$ and $\Sigma_n$ and replace $\theta^*$, $\eta^*$, $\nu^*$, $g^*$ with their estimated counterparts. For more details on the procedure and a formal result that shows that the plug-in estimator of the asymptotic covariance matrix is consistent, see Section 1 of the Supplementary Material.

\section{Kernel-Based Estimator of the Nuisance Parameter}\label{sec:kernel_estimation}


While we have stated high-level conditions for the convergence rate of the estimator of the nuisance parameter to ensure consistency (Theorem \ref{thm:consistency}) and asymptotic normality (Theorem \ref{thm:normality}) of the proposed estimator of the target parameter $\hat\vtheta$, we are unaware of any existing nonparametric estimator of $\eta$ that has been shown to have these convergence rates. In this section, we propose a kernel-based estimator of the nuisance parameter. Specifically, the expectation of the pseudo-log-likelihood in \eqref{eq:pseudo-log-likelihood} is equal to the integral of the ``covariate-conditioned expectation'' in \ref{def:conditioning}:
    \begin{align*}
        \E[\ell(\vtheta,\eta;X)] =\int_{\mathcal{Z}}\E[\ell(\vtheta,\eta;X)|\vz]\mathrm{d}\vz
    \end{align*}
This equality implies $\eta_{\vtheta}$ defined in \eqref{eq:eta_theta} is the maximizer of the covariate-conditioned expectation $\E[\ell(\vtheta,\eta;X)|\vz]$ in \eqref{eq:conditioning_covariates}. 
From this observation,
we estimate $\eta_{\vtheta}(\vz)$ by
\begin{equation}\label{eq:spatial_kernel_regression}
\hat\eta_{\vtheta}(\vz)=\arg\max_{\eta(\vz)\in \R}\widehat{\E}[\ell(\vtheta,\eta;X)|\vz],\quad \vz\in\mathcal{Z},
\end{equation}
where $\widehat{\E}[\ell(\vtheta,\eta;X)|\vz]$  is a kernel estimation of $\E[\ell(\vtheta,\eta;X)|\vz]$ defined as 
\begin{equation}\label{eq:spatial_conditioinal_kernel}
    \sum_{\vu\in X\cap A}K_h\left(\vz_{\vu}-\vz\right)\log\left[\Psi\left\{\tau_{\vtheta}(\vy_{\vu}),\eta(\vz)\right\}\right] -\int_{A} K_h\left(z_{\vu}-z\right)\Psi\left\{\tau_{\vtheta}(\vy_{\vu}),\eta(\vz)\right\}\mathrm{d}\vu.
\end{equation}
The function $K_h(\vz)=h^{-q}K\left({\vz}/{h}\right)$ is the rescaled kernel function with bandwidth $h > 0$ and the function $K(\cdot)$ is a kernel function; see Section 5.1 of the Supplementary Material for a formal definition of a kernel function.




\subsection{Convergence rate of the kernel-based  estimator}\label{eq:nuisance_convergence_rate}
To characterize the rate of convergence of the proposed kernel-based estimator above, we make assumptions about the smoothness of the intensity function, the ``joint'' Radon-Nikodym derivative, and the higher-order dependence of $X$.
\begin{assumption}[Regularity conditions for the kernel regression estimator]\label{assumption:kernel_est}
For some integers $l\geq 2$ and $m\geq 2$, the following conditions hold.

    \begin{enumerate}[label=(K\arabic*)]
        \item \label{condition:higher order smoothness} (\textit{Smoothness})  $\Psi\left\{\tau_{\vtheta^*}(\vy),\eta^*(\vz)\right\}$ and $f_n(\vy,\vz)$ are $l$-times continuously differentiable with respect to $\vz$. The kernel function $K(\cdot)$ in equation \eqref{eq:spatial_kernel_regression} is an $l$-th order kernel.
        
        \item \label{condition:kernel identification}(\textit{Identification}) 
        $\liminf_n \inf_{\vtheta,\vz}|A_n|^{-1}\partial^2/\partial\eta^2\E\left\{\ell_n(\vtheta,\eta_{\vtheta,n})|\vz\right\}>0$
        \item \label{condition:cumulant}(\textit{Higher-order weak dependence}) There exists a positive constant $C$ such that the factorial cumulant functions of $X$, denoted as $Q_{m}$, satisfy $$\sup_{\vu_1\in\R^2}\int_{\R^2}\cdots\int_{\R^2}\left|Q_{m^\prime}(\vu_1,\ldots,\vu_{m^\prime})\right|\mathrm{d}\vu_2\ldots \mathrm{d}\vu_{m^\prime}<C,\quad{} m^\prime = 2,3,\ldots,m. $$       
    \end{enumerate} 
\end{assumption}
\noindent 
Condition \ref{condition:higher order smoothness} corresponds to the usual smoothness condition in kernel regression (e.g., \citet[chap. 1.11]{li2023nonparametric}). Condition \ref{condition:kernel identification} guarantees that $\eta_{\vtheta,n}$ is unique.
Condition \ref{condition:cumulant} assumes $m$-th order weak dependence between points. 
If $X$ is Poisson, $Q_m(\vu_1,\ldots,\vu_m)=0$ for every $m$ if at least two of $\vu_1,\ldots,\vu_m$ are different. 
We remark that some works in the spatial point process literature assume a stronger $m\geq 4$ (e.g., equation (6) in \cite{guan2007thinned}). A detailed discussion of the factorial cumulant function is in Section 5.4 of the Supplementary Material.


\begin{theorem}\label{thm:kernel_estimation}
    Suppose Assumptions \ref{assumption:regularity} and \ref{assumption:kernel_est} hold. 
    Then, with an appropriately chosen bandwidth parameter (see Section 3.7 in the Supplementary Material), the estimator $\hat\eta_{\vtheta,n}(\vz)$ 
    in \eqref{eq:spatial_kernel_regression}
    and its first two derivative (i.e., $i=0,1,2$) satisfy:
\begin{equation}\label{eq:nuisance_est_thm}
\sup_{\vz\in\mathcal{Z}}\left|\frac{\partial^i}{\partial \vtheta^i}\left\{\eta^*_{\vtheta,n}(\vz)-\hat\eta_{\vtheta,n}(\vz)\right\}\right| =   o_p\left(|A_n|^{-\frac{m}{2(m+ k+q+1)}\frac{l}{l+q+1}}\right).
\end{equation}
\end{theorem}
When the dimensions of the target parameter ($k$) and the nuisance covariates ($q$) are large, we need weaker higher-order dependence (i.e., high $m$) and greater smoothness (i.e., high $l$) to ensure that the nuisance parameter converges at the necessary rate of $o_p(|A_n|^{-\frac{1}{4}})$, as defined in Assumptions \ref{assumption:np_condition_uniform} and \ref{assumption:np_condition_pointwise}. 
Also, when $X$ is Poisson process, $m \to \infty$ and so long as
the intensity function is very smooth such that $l \to \infty$, the convergence rate can be made arbitrarily close to $O_p(|A_n|^{-\frac{1}{2}})$. 
\noindent 

\section{Computational Method}\label{sec:computation}
 Directly maximizing the pseudo-log likelihood in equation \eqref{eq:pseudo-log-likelihood} for estimation is numerically intractable because of the integral.
 In this section, we propose a numerical, quadrature approximation 
to not only approximate this integral, but more importantly, to enable investigators to directly use existing R packages initially designed for semiparametric models 
from i.i.d. data. 

Let $\{\vu_j\}_{j=1}^{N}$ be the union of the data points in $X\cap A$ and points from a uniform grid on $A$. We use Dirichlet tessellation \citep{green1978computing} to generate the weight of each point $u_j$, denoted as $w_j > 0$, and approximate the pseudo-log-likelihood in \eqref{eq:pseudo-log-likelihood} by
\begin{align}
    \ell(\vtheta,\eta; X) 
    & \approx  \sum_{\vu\in X\cap A}\log\lambda(
    \vu;\vtheta,\eta) - \sum_{j=1}^N w_j \lambda(\vu_j;\vtheta,\eta)\nonumber\\
    \label{eq:pseudo_approximation_2}
    & \approx \sum_{j=1}^N (y_j\log\lambda_j- w_j \lambda_j), \quad{}  \lambda_j =  \lambda(\vu_j;\vtheta,\eta), \ y_j = \ind(\vu_j\in X\cap A).  
\end{align}

\noindent The right-hand side of \eqref{eq:pseudo_approximation_2} is equivalent to the log-likelihood of a weighted, semiparametric Poisson regression model of the form $y_j \overset{\mathrm{iid}}{\sim} \text{Poi}(\lambda_j)$ with weights $w_j$. Thus, we can use any existing methods developed for generalized partial linear models with i.i.d. data to estimate $\eta_{\vtheta}$ in step \ref{step1}. In our numerical experiments below, we use the R packages \texttt{mgcv} \citep{wood2001mgcv} and \texttt{gplm} \citep{Mueller2016gplm}. By default, \texttt{mgcv} uses penalized splines regression to estimate the nonparametric portion of the generalized partially linear model and \texttt{gplm} uses kernel regression to estimate the nonparametric portion.

\begin{remark}[Approximation based on logistic regression]
    \citet{baddeley2014logistic} proposed an alternative approximation of the pseudo-log-likelihood function based on logistic regression.
    Between the two approximations, \citet{baddeley2014logistic} generally found that the logistic approximation is less biased at the expense of conservative standard errors. We also observe this pattern for semiparametric spatial point processes.
    For further discussions, see Section 4.1 of the Supplementary Material. 
\end{remark}
\begin{remark}[Boundary bias of the kernel estimator in \texttt{gplm}] In our extended simulation studies (see Section 4.2 of the Supplementary Material), we found that the kernel estimator from \texttt{gplm} 
can sometimes be biased near the boundary of the support of the nuisance covariates. This is a well-recognized, finite-sample phenomenon among kernel-based estimators, and several solutions exist to mitigate this bias  (e.g., \citet{jones1993simple,racine2001bias}). In contrast, we generally found that the the penalized spline estimator from the R package \texttt{mgcv} suffers less from boundary-related biases. 
\end{remark}

\section{Simulation and Real Data Analysis}\label{section:data_analysis}

\subsection{Poisson spatial point processes}\label{subsection:simulation}

In this section, we evaluate the finite-sample performance of our method through simulation studies.  Our first simulation study considers Poisson spatial point processes with the following intensity function: $\lambda(\vu;\theta,\eta) = 400\exp\{\theta^* y_{\vu}+\eta(z_{\vu})\}$ and $\theta^* = 0.3$. 
The simulation settings vary: (a) the observational window $W_a$ with size $a\times a$ and $a=1,2$; (b) the true nuisance function from a linear function (i.e., $\eta^*(z)=0.3z$) to a non-linear function (i.e., $\eta^*(z)=-0.09z^2$); and (c) the target covariate $y_{\vu}$ and the nuisance covariate $z_{\vu}$ where they are either independent or dependent Gaussian random fields. 
We use the logistic approximation for numerical evaluations of the pseudo-log-likelihood; Section 4.1 of the Supplementary Material shows the results under the quadrature approximation of the pseudo-log-likelihood. We repeat the simulation 1000 times.

\begin{table}[t]
\centering
{\small
\begin{tabular}{cccccccc}
\toprule
 Window & Cov & Nuisance & $\text{Bias}_{\times 100}$ & rMSE & meanSE & CP90 & CP95 \\
\midrule
\multirow{4}{*}{$W_1$} & \multirow{2}{*}{ind} & linear & -0.2477 & 0.0467 & 0.0459 & 89.8 & 93.5 \\
 &  & poly & -0.0438 & 0.0473 & 0.0476 & 91.1 & 94.7 \\\cmidrule(lr){2-8}
 & \multirow{2}{*}{dep} & linear & -0.1318 & 0.0438 & 0.0449 & 91.2 & 95.2 \\
 &  & poly & -0.7548 & 0.0531 & 0.0532 & 89.3 & 95.3 \\\cmidrule(lr){1-8}
 \multirow{4}{*}{$W_2$} & \multirow{2}{*}{ind} & linear & 0.0317 & 0.0239 & 0.0238 & 89.3 & 94.4 \\
 &  & poly & -0.0158 & 0.0249 & 0.0254 & 91.2 & 95.7 \\\cmidrule(lr){2-8}
 & \multirow{2}{*}{dep} & linear & 0.1089 & 0.0232 & 0.0236 & 90.2 & 95.8 \\
 &  & poly & -0.0219 & 0.0266 & 0.0275 & 91.3 & 96.2 \\
\bottomrule
\end{tabular}}

\caption{\small Simulation results for Poisson spatial point processes. $W_1$ and $W_2$ represent small and large observation windows, respectively. Cov is the type of spatial covariates where ``ind'' stands for independent covariates and ``dep'' stands for dependent covariates.  Nuisance is the type of nuisance parameter (i.e. linear function or non-linear function). Bias is the bias of the estimator of $\vtheta^*$ multiplied by $100$. rMSE is the root mean square error of the estimator. meanSE is the mean of the estimated standard errors of the estimator. CP90 and CP95 are the empirical coverage probabilities of 90\% and 95\% confidence intervals, respectively.}
\label{tab:simulation_Poisson}
\end{table}
Table \ref{tab:simulation_Poisson} summarizes the results.
Our method is close to unbiased and  
the standard errors are approximately halved when the observation window increases from $W_1$ to $W_2$. 
Also, the coverage probabilities of the confidence intervals are close to nominal levels.

\subsection{Log-Gaussian Cox processes}

We also simulate log-Gaussian Cox processes (LGCP) with the conditional intensity function: 
$\Lambda(\vu;\theta,\eta) = 400\exp\left\{\theta^* y_{\vu}+\eta(z_{\vu})+G(\vu)-2/\sigma^2\right\}.  $
The term $G(\vu)$ is a mean-zero, Gaussian random field with the covariance function $C(r) = \sigma^2\exp(-r/0.2),\sigma^2=0.2$. We also consider two additional settings, the case where the pair correlation function function is known and the case where the pair correlation function is estimated using minimum constrast method in \citet{waagepetersen2009two}. The rest of the settings are identical to Section \ref{subsection:simulation}. 

\setlength{\tabcolsep}{3pt} 
\begin{table}[t]
\centering
{\small
\begin{tabular}{ccccccccccccc}
\toprule
 Window & Cov & Nuisance & $\text{Bias}_{\times 100}$ & rMSE & meanSE & meanSE* & CP90 & CP90* & CP95 & CP95* \\
\midrule
\multirow{4}{*}{$W_1$} & \multirow{2}{*}{ind} & linear & -0.3824 & 0.0628 & 0.0559 & 0.0610 & 86.4 & 88.7 & 91.5 & 94.5 \\
& & poly & 0.0534 & 0.0582 & 0.0573 & 0.0628 & 88.9 & 92.3 & 95.0 & 97.0 \\ \cmidrule(lr){2-11}
& \multirow{2}{*}{dep} & linear & -0.0493 & 0.0598 & 0.0547 & 0.0593 & 86.5 & 89.2 & 91.7 & 94.9 \\
 &  & poly & -0.5999 & 0.0673 & 0.0611 & 0.0662 & 86.1 & 89.0 & 91.9 & 94.8 \\ \midrule
\multirow{4}{*}{$W_2$} & \multirow{2}{*}{ind} & linear & 0.0645 & 0.0304 & 0.0306 & 0.0311 & 89.8 & 90.7 & 95.0 & 96.1 \\ 
 &  & poly & -0.1370 & 0.0325 & 0.0311 & 0.0317 & 89.4 & 90.4 & 93.6 & 94.6 \\ \cmidrule(lr){2-11}
 & \multirow{2}{*}{dep} & linear & 0.1506 & 0.0312 & 0.0298 & 0.0303 & 88.7 & 88.5 & 93.3 & 93.8 \\
 &  & poly & -0.2485 & 0.0328 & 0.0327 & 0.0332 & 89.8 & 90.2 & 94.3 & 94.9 \\
\bottomrule
\end{tabular}}
\caption{\small Simulation results for log-Gaussian Cox processes. 
$^{*}$ represents the case where the pair correlation function is known, and the absence of $^*$ represents the case where the pair correlation function is estimated.}
\label{tab:simulation_LGCP}
\end{table}
\setlength{\tabcolsep}{6pt} 

Table \ref{tab:simulation_LGCP} summarizes the results. Similar to Table \ref{tab:simulation_Poisson}, our method is nearly unbiased and the standard errors are approximately halved when the observation window is increased from $W_1$ to $W_2$. The coverage probabilities of the confidence intervals based on the true pair correlation function are close to the nominal levels. 
However, the confidence intervals based on the estimated pair correlation function can undercover when the observation window is small. The coverage does improve
as the observation window expands due to improvement in the estimated pair correlation function. Overall and similar to Section \ref{subsection:simulation}, the simulation study reaffirms that in finite sample, statistical inference based on the asypmtotic results in Section \ref{section:theory} is reasonable, especially for large observational windows.


\subsection{Model mis-specification} \label{subsection:sim_estimator_compare}
We also conduct a simulation study comparing different estimators of $\vtheta^*$.  
We focus on the case where the covariates are dependent, the true nuisance parameter is a non-linear function, and for LGCP, the pair correlation function is estimated. We compare three different estimators: (a) our proposed estimator where $\eta(z)$ is unknown and unspecified, (b) an estimator based on a parametric spatial point process where $\eta(z)$ is mis-specified as a linear function of $z$, and (c) an oracle estimator where $\eta(z)$ is known a priori. We repeat the simulation 1000 times.

Table \ref{tab:simulation} summarizes the results. Compared to the misspecified parametric estimator, our semiparametric estimator produces nearly unbiased estimates. Also, our estimator performs as well as the oracle estimator with respect to bias, mean squared error, and coverage probability. In other words, even without knowing the nuisance parameter a priori, our estimator can perform as well as the oracle that knows the nuisance parameter and is much better than a mis-specified parametric estimator.   

\begin{table}[t]
\centering
{\small
\begin{tabular}{cccccccc}
    \toprule
       Window & {Process} & {Estimator} & {$\text{Bias}_{\times 100}$} & {rMSE} &{meanSE} & {CP90} & {CP95} \\
    \midrule
    \multirow{6}{*}{$W_1$} & \multirow{3}{*}{Poisson}
    & Semi & -0.7548 & 0.0531 & 0.0532 & 89.3 & 95.3 \\
    & &Para &-4.1990 &0.0633 &0.0495 &78.7 & 86.3 \\
    & &Oracle & $0.1458$ & $0.0523$ & $0.0522$ & $90.5$ & $94.7$ \\  \cmidrule(lr){2-8}
    & \multirow{3}{*}{LGCP}
    & Semi & -0.6000 & 0.0673 & 0.0611 & 86.1 & 91.9  \\
    & &Para  &-4.3380&0.0726&0.0597& 82.2 & 89.3 \\
    & &Oracle & -0.2632 & 0.0644 & 0.0618 & 89.2 & 93.5 \\
    \hline
    \multirow{6}{*}{$W_2$} & \multirow{3}{*}{Poisson}
    & Semi & -0.0219 & 0.0266 & 0.0275 & 91.3 & 96.2 \\
    & &Para &-2.5624 &0.0350& 0.0257 &75.4 & 85.8 \\
    & &Oracle & $0.2058$ & $0.0268$ & $0.0275$ & $91.2$ & $95.9$  \\ \cmidrule(lr){2-8}
    & \multirow{3}{*}{LGCP}
    & Semi &  -0.2485 & 0.0328 & 0.0327 & 89.8 & 94.3 \\
    & &Para &-3.5219 & 0.0457 & 0.0316 & 70.5 &81.6 \\
    & &Oracle & -0.6360 & 0.0331 & 0.0327 & 88.8 & 94.6\\
    \bottomrule
\end{tabular}}
\caption{\small
Simulation results for comparing different estimators of $\theta$. Process is either the Poisson spatial point process (``Poisson'') or the log-Gaussian Cox process (``LGCP'') Estimator stands for different estimators where ``Semi'' is our estimator, ``Para'' is  the estimator based on a misspecified model of the nuisance function, and  ``Oracle'' is the estimator that knows the nuisance function a priori.}
\label{tab:simulation}
\end{table}

\subsection{Data analysis: Forestry}

\begin{figure}[t]
\centering
\includegraphics[width=0.9\linewidth]{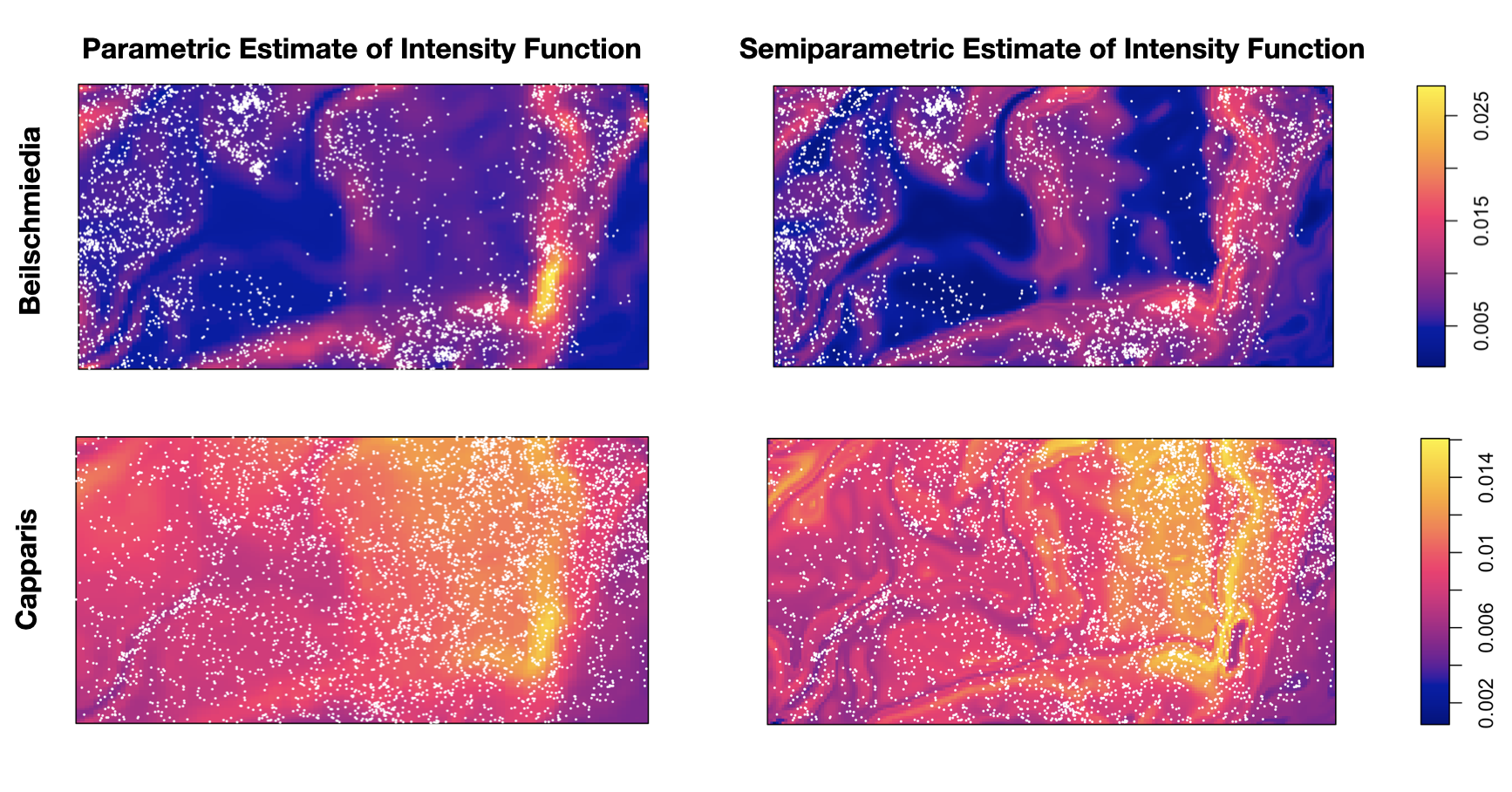}
\caption{Heat maps of the estimated intensity functions for two tree species, 
\textit{Beilschmiedia pendula} and \textit{Capparis frondosa}. Each column represents different estimators and each row represents different tree species. The white points in each plot are the observed locations of the tree species.}
\label{fig:trees_fitted}
\end{figure}

We apply our method to estimate the partial effect of elevation (i.e., the target covariate $\vy_\vu$) on the spatial distribution of two species of trees, 
\textit{Beilschmiedia pendula} and \textit{Capparis frondosa}, while nonparametrically adjusting for the gradient of the forest's topology (i.e., the nuisance covariate $\vz_\vu$). The dataset consists of a 1000 × 500 meters plot from a tropical rainforest on Barro Colorado Island; see \citet{condit1996changes}
for details. The points on Figure \ref{fig:trees_fitted} show the location of the tree species.

\begin{figure}[!t]
    \centering
    \includegraphics[width=0.5\linewidth]{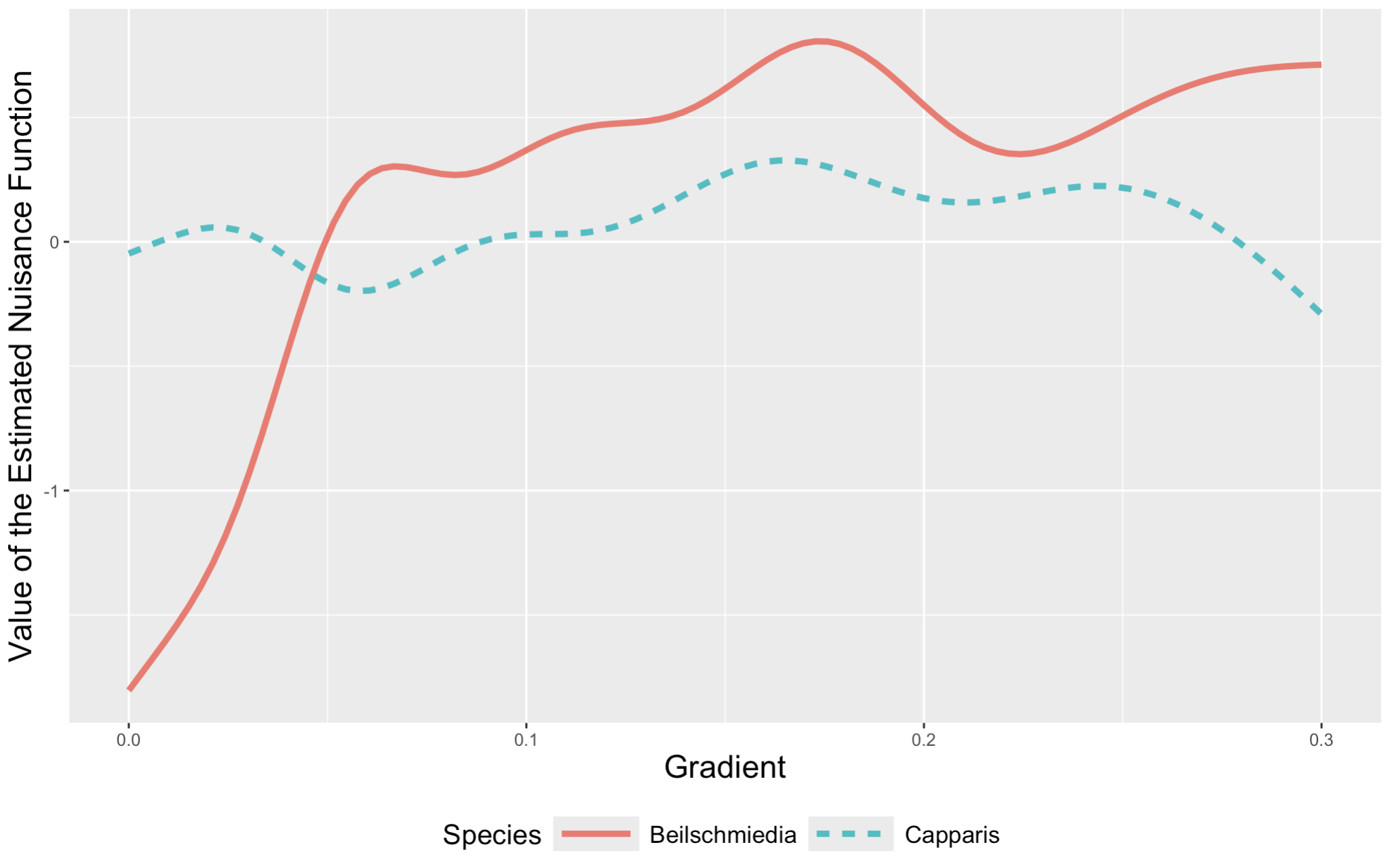}
    \caption{Estimated nuisance functions for \textit{Beilschmiedia pendula} (solid line) and \textit{Capparis frondosa} (dashed line).  
    }
    \label{fig:nuisance_estimation}
\end{figure}

For \textit{Beilschmiedia pendula}, the estimated partial effect of elevation from our semiparametric estimator is 3.136 (95\% CI: (-1.470, 7.741)) whereas the estimate from the parametric approach discussed in Section \ref{subsection:sim_estimator_compare} is 2.144 (95\% CI: (-2.453, 6.741)). To understand the difference between the two estimates, we plotted the nonparametrically estimated nuisance function in Figure \ref{fig:nuisance_estimation} and found that there is a non-linear relationship between the gradient and the spatial locations of \textit{Beilschmiedia pendula}; see Figure \ref{fig:nuisance_estimation}. From this, we suspect that the
inference based on the parametric model may suffer from bias due to model mis-specification of the nuisance parameter. We also remark that the estimated intensity map with our estimator leads to a better visual fit of the data compared to that from the parametric estimator, especially in the upper left corner of the observation window; see the top row of Figure \ref{fig:trees_fitted}.

For \textit{Capparis frondosa}, the estimated partial effect of elevation from our estimator is 2.555 (95\% CI: (-0.046, 5.151)) whereas the estimate from the parametric approach is 2.692 (95\% CI: (0.038, 5.347)). Upon examining the nonparametrically estimated nuisance function, we found that the gradient had a flat, negligible effect on the spatial distribution of \textit{Capparis frondosa}; see Figure \ref{fig:nuisance_estimation}. This explains the similarity between our estimator and the parametric estimator for this species. However, visually speaking, the estimated intensity function from our method captures more details in the observation window than that from the parametric estimator, especially in the bottom right of the observation window; see the bottom row of Figure \ref{fig:trees_fitted}.

\subsection{Data analysis: PFAS}


Per- and polyfluoroalkyl substances (PFAS) are synthetic chemicals that pose a growing risk to ecosystems and human health due to their widespread environmental presence \citep{fenton2021per,evich2022per}. While regression-based methods like Gaussian processes \citep{wiecha2025two} and random forests \citep{deluca2023using} have been effective in small, regional studies, recent research shows they are less effective at a national scale. In contrast, \cite{park2024statistical} demonstrated that spatial point processes are more suitable for national-scale PFAS analysis.


We apply our method to data from \cite{park2024statistical} and estimate the effect of potential PFAS sources (i.e., treating distances to potential sources as target covariates $\vy_u$) on PFAS contamination, specifically 
the sum of perfluorooctane sulfonic acid (PFOS) and perfluorooctanoic acid (PFOA), while nonparametrically adjusting for the socio-economic factors (i.e., including population density, median household income, and percentage of the population with a Bachelor's degree as nuisance covariates $\vz_\vu$).

\begin{figure}[!t]
    \centering
\includegraphics[width=0.9\linewidth]{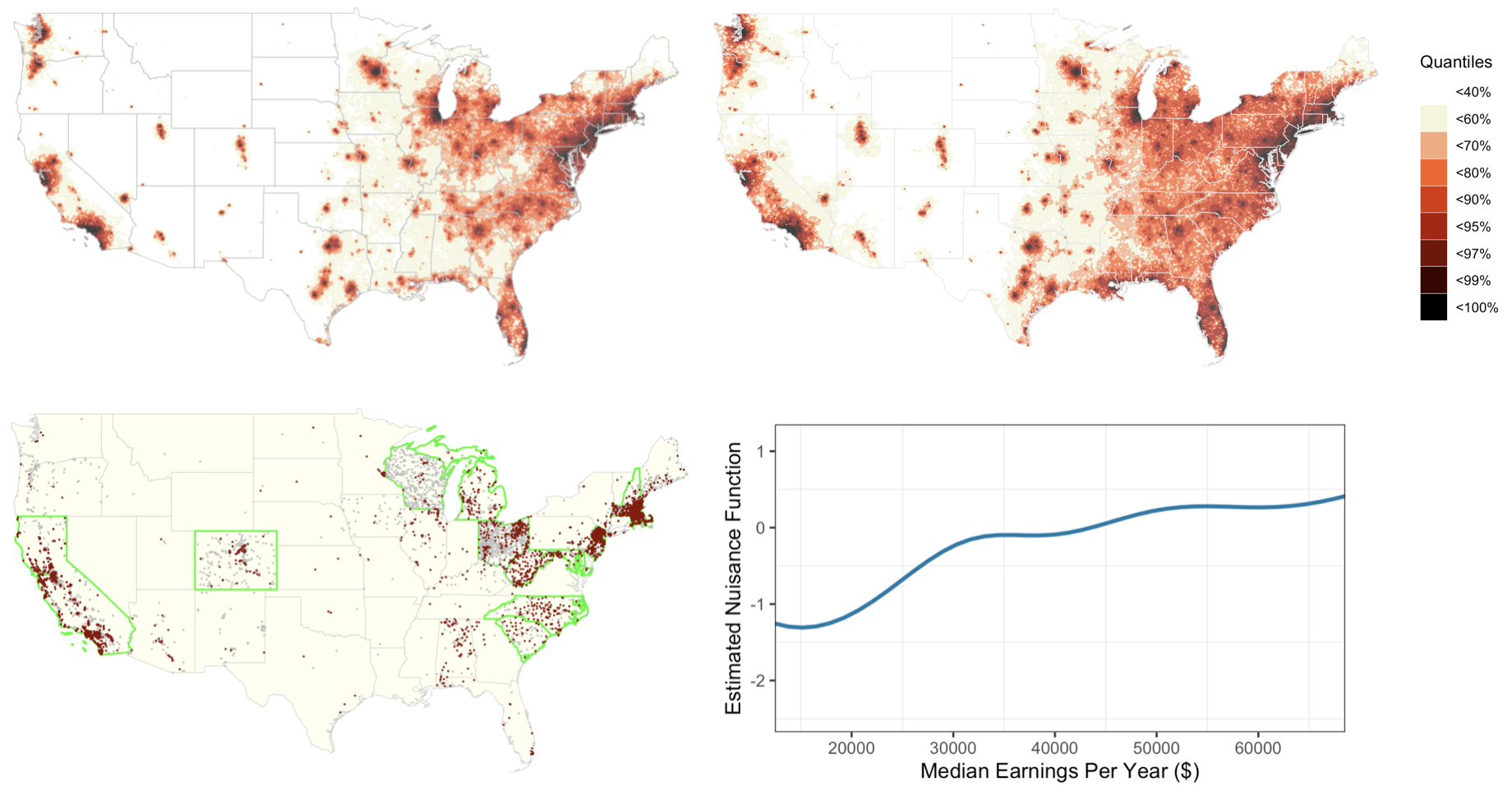}
    \caption{Top: quantile maps of fitted intensity using parametric model (left), and  semiparametric model (right). Bottom left: measurenments of PFAS; red points mark PFAS higher than 8 ppt; green boundary hightlights states with sufficient measurements. Bottom right: nonlinear effect of median earning.}
    \label{fig:PFAS}
\end{figure}

The U.S. Environmental Protection Agency (EPA) enforcement standard defines the maximum contamination level for each PFAS substance, including PFOS and PFOA, to be 4 parts per trillion (ppt). Thus, we define any location with the sum of PFOS and PFOA concentrations higher than 8 ppt as an event in a spatial point process. We fit both parametric and semiparametric models using PFAS observations from a selection of states (see Figure \ref{fig:PFAS}; additional details are in  \cite{park2024statistical}). 

\begin{table}[t]
\centering
\begin{tabular}{l cc cc}
\toprule
\multirow{2}{2cm}{Coefficient}
 & \multicolumn{2}{c}{{Semiparametric IPP}} & \multicolumn{2}{c}{{Parametric IPP}} \\
\cmidrule(lr){2-3} \cmidrule(lr){4-5}
 & {Estimate} & {S.E.} & {Estimate} & {S.E.} \\
\midrule
DistanceToAirport      & -0.1837 & (0.0224) & -0.1990 & (0.0222) \\
DistanceToMilitaryBase & -0.2067 & (0.0216) & -0.2230 & (0.0209) \\
DistanceToLandfill     & -0.1978 & (0.0209) & -0.2143 & (0.0211) \\
DistanceToApparel      & -0.0455 & (0.0198) & -0.0896 & (0.0190) \\
DistanceToFurniture    & -0.2582 & (0.0210) & -0.2142 & (0.0209) \\
\bottomrule
\end{tabular}
\caption{\small Statistical inference results for PFAS sources fitting parametric and semiparametric IPP models}
\label{tab:PFAS_inference}
\end{table}



The results of statistical inference for the target covariates, i.e., PFAS sources, are summarized in Table \ref{tab:PFAS_inference}. The estimated partial effect of \texttt{DistanceToApparel} from the semiparametric model is very different from that of the parametric model. Figure \ref{fig:PFAS} reveals a non-linear relationship between median household income and PFAS intensity. While PFAS intensity increases with median household income, the rate of increase slows as income gets higher. Since median household income is correlated with \texttt{DistanceToApparel} (correlation=-0.29), we suspect that the inference for \texttt{DistanceToApparel} in the parametric model may suffer from bias due to the misspecification of the effect of median household income.

\begin{figure}[htbp]
    \centering
    \includegraphics[width=0.95\linewidth]{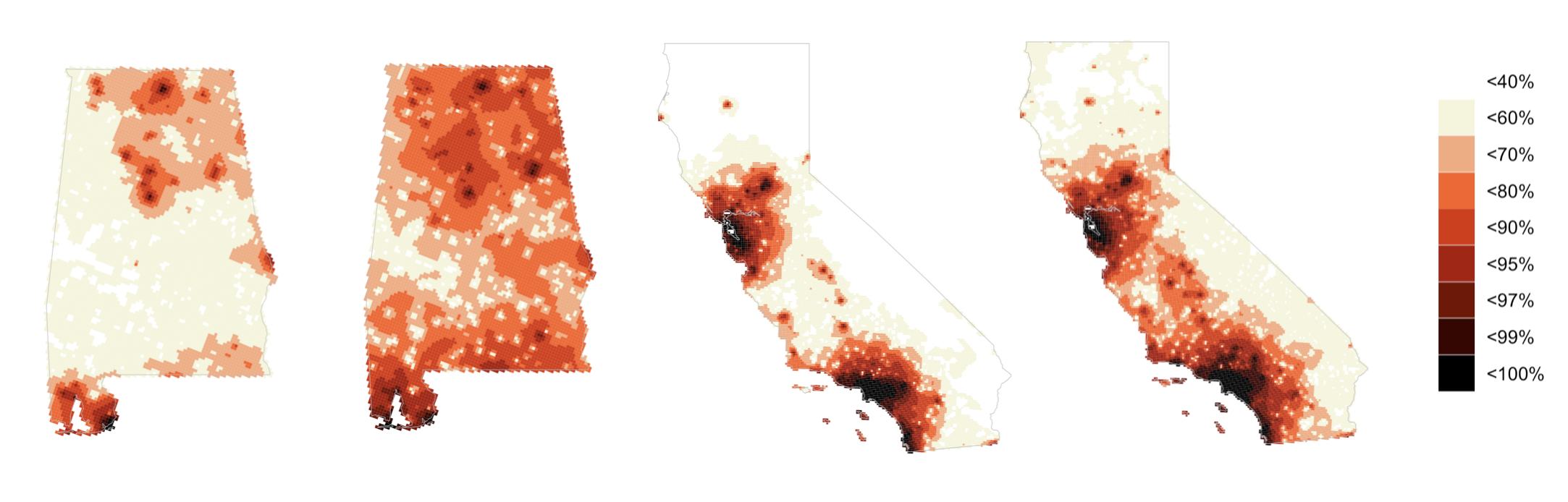}
    \caption{Quantile maps of fitted PFAS intensity in Alabama (left) and California (right).}
    \label{fig:Zoom_PFAS_Alabama_California}
\end{figure}

Figure \ref{fig:PFAS} displays quantile maps of the fitted PFAS intensity from the parametric and semiparametric models. A visual comparison shows that both models correctly identify areas with known PFAS risks and capture the broad spatial trends of contamination. 
However, as shown in Figure \ref{fig:Zoom_PFAS_Alabama_California}, the parametric model fails to identify transitional, middle-risk areas such as those in central Alabama and California.



 

\section{Discussion} \label{sec:discussion}

The paper introduces semiparametric spatial point process model featuring an intensity function with both parametric component and nonparametric component. We generalized the concept of the semiparametric efficiency lower bound in i.i.d. settings to spatial point processes. We proposed a double machine learning estimator of the parametric component that is shown to be consistenct, asymptotic normal and semiparametrically efficient. 
We also proposed a spatial kernel regression, which can achieve the desired rates of convergence for inference. 
In our simulation study, the proposed estimator is shown to exhibit the asymptotic properties in finite samples.

As laid out in Section \ref{section:theory}, 
There are some remaining challenges when $X$ is not Poisson and the intensity function is not log-linear. We conjecture that establishing asymptotic normality of $\hat{\vtheta}$ is possible under additional weak dependence restrictions on $X$ and the pair correlation function. For establishing the semiparametric efficiency lower bound, we believe that directly utilizing the
classical definition of the semiparametric efficiency lower bound (i.e. likelihood-based parametric submodels and taking the supremum of Cramer-Rao lower bounds) is insufficient. Instead, we believe either restricting the type of estimators (e.g., linear estimators) or developing a new efficiency criterion is most promising; see \citet{park2022efficient} for a related discussion under network dependence. 
Overall, we believe our work presents a new perspective on studying semiparametric point processes, and we hope the techniques in the paper are useful to both researchers in point processes and in semiparametric inference.

\section*{Funding }
This work is supported in part by the Groundwater Research and Monitoring Program from the Wisconsin Department of Agriculture, Trade and Consumer Protection.

\section*{Acknowledgment}
We would like to thank Chris Geoga, Matthias Katzfuss, Debdeep Pati, Rasmus Waagepetersen and the participants of the Online Seminar on Spatial and Spatio-temporal Point processes on March 19, 2025 for their feedback.


\bibliographystyle{abbrvnat}
\bibliography{reference}

\newpage  
\setcounter{page}{1}
\begin{center}
\Large 
    \textbf{Supplementary Material for ``Semiparametric Spatial Point Processes''}
\end{center}

The Supplementary Material includes (1) asymptotic covariance estimation, (2) technical proofs, (3) additional simulation results.

\setcounter{section}{0}
\setcounter{theorem}{0}

\renewcommand{\theequation}{A.\arabic{equation}}
\setcounter{equation}{0}

\section{Estimation of Asymptotic Covariance}

\subsection{Consistent estimation of the asymptotic covariance matrix }\label{sec:var_est}
An immediate implication of Theorem \ref{thm:normality} is that the asymptotic covariance matrix is 
$${S}_n^{-1}(\vtheta^*,\eta^*,\nu^*){\Sigma}_n(\vtheta^*,\eta^*,\nu^*,g^*){S}_n^{-1}(\vtheta^*,\eta^*,\nu^*).$$
To obtain a consistent estimator of the asymptotic covariance matrix, we use quadrature method to approximate $S_n$ and $\Sigma_n$ and replace $\theta^*$, $\eta^*$, $\nu^*$, $g^*$ with their estimated counterparts.

For the quadrature approximation, we let $\{\vu_j,w_j\}_{j=1}^N$ be a set of quadrature points and weights in the spatial domain $A_n$. Then, we can estimate the sensitivity matrix $S_n$ and the covariance matrix $\Sigma_ n$ that make up the asymptotic covariance matrix of $\hat{\vtheta}_n$ as follows: 

\textcolor{black}{\begin{align*}
    \widehat{S}_n(\vtheta^*,\eta^*,\nu^*) 
    =& \sum_{j=1}^{N}w_j \left\{\frac{\partial}{\partial\vtheta}\lambda(\vu_j;\vtheta,\eta_{\vtheta})\big|_{\vtheta = \vtheta^*}\right\}^{\otimes 2}\left\{\lambda(\vu_j;\vtheta^*,\eta^*)\right\}^{-1},  
    \\
\widehat{\Sigma}(\vtheta^*,\eta^*,\nu^*,g^*)
    =
    &\sum_{i,j=1}^{N}w_iw_j\left\{\frac{\partial}{\partial\vtheta}\lambda(\vu_i;\vtheta,\eta_{\vtheta})\big|_{\vtheta = \vtheta^*}\right\}\left\{ \frac{\partial}{\partial\vtheta}\lambda(\vu_j;\vtheta,\eta_{\vtheta})\big|_{\vtheta = \vtheta^*}\right\}^{\top}\{g(\vu_i,\vu_j)-1\}\\
    &+\widehat{S}_n(\vtheta^*,\eta^*,\nu^*).
\end{align*}}

If we have consistent estimators of $\theta^*$, $\eta^*$, $\vu^*$, $g^*$, we can replace the true parameters in the above formula with their estimated counterparts to derive a consistent estimation of the covariance matrix of $\hat\theta_n$. Specifically, the target parameter $\vtheta^*$ is replaced with  $\hat\vtheta$. The nuisance parameter $\eta^*$ is replaced with $$V^{-1}\sum_{v=1}^V\hat\eta_{\vtheta}^{(v)}\big|_{\vtheta=\hat{\vtheta}}.$$  
The estimation of pair correlation function $g^*$ for inhomogeneous point processes has been well studied in the literature \citep{baddeley2000non,illian2008statistical,waagepetersen2009two,jalilian2019orthogonal,xu2020nonparametric}. In our paper, we propose to estimate $g^*$ using the minimum contrast method detailed in \citet{waagepetersen2009two} . Specifically, we assume the pair correlation function is reweighted, isotropic, and characterized by a finite dimensional parameter $\psi$ such that $g^*(\cdot) = g(\cdot;\psi^*)$. The parameter for the pair correlation function, $\psi^*$, is then estimated by: 
\begin{equation}\label{eq:K_est}
  \hat\psi= \arg\min_{\psi \in \reals} \int_{a}^{b} \left\{K(r;\psi)-\hat K(r)\right\}^2 dr, \quad{}
\hat K(r)=\sum_{\vu,v\in X\cap A}^{\neq}\frac{I(\|\vu-v\|\leq r)}{\lambda(\vu;\hat\theta,\hat\eta)\lambda(\vu;\hat\theta,\hat\eta)|A|}. 
\end{equation}
Here, $a$ and $b$ are the range of the distances between points in $X\cap A$.  \cite{waagepetersen2009two} showed that $\hat\psi$ from \eqref{eq:K_est} is consistent under very mild conditions. 
Theorem \ref{thm:least_favorable_direction} provides the explicit form of $\nu^*$ using the ``covariate-conditioned expectation'' in Definition \ref{def:conditioning}. We let $\hat\nu$ to be the estimation of $\nu^*$ given by
\begin{equation}\label{eq:lfc_est}
    \hat\nu(\vz) = -\left[\frac{\partial^2}{\partial\eta^2}\widehat{\E}\left\{\ell(\hat\vtheta,\hat\eta;X)\mid \vz\right\}\right]^{-1}\frac{\partial^2}{\partial\vtheta\partial\eta}\widehat{\E}\left\{\ell(\hat\vtheta,\hat\eta;X)\mid \vz\right\}.
\end{equation}
where the estimation of covariate-conditioned expectation is given in \eqref{eq:spatial_conditioinal_kernel}.  Theorem \ref{thm:variance_estimation} shows our estimation of asymptotic covariance is consistent.
\begin{theorem}[Consistent Estimator of Asymptotic Covariance Matrix] \label{thm:variance_estimation}
Suppose Assumptions \ref{assumption:regularity}  -  \ref{assumption:np_condition_pointwise} hold. Then, we have 
\begin{align*}
&\hat{S}_n^{-1}(\hat\vtheta_n,\hat\eta_n,\hat{\nu}_n){\Sigma}_n(\hat\vtheta_n,\hat\eta_n,\hat{\nu}_n,\hat\psi_n)\hat{S}_n^{-1}(\hat\vtheta_n,\hat\eta_n,\hat{\nu}_n) \\ 
=& {S}_n^{-1}(\vtheta^*,\eta^*,\nu^*){\Sigma}_n(\vtheta^*,\eta^*,\nu^*,\psi^*){S}_n^{-1}(\vtheta^*,\eta^*,\nu^*)+o_p(|A_n|^{-1})
\end{align*}
\end{theorem}

\noindent 

\noindent Theorem \ref{thm:variance_estimation} does not require that $X$ is a Poisson spatial point process or that the intensity function is log-linear.  In other words, if investigators simply wish to characterize the uncertainty of the proposed estimate $\hat{\vtheta}$ with standard errors, $X$ does not have to be Poisson. 

If $X$ is a Poisson spatial point process, the pair correlation function becomes $g(u,v;\psi^*) = 1$ and consequently, ${S}^{-1}(\vtheta^*,\eta^*,\nu^*)$ is the asymptotic covariance matrix of $\hat{\vtheta}$. Therefore, we can simply use $\widehat{S}^{-1}_n(\hat\vtheta_n,\hat\eta_n,\hat{\nu}_n)$ as the estimator of asymptotic covariance and 
we do not need to estimate $\psi^*$.


\section{Proof of Main Theorem}

\subsection{Proof of Semiparametric Efficiency Lower Bound
(Theorem \ref{thm:least_favorable_direction} in main manuscript)
}

Let $L_2(A)$ be the linear space consisting of all bounded functions from $A$ to $\R$, and let $L_2^{(k)}(A)$ be its $k$-product space, i.e., $L_2^{(k)}(A) =L_2(A)\times \ldots L_2(A)$. For any $f,g\in L_2^{(k)}(A)$, we define their inner-product as
\begin{equation}\label{eq:inner_product}
    \langle f,g\rangle_A := \int_A \lambda(\vu;\vtheta^*,\eta^*)f(\vu)^\top g(\vu) \mathrm{d}\vu.
\end{equation}
We define their outer-product as
\begin{equation}\label{eq:outer_product}
   \left|f\rangle \langle g \right|_A := \int_A \lambda(\vu;\vtheta^*,\eta^*)f(\vu)g(\vu)^\top \mathrm{d}\vu.
\end{equation}
For brevity, we further denote 
\begin{gather*}
    f_1(\vu):=\frac{\partial}{\partial\vtheta}\log\lambda(\vu;\vtheta^*,\eta^*), \quad{}g_1(\vy,\vz) := \frac{\partial}{\partial\vtheta}\log\Psi(\tau_{\vtheta^*}(\vy),\eta(\vz)),\\ 
    f_2(\vu):=\frac{\partial}{\partial\eta}\left\{\log\lambda[\vu;\vtheta^*,\eta^*]\right\},\quad{} g_2(\vy,\vz) := \frac{\partial}{\partial\eta}\log\left\{\Psi[\tau_{\vtheta^*}(\vy),\eta(\vz)]\right\}.
\end{gather*}
Then, the sensitivity matrix \eqref{eq:sensitivity_submodel} can be expressed in terms of the outer-product as
$$S(\vtheta^*,\eta^*,\nu)=\left|f_1+f_2\cdot \nu\rangle \langle f_1+f_2\cdot \nu \right|_A.$$
We proceed the proof in two steps. In step 1, we show that $\nu^*$ minimizes the inner product of $f_1+f_2\cdot\nu$ over all $\nu\in \mathcal{H}^k$, i.e., the $k$-product space of the nuisance parameter space. In step 2, we show that if $\nu^*$ minimizes the inner product of $f_1+f_2\cdot\nu$, it minimizes the outer product of $f_1+f_2\cdot\nu$, i.e., the sensitivity matrix \eqref{eq:sensitivity_submodel}, as well.


\textit{Step 1:} 
Since $\mathcal{H}^k$ is linear, $\{f_2\cdot \nu:\nu\in\mathcal{H}^k\}$ is a linear subspace of $L_2^{(k)}(A)$. By the projection Theorem (Theorem 2.1 in \cite{tsiatis2006semiparametric}), the infimum of the inner-product of $f_1+f_2\cdot\nu$ is attained at a unique $\nu^*\in\mathcal{H}^k$, and $-f_2\cdot\nu^*$ is the the linear projection of $f_1$ onto $\{f_2\cdot \nu:\nu\in\mathcal{H}^k\}$ satisfying
\begin{equation}\label{eq:orthogonality}
    \left\langle f_1+f_2\cdot\nu^*,f_2\cdot\nu\right\rangle_A=0, \quad{}\forall \nu\in\mathcal{H}^k.
\end{equation}
By applying the change of variable to the left-hand side of \eqref{eq:orthogonality}, we have
\begin{align*}
&\left\langle f_1+f_2\cdot\nu^*,f_2\cdot\nu\right\rangle_A\\
=& \int_A \lambda(\vu;\vtheta^*,\eta^*)\left[f_1(\vu)+f_2(\vu)\nu^*(\vz_\vu)\right]^\top f_2(\vu)\nu(\vz_\vu)\mathrm{d}\vu\\ \nonumber
=& \int_{\mathcal{Z}}\int_{\mathcal{Y}}
\Psi[\tau_{\vtheta^*}(\vy),\eta^*(\vz)]\left[g_1(\vy,\vz)+g_2(\vy,\vz)\nu^*(\vz)\right]^\top g_2(\vy,\vz)\nu(\vz)f(\vy,\vz)\mathrm{d}\vy\mathrm{d}\vz. \nonumber
\end{align*}
Then, condition \eqref{eq:orthogonality} is satisfied if for every $\vz\in\mathcal{Z}$,
$$\int_{\mathcal{Y}}
\Psi[\tau_{\vtheta^*}(\vy),\eta^*(\vz)]\left[g_1(\vy,\vz)+g_2(\vy,\vz)\nu^*(\vz)\right]^\top g_2(\vy,\vz)f(\vy,\vz)\mathrm{d}\vy=0,$$
which gives us the expression of $\nu^*$ as follows:
\begin{align*}
    \nu^*(\vz) &= -  \frac{\int_{\mathcal{Y}}
\Psi[\tau_{\vtheta^*}(\vy),\eta^*(\vz)]g_1(\vy,\vz) g_2(\vy,\vz)f(\vy,\vz)\mathrm{d}\vy}{\int_{\mathcal{Y}}
\Psi[\tau_{\vtheta^*}(\vy),\eta^*(\vz)]g_2(\vy,\vz)g_2(\vy,\vz)f(\vy,\vz)\mathrm{d}\vy},
\end{align*}
which is formula \eqref{eq:least_favorable_curve} defined in Theorem \ref{thm:least_favorable_direction}.

\textit{Step 2:} In this step, we show that the outer-product of $f_1+f_2\cdot\nu$ is attained by $\nu^*$ as well.
By the orthogonal condition \eqref{eq:orthogonality}, $f_1+f_2\cdot\nu^*$ is orthogonal to $\{f_2\cdot \nu:\nu\in\mathcal{H}^k\}$. Further, $\{f_2\cdot \nu:\nu\in\mathcal{H}^k\} = \{f_2\cdot \nu:\nu\in\mathcal{H}\}^k$ and $\{f_2\cdot \nu:\nu\in\mathcal{H}\}$ is a linear subspace of $L_2(A)$. Thus, by Lemma \ref{thm:pythagorean}, for any $\nu\in\mathcal{H}^k$,
\begin{align*}
     S(\vtheta^*,\eta^*,\nu)
     =& \left|f_1+f_2\cdot\nu\right\rangle \left\langle f_1+f_2\cdot\nu\right|_A \\
    =&\left|f_1+f_2\cdot\nu^*+f_2\cdot(\nu-\nu^*)\right\rangle\left\langle f_1+f_2\cdot\nu^*+f_2\cdot(\nu-\nu^*)\right|_A\\
     =& \left|f_2\cdot(\nu-\nu^*)\right\rangle\left\langle f_2\cdot(\nu-\nu^*)\right|_A +  S(\vtheta^*,\eta^*,\nu^*),
\end{align*}
which simplifies to 
\begin{equation*}
    S(\vtheta^*,\eta^*,\nu)-S(\vtheta^*,\eta^*,\nu^*) = \left|f_2\cdot(\nu-\nu^*)\right\rangle\left\langle f_2\cdot(\nu-\nu^*)\right|_A.
\end{equation*}
The right-hand-side of the above equality is always positive definite when $\nu^*$ is not equal to $\nu$, so $\nu^*$ minimizes $S(\vtheta^*,\eta^*,\nu)$.

\subsection{Proof of Random Thinning Properties (Proposition \ref{proposition:thinning} in main manuscript})
    
\textit{Proof of result (i):}
First, we prove the result (i)
For every point $\vu\in X$, we let $c_{\vu}:=\sum_{j=1}^V j\cdot \chi(\vu\in X_j)$ indicating which subprocess $\vu$ belongs to. Then, for any $A\subset\R^2$ and any $I\subset[V]$, the intensity function of $ X^{(I)}$ is
\begin{align*}
    &\E\left[\sum_{\vu\in X^{(I)}}{1}(\vu\in A)\right]\\
    =&  \E\left[\sum_{\vu\in X}{1}(\vu\in A){1}(c_{\vu}\in I)\right]\\
    =&  \E\left[\E\left[\sum_{\vu\in X}{1}(\vu\in A){1}(c_{\vu}\in I)\bigg|X\right]\right]\\
    =&  \E\left[\sum_{\vu\in X}\E\left[{1}(\vu\in A){1}(c_{\vu}\in I)\bigg|X\right]\right]\\
     =&  \E\left[\sum_{\vu\in X}{1}(\vu\in A)\E\left[{1}(c_{\vu}\in I)\bigg|X\right]\right]\\
     =&  \E\left[\sum_{\vu\in X}{1}(\vu\in A)\frac{|I|}{V}\right]\\
     =& \frac{|I|}{V} \int_{A}\lambda(\vu)\mathrm{d}\vu.
\end{align*}

\textit{Proof of Result (ii):} The true parameters $\vtheta^*,\eta^*$ maximize  the pseudo-log-likelihood of the original point process $X$, i.e., $\E\left[\ell(\vtheta,\eta;X)\right]$. Therefore, in order to show $\vtheta^*,\eta^*$ maximizes $\ell(\vtheta,\eta;X^{(I)})$,
it suffices to show that $$\arg\max_{\vtheta\in\Theta,\eta\in \mathcal{H}}\E\left[\ell(\vtheta,\eta;X^{(I)})\right] = \arg\max_{\vtheta\in\Theta,\eta\in \mathcal{H}}\E\left[\ell(\vtheta,\eta;X)\right],$$
which is satisfies as follows:
\begin{align*}
    &\arg\max_{\vtheta\in\Theta,\eta\in \mathcal{H}}\E\left[\ell(\vtheta,\eta;X^{(I)})\right]\\    =&\arg\max_{\vtheta\in\Theta,\eta\in \mathcal{H}} \left\{\int_A \log\left[\frac{|I|}{V}\lambda(\vu;\vtheta,\eta)\right]\frac{|I|}{V}\lambda(\vu;\vtheta^*,\eta^*)\mathrm{d}\vu-\int_A \frac{|I|}{V}\lambda(\vu;\vtheta,\eta)\mathrm{d}{u}\right\}\\    =&\arg\max_{\vtheta\in\Theta,\eta\in \mathcal{H}} \left\{\int_A \left(\log\left[\lambda(\vu;\vtheta,\eta)\right]-\log\left[\frac{|I|}{V}\right]\right)\frac{|I|}{V}\lambda(\vu;\vtheta^*,\eta^*)\mathrm{d}\vu-\int_A \frac{|I|}{V}\lambda(\vu;\vtheta,\eta)\mathrm{d}{u}\right\}\\    =&\arg\max_{\vtheta\in\Theta,\eta\in \mathcal{H}} \left\{\int_A \log\left[\lambda(\vu;\vtheta,\eta)\right]\frac{|I|}{V}\lambda(\vu;\vtheta^*,\eta^*)\mathrm{d}\vu-\int_A \frac{|I|}{V}\lambda(\vu;\vtheta,\eta)\mathrm{d}{u}\right\}\\
    =&\arg\max_{\vtheta\in\Theta,\eta\in \mathcal{H}} \left\{\int_A \log\left[\lambda(\vu;\vtheta,\eta)\right]\lambda(\vu;\vtheta^*,\eta^*)\mathrm{d}\vu-\int_A \lambda(\vu;\vtheta,\eta)\mathrm{d}{u}\right\}\\
    =&\arg\max_{\vtheta\in\Theta,\eta\in \mathcal{H}}\E\left[\ell(\vtheta,\eta;X)\right].
\end{align*}

Similarly, in order to show  $\eta_{\vtheta}$ maximizes $\ell(\vtheta,\eta;X^{(I)})$ for any fixed $\vtheta$, it suffices to show that for any $\vtheta$,
$$\arg\max_{\eta\in \mathcal{H}}\E\left[\ell(\vtheta,\eta;X^{(I)})\right] = \arg\max_{\eta\in \mathcal{H}}\E\left[\ell(\vtheta,\eta;X)\right],$$
which is satisfied as follows:
\begin{align*}
    &\arg\max_{\eta\in \mathcal{H}}\E\left[\ell(\vtheta,\eta;X^{(I)})\right]\\    =&\arg\max_{\eta\in \mathcal{H}} \left\{\int_A \log\left[\frac{|I|}{V}\lambda(\vu;\vtheta,\eta)\right]\frac{|I|}{V}\lambda(\vu;\vtheta^*,\eta^*)\mathrm{d}\vu-\int_A \frac{|I|}{V}\lambda(\vu;\vtheta,\eta)\mathrm{d}{u}\right\}\\    =&\arg\max_{\eta\in \mathcal{H}} \left\{\int_A \left(\log\left[\lambda(\vu;\vtheta,\eta)\right]-\log\left[\frac{|I|}{V}\right]\right)\frac{|I|}{V}\lambda(\vu;\vtheta^*,\eta^*)\mathrm{d}\vu-\int_A \frac{|I|}{V}\lambda(\vu;\vtheta,\eta)\mathrm{d}{u}\right\}\\   
    =&\arg\max_{\eta\in \mathcal{H}} \left\{\int_A \log\left[\lambda(\vu;\vtheta,\eta)\right]\lambda(\vu;\vtheta^*,\eta^*)\mathrm{d}\vu-\int_A \lambda(\vu;\vtheta,\eta)\mathrm{d}{u}\right\}\\
    =&\arg\max_{\eta\in \mathcal{H}}\E\left[\ell(\vtheta,\eta;X)\right].
\end{align*}

\textit{Proof of result (iii):} 
It suffices to show that for any $0<q<1$, randomly thinning $X$ into two subprocesses $X_1$ and $X_2$ with probabilities $q$ and $1-q$, respectively, gives us two independent Poisson spatial point processes. To do so, we proceed in two steps. Step 1 shows $X_1$ and $X_2$ are Poisson processes by deriving their density functions. Step 2 shows their independence by illustrating the joint density of $X_1$ and $X_2$ equals the product of the marginal densities of $X_1$ and $X_2$.

\textit{Step 1:}
For brevity, we denote ${card}(X)$  as the cardinality of $X$, i.e. the number of observed points, $X=\{\vu_1,\ldots,\vu_{{Card}(X)}\}$, and $\mu(A) = \int_A \lambda(\vu)\mathrm{d}\vu$. 

When $X$ is Poisson, ${Card}(X)$ follows the Poisson distribution with expectation $\mu(A)$. Also, the points $\{\vu_1,\ldots,\vu_{{Card}(X)}\}$ are i.i.d. The conditional density of $\vu$ given ${Card}(X)$ is $\lambda(\vu)/\mu(A)$ and the density function of the entire $X$ is
\begin{align*}
    p(X)=& p\left[X\mid {Card}(X)\right]\pr\left[{Card}(X)\right]\\
=&\left[\prod_{i=1}^{{Card}(X)}\frac{\lambda(\vu_i)}{\mu(A)}\right] \frac{\exp[-\mu(A)]\mu(A)^{{Card}(X)}}{{Card}(X)!}.
\end{align*}
We will show that the probability density functions of $X_1$ and $X_2$ are consistent with the above form. First, consider the probability mass function of ${Card}(X_1)$ 
\begin{align*}
    \pr\left[{Card}(X_1)\right]&=\sum_{n={Card}(X_1)}^{\infty} \pr\left[{Card}(X_1)\mid {Card}(X)=n\right]\pr[{Card}(X)=n]\\
    &=\sum_{n={Card}(X_1)}^{\infty} \binom{n}{{Card}(X_1)}q^{{Card}(X_1)}(1-q)^{[n-{Card}(X_1)]}\frac{\exp{[-\mu(A)]}[\mu(A)]^n}{n!}\\
    & = \frac{\exp{[-\mu(A)]}}{{Card}(X_1)!}\left[\mu(A)q\right]^{{Card}(X_1)}\sum_{n={Card}(X_1)}^{\infty}\frac{\left[(1-q)\mu(A)\right]^{[n-{Card}(X_1)]}}{[n-{Card}(X_1)]!}\\
    & = \frac{\exp{[-q\mu(A)]}}{{Card}(X_1)!}\left[q\mu(A)\right]^{{Card}(X_1)}.
\end{align*}
By the result in (i), the intensity function of $X_1$ is $q\lambda(\vu)$. Since the thinning is independent with points in $X$, the retained points in $X_1$ are still independent and are identically distributed with the conditional density $\lambda(\vu_i)/q\mu(A)$ given ${Card}(X_1)$. Therefore,
\begin{align*}
    p(X_1) = &p[X_1\mid{Card}(X_1)]\pr[{Card}(X_1)]\\
=&\left[\prod_{i=1}^{{Card}(X_1)}\frac{\lambda(\vu_i)}{q\mu(A)}\right]\frac{\exp[-q\mu(A)][q\mu(A)]^{{Card}(X_1)}}{{Card}(X_1)!}.
\end{align*}
Similarly, we have
\begin{align*}
    p(X_2) = &p[X_2\mid{Card}(X_2)]\pr({Card}(X_2))\\
=&\left[\prod_{i=1}^{{Card}(X_2)}\frac{\lambda(\vu_i)}{(1-q)\mu(A)}\right]\frac{\exp[-(1-q)\mu(A)][(1-q)\mu(A)]^{{Card}(X_2)}}{{Card}(X_2)!}.
\end{align*}
Thus, $X_1$ and $X_2$ are Poisson spatial point processes.

\textit{Step 2: }We consider the joint probability mass function of ${Card}(X_1)$ and ${Card}(X_2)$,
\begin{align*}
    & \pr[{Card}(X_1),{Card}(X_2)] \\
    =& \pr[{Card}(X_1),{Card}(X_2)\mid {Card}(X)]\pr[{Card}(X)]\\ =&\pr[{Card}(X_1)\mid {Card}(X)]\pr[{Card}(X)]\\ 
    =&\binom{{Card}(X)}{{Card}(X_1)}q^{{Card}(X_1)}(1-q)^{{Card}(X_2)} \frac{\exp{[-\mu(A)]}[\mu(A)]^{|X_1+X_2|}}{|X_1+X_2|!} \\
    =&\frac{{Card}(X)!}{{Card}(X_1)!{Card}(X_2)!}q^{{Card}(X_1)}(1-q)^{{Card}(X_2)} \frac{\exp{[-\mu(A)]}[\mu(A)]^{{Card}(X_1+X_2)}}{|X_1+X_2|!}\\
    =&\frac{\exp[-q\mu(A)][q\mu(A)]^{{Card}(X_1)}}{{Card}(X_1)!}\cdot  \frac{\exp[-(1-q)\mu(A)][(1-q)\mu(A)]^{{Card}(X_2)}}{{Card}(X_2)!}\\
    =& \pr[{Card}(X_1)]\pr[{Card}(X_2)].
\end{align*}
Conditioning on ${Card}(X_1)$ and ${Card}(X_2)$, the points in $X_1$ and points in $X_2$ are independent. Thus, 
\begin{align*}
p(X_1,X_2)=&p[X_1,X_2\mid{Card}(X_1),{Card}(X_2)]\pr[{Card}(X_1),{Card}(X_2)]\\
=&p[X_1\mid{Card}(X_1)]p(X_2\mid{Card}(X_2)]\pr[{Card}(X_1)]\pr[{Card}(X_2)]\\
=&p(X_1)p(X_2).
\end{align*}
Therefore, $X_1$ and $X_2$ are independent Poisson spatial point processes.

\subsection{Proof of Consistency (Theorem \ref{thm:consistency} in main manuscript)}

Since $\hat{\vtheta}_n = V^{-1}\sum_{v=1}^V\hat\vtheta_n^{(v)}$, it suffices to show that for every $v\in[V]$, $\hat{\vtheta}^{(v)}_n$ obtained via
$$ \hat\vtheta_n^{(v)} =\arg\max_{\vtheta\in \Theta}\ell_n(\vtheta,\hat{\eta}^{(v)}_{\vtheta,n};X_v),$$ 
It is a consistent estimator of $\vtheta^*$. 
For the convenience of the proof, we redefine $\hat\vtheta_n$ as
\begin{equation}\label{eq:thm_consistent_redef}
\hat\vtheta_n:=\arg\max_{\vtheta\in\Theta}\ell_n(\vtheta,\hat\eta_{\vtheta,n}),
\end{equation}
where $\hat\eta_{\vtheta,n}$ is an estimators of $\eta_{\vtheta,n}$. 

By Lemma \ref{proposition:thinning}(ii), the pseudo-log-likelihood of the subprocess $X_v$ and the original process $X$ have the same first-order property. Thus, it suffices to show that $\hat\vtheta_n$ defined in \eqref{eq:thm_consistent_redef} is a consistent estimator of $\vtheta^*$ if $\hat\eta_{\vtheta,n}$ satisfies Assumption \ref{assumption:np_condition_uniform}.


To do so, we proceed in four steps. Step 1 shows that $|\hat\vtheta_n-\vtheta^*|$ can be bounded by$$\E\left[\ell(\vtheta^*,\eta_{\vtheta^*,n})\right]-\E\left[\ell(\hat\vtheta_n,\hat\eta_{\hat\vtheta_n,n})\right],$$ 
which is decomposed into three remainder terms. Steps 2-4 bound the remainder terms with auxiliary lemmas.

\textit{Step 1:}  Note that $\eta_{\vtheta^*,n}=\eta^*$. Thus, when condition \ref{condition:smoothness}, \ref{condition:Sufficient Separation}, \ref{condition:boundedness} in Assumption \ref{assumption:regularity} are satisfied, we can apply Lemma \ref{lemma:sufficient separation} for $(\hat\vtheta_n,\hat\eta_{\hat\vtheta_n,n})\neq (\vtheta^*,\eta^*)$, which gives us
$$ \E\left[\ell(\vtheta^*,\eta_{\vtheta^*,n})\right]-\E\left[\ell(\hat\vtheta_n,\hat\eta_{\hat\vtheta_n,n})\right]=\Theta(|A_n|)\min(1,|\hat\vtheta_n-\vtheta^*|).$$
Thus, it suffices to show that
\begin{equation*}
    \E\left[\ell(\vtheta^*,\eta_{\vtheta^*,n})\right]-\E\left[\ell(\hat\vtheta_n,\hat\eta_{\hat\vtheta_n,n})\right] = o_p(|A_n|).
\end{equation*}
Denote
\begin{gather*}
    R_1:=\left|\E\left[\ell(\hat\vtheta_n,\hat\eta_{\hat\vtheta_n,n})\right]-\ell(\hat\vtheta_n,\hat\eta_{\hat\vtheta_n,n})\right|,\\
    R_2:=\left|\ell(\hat\vtheta_n,\hat\eta_{\hat\vtheta_n,n})-\sup_{\vtheta\in\Theta}\ell(\vtheta,\eta_{\vtheta,n})\right|,\\
    R_3:=\left|\E\left[\ell(\vtheta^*,\eta_{\vtheta^*,n})\right]-\sup_{\vtheta\in\Theta}\ell(\vtheta,\eta_{\vtheta,n})\right|.
\end{gather*}
Then, by triangular inequality,
$$\E\left[\ell(\vtheta^*,\eta_{\vtheta^*,n})\right]-\E\left[\ell(\hat\vtheta_n,\hat\eta_{\hat\vtheta_n,n})\right]\leq R_1+R_2+R_3.$$
Thus, it suffices to show that
\begin{gather}
    R_1 = o_p(|A_n|)\label{eq:consist_R_1},\\
    R_2 = o_p(|A_n|)\label{eq:consist_R_2},\\
    R_3 = o_p(|A_n|)\label{eq:consist_R_3}.
\end{gather}

\textit{Step 2:} We establish \eqref{eq:consist_R_1} in this step. When condition \ref{condition:smoothness}, \ref{condition:boundedness}, \ref{condition:pair_correlation} are satisfied, by equation \eqref{eq:rate of likelihood bias} in Corollary \ref{corollary:rate of bias},
$$\sup_{\vtheta\in\vtheta,\eta\in\mathcal{H}}\left|\ell_n(\vtheta,\eta)-\E\left[\ell_n(\vtheta,\eta)\right]\right|=O_p(|A_n|^{\frac{1}{2}}).$$
Thus, 
\begin{align*}
R_1
 \leq& \sup_{\vtheta\in\vtheta,\eta\in\mathcal{H}}\left|\ell_n(\vtheta,\eta)-\E\left[\ell_n(\vtheta,\eta)\right]\right|\\
 =&O_p(|A_n|^{\frac{1}{2}})=o_p(|A_n|).
\end{align*}
\textit{Step 3:} We establish \eqref{eq:consist_R_2} in this step. Under Assumption \ref{assumption:np_condition_uniform} and conditions \ref{condition:smoothness}, \ref{condition:boundedness}, \ref{condition:pair_correlation}, Lemma \ref{lemma:first order plug-in error} shows that
$$\sup_{\vtheta\in\Theta}\left|\ell_n(\vtheta,\hat{\eta}_{{\vtheta},n}) -\ell_n(\vtheta,{\eta}_{{\vtheta},n})\right| =o(1).$$
Since $\ell(\hat\vtheta_n,\hat\eta_{\hat\vtheta_n,n}) =\sup_{\vtheta\in\Theta}\ell(\vtheta,\hat\eta_{\vtheta,n})$,
\begin{align*}
    R_2
    = & \left|\sup_{\vtheta\in\Theta}\ell(\vtheta,\hat\eta_{\vtheta,n})-\sup_{\vtheta\in\Theta}\ell(\vtheta,\eta_{\vtheta,n})\right|\\
    \leq &\sup_{\vtheta\in\Theta}\left|\ell(\vtheta,\hat\eta_{\vtheta,n})-\ell(\vtheta,\eta_{\vtheta,n})\right|\\
    =&O_p(|A_n|e_{n}^{(1)})=o_p(|A_n|).
\end{align*}

\textit{Step 4:} We establish \eqref{eq:consist_R_3} in this step. When condition \ref{condition:smoothness}, \ref{condition:boundedness}, \ref{condition:pair_correlation} are satisfied,
by equation \eqref{eq:rate of likelihood bias} in Corollary \ref{corollary:rate of bias},
$$\sup_{\vtheta\in\vtheta,\eta\in\mathcal{H}}\left|\ell_n(\vtheta,\eta)-\E\left[\ell_n(\vtheta,\eta)\right]\right|=O_p(|A_n|^{\frac{1}{2}})$$
Since $\E\left[\ell(\vtheta^*,\eta_{\vtheta^*,n})\right] = \sup_{\vtheta\in\Theta}\E\left[\ell(\vtheta,\eta_{\vtheta,n})\right]$, 
\begin{align*}
R_3
=& \left|\sup_{\vtheta\in\Theta}\E\left[\ell(\vtheta,\eta_{\vtheta,n})\right]-\sup_{\vtheta\in\Theta}\ell(\vtheta,\eta_{\vtheta,n})\right|\\
\leq & \sup_{\vtheta\in\Theta}\left|\E\left[\ell(\vtheta,\eta_{\vtheta,n})\right]-\ell(\vtheta,\eta_{\vtheta,n})\right|\\
\leq & \sup_{\vtheta\in\vtheta,\eta\in\mathcal{H}}\left|\ell_n(\vtheta,\eta)-\E\left[\ell_n(\vtheta,\eta)\right]\right|\\
=& O_p(|A_n|^{\frac{1}{2}}) =o_p(|A_n|).
\end{align*}

\subsection{Proof of Asymptotic Normality (Theorem \ref{thm:normality} in main manuscript)}

Lemma \ref{lemma:central limit theorem} states that
under condition \ref{condition:smoothness}, \ref{condition:nonsigularity}, \ref{condition:nonsingular_covariance}, \ref{condition:alpha-mixing}, 
$$|A_n|^{-\frac{1}{2}}\bar{\Sigma}^{-\frac{1}{2}}_n(\vtheta^*,\eta^*,\nu^*,\psi^*)\frac{\partial}{\partial\vtheta}\ell_n(\vtheta,\eta_{\vtheta,n})\bigg|_{\vtheta=\vtheta^*}\rightarrow_d N({0},{I}_k)$$
Therefore, it suffices to show that
\begin{equation}\label{eq:normal_approx}
    |A_n|^{\frac{1}{2}}(\hat\vtheta_n-\vtheta^*) =- |A_n|^{-\frac{1}{2}}\bar{S}_n^{-1}(\vtheta^*,\eta^*,\nu^*)\frac{\partial}{\partial\vtheta}\ell_n(\vtheta,\eta_{\vtheta,n})\bigg|_{\vtheta=\vtheta^*}+o_p(1)
\end{equation}

To do so, we proceed in six steps. Steps 1-2 shows the main argument, and steps 3–6 present auxiliary calculations.

\textit{Step 1:} In this step, we show that the analogue of \eqref{eq:normal_approx} holds for every subprocess. By Taylor Theorem, for every $v\in[V]$, there exists a sequence $\tilde{\vtheta}^{(v)}_n\in [\vtheta^*,\hat\vtheta_n^{(v)}]$ that
\begin{align}
    0 & = \frac{\partial}{\partial\vtheta}\ell_n(\vtheta,\hat{\eta}^{(v)}_{\vtheta,n};X_v)\bigg|_{\vtheta=\hat\vtheta^{(v)}_n}\nonumber\\ 
    & =   \frac{\partial}{\partial\vtheta}\ell_n(\vtheta,\hat{\eta}^{(v)}_{\vtheta,n};X_v)\bigg|_{\vtheta=\vtheta^*}+\frac{\partial^2}{\partial\vtheta^2}\ell_n(\vtheta,\hat{\eta}^{(v)}_{\vtheta,n};X_v)\bigg|_{\vtheta=\tilde\vtheta^{(v)}_n}(\hat\vtheta^{(v)}_n-\vtheta^*).\label{eq:thm_normality_taylor}
\end{align}
Denote 
\begin{gather*}
    \widehat{S}_n^{(v)} := |A_n|^{-1}\frac{\partial^2}{\partial\vtheta^2}\ell_n(\vtheta,\hat{\eta}^{(v)}_{\vtheta,n};X_v)\bigg|_{\vtheta=\tilde\vtheta^{(v)}_n},\\
    \bar{S}_n^{(v)}:=|A_n|^{-1}\E\left[\frac{\partial^2}{\partial\vtheta^2}\ell_n(\vtheta,\eta_{\vtheta,n};X_v)\bigg|_{\vtheta=\vtheta^*}\right],\\
    R_{1,n}^{(v)} := \widehat{S}_n^{(v)} - \bar{S}_n^{(v)},\\
    R_{2,n}^{(v)} := \frac{\partial}{\partial\vtheta}\left[\ell_n(\vtheta,\hat{\eta}^{(v)}_{\vtheta,n};X_v)-\ell_n(\vtheta,\eta_{\vtheta,n};X_v)\right]\bigg|_{\vtheta=\vtheta^*}.
\end{gather*}
In Steps 3-6 respectively, we will show that
\begin{gather}
     |R_{1,n}^{(v)}| = o_p(1)\label{eq:normality_R_1},\\
     |R_{2,n}^{(v)}| = o_p(|A_n|^{\frac{1}{2}})\label{eq:normality_R_2},\\
     \left||A_n|^{-\frac{1}{2}}\frac{\partial}{\partial\vtheta}\ell_n(\vtheta,\hat{\eta}^{(v)}_{\vtheta,n};X_v)\bigg|_{\vtheta=\vtheta^*}\right|=O_p(1)\label{eq:normality_partial_ell},\\
     \left(\bar{S}_n^{(v)}\right)^{-1}=O(1). \label{eq:normality_partial_second_ell}
\end{gather}

By Condition \ref{condition:nonsigularity}, all singular values of $\bar{S}_n(\vtheta^*,\eta^*,\nu^*)$ is bounded below from zero when $n$ is sufficiently large. By Lemma \ref{proposition:thinning}(i), the intensity function of $X_v$ is $V^{-1}\lambda(\vu;\vtheta^*,\eta^*)$. Then, the sensitivity matrix of the subprocesses satisfies
\begin{equation}\label{eq:sub_sum_sensitivity}
\bar{S}_n^{(v)}=V^{-1}\bar{S}_n(\vtheta^*,\eta^*,\nu^*).
\end{equation}
so all singular values of $\bar{S}_n^{(v)}$ is bounded below from zero when $n$ is sufficiently large. Therefore, when condition \eqref{eq:normality_R_1} holds, the inverse of $\widehat{S}_n^{(v)}$ exists with probability $1-o(1)$, and \eqref{eq:thm_normality_taylor} can be written as 
\begin{align}
    |A_n|^{\frac{1}{2}}(\hat\vtheta^{(v)}_n-\vtheta^*) =&-\left\{|A_n|^{-1}\frac{\partial^2}{\partial\vtheta^2}\ell_n(\vtheta,\hat{\eta}^{(v)}_{\vtheta,n};X_v)\bigg|_{\vtheta=\tilde\vtheta^{(v)}_n}\right\}^{-1} \frac{\partial}{\partial\vtheta}\ell_n(\vtheta,\hat{\eta}^{(v)}_{\vtheta,n};X_v)\bigg|_{\vtheta=\vtheta^*}\nonumber\\ 
    =&- \left(\bar{S}_n^{(v)}+R_{1,n}^{(v)}\right)^{-1}|A_n|^{-\frac{1}{2}}\left(\frac{\partial}{\partial\vtheta}\ell_n(\vtheta,\eta_{\vtheta,n};X_v)\bigg|_{\vtheta=\vtheta^*}+R_{2,n}^{(v)}\right).\label{eq:normality_original}
\end{align}
Furthermore, by Woodbury matrix identity, 
\begin{equation*}
\left(\bar{S}_n^{(v)}+R_{1,n}^{(v)}\right)^{-1}-\left(\bar{S}_n^{(v)}\right)^{-1}=-\left(\bar{S}_n^{(v)}+R_{1,n}^{(v)}\right)^{-1}R_{1,n}^{(v)}\left(\bar{S}_n^{(v)}\right)^{-1}.
\end{equation*}
Then, it follows from \eqref{eq:normality_R_1} that
\begin{align}
&\left|\left(\bar{S}_n^{(v)}+R_{1,n}^{(v)}\right)^{-1}-\left(\bar{S}_n^{(v)}\right)^{-1}\right|\nonumber\\
\leq&\left|\left(\bar{S}_n^{(v)}+R_{1,n}^{(v)}\right)^{-1}\right|\times\left|R_{1,n}^{(v)}\right|\times\left|\left(\bar{S}_n^{(v)}\right)^{-1}\right|\nonumber\\
=&O_p(1)\times o_p(1)\times O(1)=o_p(1).\label{eq:normality_denominator}
\end{align}
Additionally, by condition \eqref{eq:normality_R_2} and \eqref{eq:normality_partial_ell},
\begin{align}
     |A_n|^{-\frac{1}{2}} \left|\frac{\partial}{\partial\vtheta}\ell_n(\vtheta,\eta_{\vtheta,n};X_v)\bigg|_{\vtheta=\vtheta^*}+R_{2,n}^{(v)}\right|\leq & |A_n|^{-\frac{1}{2}} \left|\frac{\partial}{\partial\vtheta}\ell_n(\vtheta,\eta_{\vtheta,n})\bigg|_{\vtheta=\vtheta^*}\right|+|A_n|^{-\frac{1}{2}} \left|R_2\right|\nonumber\\
     =&O_p(1) +o_p(1)=O_p(1). \label{eq:normality_numerator}
\end{align}
Combining \eqref{eq:normality_denominator} and \eqref{eq:normality_numerator} gives
\begin{align*}
&\left|\left(\left(\bar{S}_n^{(v)}+R_{1,n}^{(v)}\right)^{-1}-\left(\bar{S}_n^{(v)}\right)^{-1}\right)|A_n|^{-\frac{1}{2}}\left(\frac{\partial}{\partial\vtheta}\ell_n(\vtheta,\eta_{\vtheta,n};X_v)\bigg|_{\vtheta=\vtheta^*}+R_{2,n}^{(v)}\right) \right|. \\
\leq &\left|\left(\bar{S}_n^{(v)}+R_{1,n}^{(v)}\right)^{-1}-\left(\bar{S}_n^{(v)}\right)^{-1}\right|\times|A_n|^{-\frac{1}{2}}\times \left|\frac{\partial}{\partial\vtheta}\ell_n(\vtheta,\eta_{\vtheta,n};X_v)\bigg|_{\vtheta=\vtheta^*}+R_{2,n}^{(v)}\right|\\
=& o_p(1)\times O_p(1)=o_p(1)
\end{align*}
Next, substituting the above inequality into \eqref{eq:normality_original} yields
\begin{align}
    |A_n|^{\frac{1}{2}}(\hat\vtheta^{(v)}_n-\vtheta^*) =&- \left(\bar{S}_n^{(v)}\right)^{-1}|A_n|^{-\frac{1}{2}}\left(\frac{\partial}{\partial\vtheta}\ell_n(\vtheta,\eta_{\vtheta,n};X_v)\bigg|_{\vtheta=\vtheta^*}+R_{2,n}^{(v)}\right)+o_p(1)\nonumber\\
    =& \left(\bar{S}_n^{(v)}\right)^{-1}|A_n|^{-\frac{1}{2}}\frac{\partial}{\partial\vtheta}\ell_n(\vtheta,\eta_{\vtheta,n};X_v)\bigg|_{\vtheta=\vtheta^*}+\left(\bar{S}_n^{(v)}\right)^{-1}|A_n|^{-\frac{1}{2}}R_{2,n}^{(v)}+o_p(1)\nonumber\\
    =&\left(\bar{S}_n^{(v)}\right)^{-1}|A_n|^{-\frac{1}{2}}\frac{\partial}{\partial\vtheta}\ell_n(\vtheta,\eta_{\vtheta,n};X_v)\bigg|_{\vtheta=\vtheta^*}+o_p(1).\label{eq:sub_normal_approx}
\end{align}

\textit{Step 2:} In this step, we show that if the condition \eqref{eq:sub_normal_approx} holds for for every subprocess, condition \eqref{eq:normal_approx} will be satisfied. Step \ref{step2} states that $\hat{\vtheta}_n = V^{-1}\sum_{v=1}^V\hat\vtheta_n^{(v)}$, and $\hat\vtheta_n^{(v)}$ is obtained via
\begin{gather*}
    \hat\vtheta_n^{(v)} =\arg\max_{\vtheta\in \Theta}\ell_n(\vtheta,\hat{\eta}^{(v)}_{\vtheta,n};X_v),
\end{gather*}
$\ell_n(\vtheta,\hat{\eta}^{(v)}_{\vtheta,n};X_v)$ is the log-pseudo-likelihood of the subprocess $X_v$. $\hat{\eta}^{(v)}_{\vtheta}$ is an estimator of $\eta_{\vtheta,n}$ 
Moreover, since $X=\bigcup_{v=1}^V X_v$, 
\begin{equation}\label{eq:sub_sum_pseudo}
\sum_{v=1}^V\ell_n(\vtheta,\eta_{\vtheta,n};X_v) = \ell_n(\vtheta,\eta_{\vtheta,n}).
\end{equation}
Combining equation \eqref{eq:sub_sum_sensitivity} and \eqref{eq:sub_sum_pseudo} with \eqref{eq:sub_normal_approx} gives us \eqref{eq:normal_approx} as follows:
\begin{align*}
    |A_n|^{\frac{1}{2}}(\hat\vtheta_n-\vtheta^*)=&|A_n|^{\frac{1}{2}}\left( V^{-1}\sum_{v=1}^V\hat\vtheta_n^{(v)}-\vtheta^*\right)
    =V^{-1}\sum_{v=1}^V|A_n|^{\frac{1}{2}}( \hat\vtheta_n^{(v)}-\vtheta^*)\\
=&\left(V\bar{S}_n^{(v)}\right)^{-1}|A_n|^{-\frac{1}{2}}\frac{\partial}{\partial\vtheta}\left(\sum_{v=1}^V\ell_n(\vtheta,\eta_{\vtheta,n};X_v)\bigg|_{\vtheta=\vtheta^*}\right)+o_p(1)\\
=& \left(\bar{S}_n(\vtheta^*,\eta^*,\nu^*)\right)^{-1} |A_n|^{-\frac{1}{2}}\frac{\partial}{\partial\vtheta}\ell_n(\vtheta,\eta_{\vtheta,n})\bigg|_{\vtheta=\vtheta^*}+o_p(1)
\end{align*}

\textit{Step 3:} In this step, we establish \eqref{eq:normality_R_1}. By triangle inequality, 
\begin{equation}\label{eq:thm_normality_R_1_2_decpm}
    |R_{1,n}^{(v)}|\leq \mathcal{I}_{1,n}^{v}+\mathcal{I}_{2,n}^{v}+\mathcal{I}_{3,n}^{v},
\end{equation}
where
\begin{gather*}
    \mathcal{I}_{1,n}^{v}:=|A_n|^{-1}\times\left|\frac{\partial^2}{\partial\vtheta^2}\ell_n(\vtheta,\hat\eta_{\vtheta,n}^{(v)};X_v)\bigg|_{\vtheta=\tilde\vtheta_n^{(v)}} - \E\left[ \frac{\partial^2}{\partial\vtheta^2}\ell_n(\vtheta,\hat\eta^{(v)}_{\vtheta,n};X_v)\bigg|_{\vtheta=\tilde\vtheta_n^{(v)}}\right]\right|,\\
    \mathcal{I}_{2,n}^{v}:=|A_n|^{-1}\times\left|\E\left[ \frac{\partial^2}{\partial\vtheta^2}\ell_n(\vtheta,\hat\eta^{(v)}_{\vtheta,n};X_v)\bigg|_{\vtheta=\tilde\vtheta_n^{(v)}}\right] -\E\left[ \frac{\partial^2}{\partial\vtheta^2}\ell_n(\vtheta,\eta_{\vtheta,n};X_v)\bigg|_{\vtheta=\tilde\vtheta_n^{(v)}}\right]\right|,\\
    \mathcal{I}_{3,n}^{v}:=|A_n|^{-1}\times\left|\E\left[ \frac{\partial^2}{\partial\vtheta^2}\ell_n(\vtheta,\eta_{\vtheta,n};X_v)\bigg|_{\vtheta=\tilde\vtheta_n^{(v)}}\right]-\E\left[\frac{\partial^2}{\partial\vtheta^2}\ell_n(\vtheta,\eta_{\vtheta,n};X_v)\bigg|_{\vtheta=\vtheta^*}\right]\right|.
\end{gather*}
To bound $\mathcal{I}_{1,n}^{v}$, we further denote 
\begin{gather*}
    \hat\eta^{(v)}_n := \hat\eta_{\vtheta,n}^{(v)}\bigg|_{\vtheta=\hat\vtheta_{n}^{(v)}},\\
    \hat{\nu}_n^{(v)} := \frac{\partial}{\partial\vtheta}\hat\eta_{\vtheta,n}^{(v)}\bigg|_{\vtheta=\hat\vtheta_{n}^{(v)}},\\
    \hat{\nu}_n^{(v)\prime} := \frac{\partial^2}{\partial\vtheta^2}\hat\eta_{\vtheta,n}^{(v)}\bigg|_{\vtheta=\hat\vtheta_{n}^{(v)}}.
\end{gather*}
By the chain rule, 
\begin{multline}\label{eq:I_1_chain_rule}
   \frac{\partial^2}{\partial\vtheta^2}\ell_n(\vtheta,\hat\eta_{\vtheta,n}^{(v)};X_v)\bigg|_{\vtheta=\tilde\vtheta_n^{(v)}}=\frac{\partial^2}{\partial\vtheta^2}\ell_n(\vtheta,\hat\eta^{(v)}_n;X_v)\bigg|_{\vtheta=\tilde\vtheta_n^{(v)}}\\
   +\frac{\partial^2}{\partial\vtheta\partial\eta}\ell_n(\vtheta,\hat\eta^{(v)}_n;X_v)\bigg|_{\vtheta=\tilde\vtheta_n^{(v)}}\hat{\nu}_n^{(v)}+\frac{\partial}{\partial\eta}\ell_n(\vtheta,\hat\eta_{\vtheta,n}^{(v)};X_v)\bigg|_{\vtheta=\tilde\vtheta_n^{(v)}}\hat{\nu}_n^{(v)\prime}. 
\end{multline}
If we substitute \eqref{eq:I_1_chain_rule} into $ \mathcal{I}_{1,n}^{v}$, we have
\begin{align*}
    \mathcal{I}_{1,n}^{v}\leq &|A_n|^{-1}\times\sup_{\vtheta\in\Theta,\eta\in\mathcal{H}}\left|\frac{\partial^2}{\partial\vtheta^2}\ell_n(\vtheta,\eta;X_v)- \E\left[\frac{\partial^2}{\partial\vtheta^2}\ell_n(\vtheta,\eta;X_v)\right]\right|\\
    & +|A_n|^{-1}\times\sup_{\vtheta\in\Theta,\eta\in\mathcal{H}}\left|\frac{\partial^2}{\partial\vtheta\partial\eta}\ell_n(\vtheta,\eta;X_v) - \E\left[\frac{\partial^2}{\partial\vtheta\partial\eta}\ell_n(\vtheta,\eta;X_v)\right]\right|\times \sup_{\nu\in\mathcal{H}^k}|\nu|\\
&+|A_n|^{-1}\times\sup_{\vtheta\in\Theta,\eta\in\mathcal{H}}\left|\frac{\partial}{\partial\eta}\ell_n(\vtheta,\eta;X_v) - \E\left[\frac{\partial}{\partial\eta}\ell_n(\vtheta,\eta;X_v)\right]\right|\times \sup_{\nu^\prime\in\mathcal{H}^k\times \mathcal{H}^k}|\nu^\prime|.
\end{align*}
By Corollary \ref{corollary:rate of bias}, the supremum terms appear in the above inequality are all of the rate $O_p(|A_n|^{\frac{1}{2}})$. Moreover, the nuisance parameters in $\mathcal{H}$ are bounded functions. Thus,$$\mathcal{I}_{1,n}^{v}=O_p(|A_n|^{-\frac{1}{2}}).$$
It follows from Lemma \ref{lemma:first order plug-in error} that when $\hat\eta_{\vtheta,n}^{(v)}$ satisfies Assumption \ref{assumption:np_condition_uniform},
$$\sup_{\vtheta\in \Theta}\left|\E\left[ \frac{\partial^2}{\partial\vtheta^2}\ell_n(\vtheta,\hat\eta^{(v)}_{\vtheta,n};X_v)\right] -\E\left[ \frac{\partial^2}{\partial\vtheta^2}\ell_n(\vtheta,\eta_{\vtheta,n};X_v)\right]\right|= o(|A_n|).$$
Thus, $$\mathcal{I}_{2,n}^{v}=o(1).$$

Next, we bound $\mathcal{I}_{3,n}^{v}$. Note that $\ell_n(\vtheta,\eta_{\vtheta,n})$ is twice continuously differentiable with respect to $\vtheta\in\Theta$ and $\Theta$ is compact. Thus, for any $\vtheta\in\Theta$, there exists a positive constant $C$ such that 
$$\mathcal{I}_{3,n}^{v}\leq|A_n|^{-1}\times\left|\E\left[\frac{\partial^2}{\partial\vtheta^2}\ell_n(\vtheta,\eta_{\vtheta,n};X_v)\bigg|_{\vtheta=\vtheta^*}\right]\right|\times C \left|\vtheta^*-\tilde\vtheta_n^{(v)}\right|.$$
Thus, $\mathcal{I}_{3,n}^{v}=o_p(1)$ follows from the consistency of $\tilde\vtheta_n^{(v)}$ and the equation \eqref{eq:rate of likelihood bias} in Corollary \ref{corollary:rate of bias}. 

\textit{Step 4:} In this step, we establish \eqref{eq:normality_R_2}. To bound $R_{2,n}^{(v)}$,  We denote
\begin{gather*}
    r_n^{(2,v)}(\vtheta) := \ell_n(\vtheta,\hat{\eta}^{(v)}_{\vtheta,n};X_v)-\ell_n(\vtheta,\eta_{\vtheta,n}) -\frac{\partial}{\partial\eta}\ell_n(\vtheta,\eta_{\vtheta,n};X_v))[\hat\eta^{(v)}_{\vtheta,n}-\eta_{\vtheta,n}],\\
    \mathcal{I}_{4,n}^{v}:=\frac{\partial^2}{\partial\vtheta\partial\eta}  \ell_n(\vtheta,\eta_{\vtheta,n};X_v)\bigg|_{\vtheta=\vtheta^*}[\hat\eta^{(v)}_{\vtheta^*,n}-\eta_{\vtheta^*,n}^*],\\
    \mathcal{I}_{5,n}^{v}:=\frac{\partial}{\partial\eta}\ell_n(\vtheta^*,\eta_{\vtheta^*,n};X_v)\left[\frac{\partial}{\partial\vtheta}\hat\eta_{\vtheta^*,n}^{(v)}-\frac{\partial}{\partial\vtheta}\eta_{\vtheta^*,n}^*\right].
\end{gather*}
By the chain rule, 
\begin{align}
    R_{2,n}^{(v)} =& \frac{\partial}{\partial\vtheta}\left(\frac{\partial}{\partial\eta}\ell_n(\vtheta,\eta_{\vtheta,n};X_v)[\hat\eta^{(v)}_{\vtheta,n}-\eta_{\vtheta,n}]+r_n^{(2,v)}(\vtheta)\right)\bigg|_{\vtheta=\vtheta^*}\nonumber\\
    =&\frac{\partial}{\partial\vtheta}\left(  \frac{\partial}{\partial\eta}\ell_n(\vtheta,\eta_{\vtheta,n};X_v)[\hat\eta^{(v)}_{\vtheta,n}-\eta_{\vtheta,n}]\right)\bigg|_{\vtheta=\vtheta^*}+\frac{\partial}{\partial\vtheta}r_n^{(2,v)}(\vtheta)\bigg|_{\vtheta=\vtheta^*}\nonumber\\
    =& \mathcal{I}_{4,n}^{v}+\mathcal{I}_{5,n}^{v}
      +\frac{\partial}{\partial\vtheta}r_n^{(2,v)}(\vtheta)\bigg|_{\vtheta=\vtheta^*}.\label{eq:normality_chain_rule_R_2}
\end{align}
Lemma \ref{lemma:second-order plug-in error} shows that when Assumption \ref{assumption:np_condition_pointwise} is satisfied, 
\begin{equation}\label{eq:normality_second_order_plugin}
    \frac{\partial}{\partial\vtheta}r_n^{(2,v)}(\vtheta)\bigg|_{\vtheta=\vtheta^*} = o_p(|A_n|^{\frac{1}{2}}).
\end{equation}
By Lemma \ref{proposition:thinning}(iii), $\hat\eta_{\vtheta,n}^{(v)}$ is independent of $X_v$ when $X$ is a Poisson spatial point process. Moreover, Lemma \ref{lemma:empirical process} shows that if either the intensity function is log-linear or $\hat\eta_{\vtheta,n}^{(v)}$ is independent of $X_v$,
\begin{equation}\label{eq:normality_I_4_5}
    \mathcal{I}_{4,5}^{(v)} = o_p(|A_n^{\frac{1}{2}}|),\quad{}\mathcal{I}_{4,5}^{(v)} = o_p(|A_n^{\frac{1}{2}}|).
\end{equation}
Substituting \eqref{eq:normality_second_order_plugin} and 
\eqref{eq:normality_I_4_5} into \eqref{eq:normality_chain_rule_R_2} gives us
\eqref{eq:normality_R_2}

\textit{Step 5:} In this step, we establish \eqref{eq:normality_partial_ell}. Denote
\begin{gather*}
    \hat\eta^{(v)}_n := \hat\eta_{\vtheta,n}^{(v)}\bigg|_{\vtheta=\hat\vtheta_{n}^{(v)}},\\
    \hat{\nu}_n^{(v)} := \frac{\partial}{\partial\vtheta}\hat\eta_{\vtheta,n}^{(v)}\bigg|_{\vtheta=\hat\vtheta_{n}^{(v)}}.
\end{gather*}
By the chain rule and taking supremum over $\vtheta$ and $ \hat\eta^{(v)}_n$, 
\begin{align*}
    \left||A_n|^{-\frac{1}{2}}\frac{\partial}{\partial\vtheta}\ell_n(\vtheta,\hat{\eta}^{(v)}_{\vtheta,n};X_v)\bigg|_{\vtheta=\vtheta^*}\right|\leq & \left|\frac{\partial}{\partial\vtheta}\ell_n(\vtheta,\hat\eta^{(v)}_n;X_v)\bigg|_{\vtheta=\tilde\vtheta_n^{(v)}}\right|
   +\left|\frac{\partial}{\partial\eta}\ell_n(\vtheta,\hat\eta^{(v)}_n;X_v)\bigg|_{\vtheta=\tilde\vtheta_n^{(v)}}\hat{\nu}_n^{(v)}\right|\\
   \leq & \sup_{\vtheta\in\Theta,\eta\in\mathcal{H}}\left|\frac{\partial}{\partial\vtheta}\ell_n(\vtheta,\eta;X_v)\right|
   +\sup_{\vtheta\in\Theta,\eta\in\mathcal{H}}\left|\frac{\partial}{\partial\eta}\ell_n(\vtheta,\eta;X_v)\right|\times\sup_{\nu\in\mathcal{H}^k}\left|\nu\right|.
\end{align*}
Then, \eqref{eq:normality_partial_ell} follows from Corollary \ref{corollary:rate of bias} and the fact that all functions in $\mathcal{H}$ is bounded.

\textit{Step 6:} In this step, we establish \eqref{eq:normality_partial_second_ell}. By equation \eqref{eq:sub_sum_sensitivity},
equation \eqref{eq:normality_partial_second_ell} directly follows from condition \ref{condition:nonsigularity}.

\subsection{Proof of Consistent Variance Estimation (Theorem \ref{thm:variance_estimation} in Supplement Material)} 
 It suffices to show that 
 \begin{gather*}
     \left||A_n|^{-1}\widehat{S}_n(\hat\vtheta_n,\hat\eta_n,\hat{\nu}_n)-\bar{S}_n(\vtheta^*,\eta^*,\nu^*)\right|\rightarrow_p 0,\\
     \left||A_n|^{-1}\widehat{\Sigma}_n(\hat\vtheta_n,\hat\eta_n,\hat{\nu}_n,\hat\psi_n)-\bar{\Sigma}_n(\vtheta^*,\eta^*,\nu^*,\psi^*)\right|\rightarrow_p 0.
 \end{gather*}
 Note that $\widehat{S} _n(\vtheta,\eta,\nu)$ and $\widehat{\Sigma}_n(\vtheta,\eta,\nu,\psi)$ are obtained by quadrature approximation. According to \citet[sec. 4, sec. 5]{baddeley2000practical}, the quadrature approximation error is controlled by the number of the dummy points, i.e, the density of the uniform grid. In this proof, we simply assume that sufficiently many dummy points are generated according to the required estimation accuracy. 

Moreover, since $\widehat{S} _n(\vtheta,\eta,\nu)$ is absolutely continuous with respect to $\vtheta,\eta,\nu$; $\widehat{\Sigma}_n(\vtheta,\eta,\nu,\psi)$  is absolutely continuous with respect to $\vtheta,\eta,\nu,\psi$; $\hat\vtheta_n$ is a consistent estimator of $\vtheta^*$ by Theorem \ref{thm:consistency}; $\hat\eta_n$ is a consistent estimator of $\eta^*$ by Theorem \ref{thm:kernel_estimation}; $\hat\psi_n$ is a  consistent estimators of $\psi^*$, it suffices to show that $\hat{\nu}_n$ defined as \eqref{eq:lfc_est} 
satisfies
\begin{equation}\label{eq:var_nu_consist}
\sup_{\vz\in\mathcal{Z}}|\hat{\nu}_n(\vz)-\nu^*(\vz)|\rightarrow_p 0.
\end{equation} According to Lemma \ref{lemma:bias rate}, Condition \ref{condition:nonsigularity}, \ref{condition:nonsingular_covariance}, and \ref{condition:pair_correlation}. The increasing rate of both the sensitivity matrix and covariance matrix are the same $|A_n|$. Therefore, it suffices to show that 
\begin{gather}
    \left|\frac{\partial^2}{\partial\eta^2}\widehat{\E}\left[\ell_n(\vtheta^*,\eta)\mid \vz\right]\bigg|_{\eta=\eta^*}-\frac{\partial^2}{\partial\eta^2}{\E}\left[\ell_n(\vtheta^*,\eta)\mid \vz\right]\bigg|_{\eta=\eta^*}\right|=o_p(1),\label{eq:thm_var_denom_rate}\\
    \left|\frac{\partial^2}{\partial\vtheta\partial\eta}\widehat{\E}\left[\ell_n(\vtheta^*,\eta)\mid \vz\right]\bigg|_{\eta=\eta^*}-\frac{\partial^2}{\partial\vtheta\partial\eta}{\E}\left[\ell_n(\vtheta^*,\eta)\mid \vz\right]\bigg|_{\eta=\eta^*}\right|=o_p(1), \label{eq:thm_var_num_rate}\\
    \frac{\partial^2}{\partial\eta^2}{\E}\left[\ell_n(\vtheta^*,\eta)\mid \vz\right]\bigg|_{\eta=\eta^*}=\Omega(|A_n|),\label{eq:thm_var_denom_Omega}\\
    \frac{\partial^2}{\partial\vtheta\partial\eta}{\E}\left[\ell_n(\vtheta^*,\eta)\mid \vz\right]\bigg|_{\eta=\eta^*}=O(|A_n|).\label{eq:thm_var_num_big_O}
\end{gather}
Equation \eqref{eq:thm_var_denom_rate} and \eqref{eq:thm_var_num_rate} follow from Lemma \ref{lemma:kernel objective function rate}. Equation \eqref{eq:thm_var_denom_Omega} follows from condition \ref{condition:kernel identification}. Now we consider equation \eqref{eq:thm_var_num_big_O}. By the definition of the Radon–Nikodym derivative $f_n(\vy,\vz)$,
$$\int_{\mathcal{Y}}\int_{\mathcal{Z}} f_n(\vy,\vz)\mathrm{d}\vy\mathrm{d}\vz=|A_n|. $$
Since $\mathcal{Z}$ is compact, we have 
$$\int_{\mathcal{Y}}f_n(\vy,\vz)\mathrm{d}\vy=O(|A_n|),$$
almost surely for $\vz\in\mathcal{Z}$. By condition \ref{condition:smoothness}, $\Psi[\tau_{\vtheta^*}(\vy),\eta^*(\vz)]$ and its derivatives are bounded. Thus, equation \eqref{eq:thm_var_num_big_O} follows from
\begin{align*}
    &\frac{\partial^2}{\partial\vtheta\partial\eta}{\E}\left[\ell_n(\vtheta^*,\eta^*)\mid \vz\right]\\
    =&\int_{\mathcal{Y}}\frac{\partial}{\partial\vtheta}\Psi[\tau_{\vtheta^*}(\vy),\eta^*(\vz)]\frac{\partial}{\partial\eta}\Psi[\tau_{\vtheta^*}(\vy),\eta^*]\Psi[\tau_{\vtheta^*}(\vy),\eta^*(\vz)]f_n(\vy,\vz)\mathrm{d}\vy\\
   =&O(|A_n|).
\end{align*}

\subsection{Proof of Nuisance Estimation Rate (Theorem \ref{thm:kernel_estimation})} \label{proof:prop:kernel_estimation}
The estimator of the nuisance function $\hat\eta_{\vtheta}^{(v)}(\vz)$ is obtained via
$$\hat\eta_{\vtheta}^{(v)}(\vz)=\arg\max_{M} \widehat\E\left[\ell(\vtheta,\eta;X_v^c)|{z}\right].$$
By proposition \ref{proposition:thinning}(ii), it suffices to show that $\hat\eta_{\vtheta,n}(\vz)$ obtained by
$$\hat\eta_{\vtheta,n}(\vz) :=\arg\max_{M}\widehat\E[\ell_n(\vtheta,\eta;X)|\vz]$$ 
using the full process $X$ satisfies the error rate in \eqref{eq:nuisance_est_thm}. To do that, we proceed in three steps. Step 1 shows that $\hat\eta_{\vtheta,n}(\vz)$ is a uniformly consistent estimator of $\eta_{\vtheta,n}(\vz)$. Step 2 shows equation \eqref{eq:nuisance_est_thm} for $j=0$. Step 3 shows equation \eqref{eq:nuisance_est_thm} for $j=1,2$.

\textit{Step 1:}  In this step, we show that $\hat\eta_{\vtheta,n}(\vz)$ is a uniformly consistent estimator of $\eta_{\vtheta,n}(\vz)$ obtained via
$$\eta_{\vtheta,n}(\vz) =\arg\max_{M}\E[\ell_n(\vtheta,\eta)|\vz].$$
When condition \ref{condition:kernel identification} is satisfied, it follows from the Implicit Function Theorem \ref{thm:implicit function theorem} that for every $\vtheta\in\Theta$ and $\vz\in\mathcal{Z}$, $\eta_{\vtheta,n}(\vz)$ is the unique solution of $\partial/\partial\eta\E[\ell_n(\vtheta,\eta)|\vz]=0$  when $n$ is sufficiently large. Combining with Corollary \eqref{corollary:rate of bias}, there exists a positive constant $c$ such that
$$\left||A_n|^{-1}\frac{\partial}{\partial\eta}\E[\ell_n(\vtheta,\eta)|\vz]
\bigg|_{\eta=\hat\eta_{\vtheta,n}}\right|\geq c\left|\hat\eta_{\vtheta,n}(\vz)-\eta_{\vtheta,n}(\vz)\right|.$$
Therefore, for any $\epsilon>0$, there exists $\delta>0$ such that 
\begin{align}
&\pr\left(\sup_{\vtheta\in\Theta,\vz\in\mathcal{Z}}\left|\hat\eta_{\vtheta,n}(\vz)-\eta_{\vtheta,n}(\vz)\right|>\epsilon\right)\nonumber \\
\leq & \pr\left(\sup_{\vtheta\in\Theta,\vz\in\mathcal{Z}}\left||A_n|^{-1}\frac{\partial}{\partial\eta}\E[\ell_n(\vtheta,\eta)|\vz]\bigg|_{\eta=\hat\eta_{\vtheta,n}}\right|>\delta\right)\nonumber\\
\leq & \pr\left(\sup_{\vtheta\in\Theta,\vz\in\mathcal{Z}}|A_n|^{-1}\left|\frac{\partial}{\partial\eta}\widehat\E[\ell_n(\vtheta,\eta)|\vz]\bigg|_{\eta=\hat\eta_{\vtheta,n}}-\frac{\partial}{\partial\eta}\E[\ell_n(\vtheta,\eta)|\vz]\bigg|_{\eta=\hat\eta_{\vtheta,n}}\right|>\delta\right)\nonumber\\
\leq &\pr\left(\sup_{\vtheta\in\Theta,\vz\in\mathcal{Z},\eta\in \mathcal{H}}|A_n|^{-1}\left|\frac{\partial}{\partial\eta}\widehat\E[\ell_n(\vtheta,\eta)|\vz]-\frac{\partial}{\partial\eta}\E[\ell_n(\vtheta,\eta)|\vz]\right|>\delta\right).\label{eq:thm_kernel_consist_align}
\end{align}
Moreover, Lemma \ref{lemma:kernel objective function rate} indicates that 
\begin{equation}\label{eq:thm_kernel_obj_rate_1}
\sup_{\vtheta\in\Theta,\vz\in\mathcal{Z},\eta\in\mathcal{H}}|A_n|^{-1}\left|\frac{\partial}{\partial\eta}\widehat\E[\ell_n(\vtheta,\eta)|\vz]-\frac{\partial}{\partial\eta}\E[\ell_n(\vtheta,\eta)|\vz]\right|=o_p(1).
\end{equation}
If we combine \eqref{eq:thm_kernel_consist_align} and \eqref{eq:thm_kernel_obj_rate_1}, we know that for any $\epsilon>0$,
$$\pr\left(\sup_{\vtheta\in\Theta,\vz\in\mathcal{Z}}\left|\hat\eta_{\vtheta,n}(\vz)-\eta_{\vtheta,n}(\vz)\right|>\epsilon\right)\rightarrow 0.$$
We can thus have the uniform consistency of nuisance estimation:
\begin{equation}\label{eq:thm_kernel_consistent}
\sup_{\vtheta\in\Theta,\vz\in\mathcal{Z}}\left|\hat\eta_{\vtheta,n}(\vz)-\eta_{\vtheta,n}(\vz)\right|=o_p(1).
\end{equation}

\textit{Step 2:} In this step, we establish \eqref{eq:nuisance_est_thm} for $j=0$. 
We denote
\begin{gather*}
    r_n(\vtheta,\vz) := \frac{\partial}{\partial\eta}\widehat\E[\ell_n(\vtheta,\eta)|\vz]\bigg|_{\eta=\hat\eta_{\vtheta,n}}-\frac{\partial}{\partial\eta}\E[\ell_n(\vtheta,\eta)|\vz]\bigg|_{\eta=\hat\eta_{\vtheta,n}},\\
    d_n(\vtheta,\vz):=\int_0^1 \frac{\partial^2}{\partial\eta^2}\E[\ell_n(\vtheta,\eta)|\vz]\bigg|_{\eta=t\hat\eta_{\vtheta,n}+(1-t)\eta_{\vtheta,n}}\mathrm{d}t.
\end{gather*}
Since $\hat\eta_{\vtheta,n}(\vz)$ and $\eta_{\vtheta,n}(\vz)$ satisfy
\begin{gather*}
     \frac{\partial}{\partial\eta}\widehat\E[\ell_n(\vtheta,\eta)|\vz]\bigg|_{\eta=\hat\eta_{\vtheta,n}}=0,\\
     \frac{\partial}{\partial\eta}\E[\ell_n(\vtheta,\eta)|\vz]\bigg|_{\eta=\eta_{\vtheta,n}}=0.
\end{gather*}
By Taylor's Theorem, we have
\begin{align}
     0 = &\frac{\partial}{\partial\eta}\widehat\E[\ell_n(\vtheta,\eta)|\vz]\bigg|_{\eta=\hat\eta_{\vtheta,n}} - \frac{\partial}{\partial\eta}\E[\ell_n(\vtheta,\eta)|\vz]\bigg|_{\eta=\eta_{\vtheta,n}} \nonumber\\
     =& r_n(\vtheta,\vz) +d_n(\vtheta,\vz) \left[\hat\eta_{\vtheta,n}(\vz)-\eta_{\vtheta,n}(\vz)\right]\label{eq:kernel taylor}
\end{align}
By the identification condition \ref{condition:kernel identification} and the uniform consistency of $\hat\eta_{\vtheta,n}(\vz)$ in \eqref{eq:thm_kernel_consistent},
\begin{align}
    \liminf_n\inf_{\vtheta,\vz} d_n(\vtheta,\vz)=&\liminf_n\inf_{\vtheta,\vz} \int_0^1 \frac{\partial^2}{\partial\eta^2}\E[\ell_n(\vtheta,\eta)|\vz]\bigg|_{\eta=t\hat\eta_{\vtheta,n}+(1-t)\eta_{\vtheta,n}}\mathrm{d}t\nonumber\\
    =&\inf_{\vtheta,\vz} \int_0^1 \liminf_n\frac{\partial^2}{\partial\eta^2}\E[\ell_n(\vtheta,\eta)|\vz]\bigg|_{\eta=t\hat\eta_{\vtheta,n}+(1-t)\eta_{\vtheta,n}}\mathrm{d}t\nonumber\\
    =&\inf_{\vtheta,\vz} \frac{\partial^2}{\partial\eta^2}\E[\ell_n(\vtheta,\eta)|\vz]\bigg|_{\eta=\eta_{\vtheta,n}}=\Omega(|A_n|). \label{eq:thm_kernel_rate_d}
\end{align}
Moreover, Lemma \ref{lemma:kernel objective function rate} shows that 
\begin{equation}\label{eq:thm_kernel_rate_r}
    \sup_{\vtheta,\vz}\left|r_n(\vtheta,\vz)\right|=o_p\left(|A_n|^{1-\frac{m}{2(m+k+q+1)}\cdot \frac{l}{l+q+1}}\right).
\end{equation}
Substituting \eqref{eq:thm_kernel_rate_d}, \eqref{eq:thm_kernel_rate_r} into  \eqref{eq:kernel taylor} gives us
\begin{equation}\label{eq:thm_kernel_eta_consistency}
    \sup_{\vtheta\in\Theta,\vz\in\mathcal{Z}}\left|\hat\eta_{\vtheta,n}(\vz)-\eta_{\vtheta,n}(\vz)\right| = o_p\left(|A_n|^{-\frac{m}{2(m+k+q+1)}\cdot \frac{l}{l+q+1}}\right).
\end{equation}

\textit{Step 3:} In this step, we establish \eqref{eq:nuisance_est_thm} for $j=1,2$. Differentiating equation \eqref{eq:kernel taylor} with respect to $\vtheta$ yields
\begin{equation}\label{eq:thm_kernel_taylor_diff}
    0 = \frac{\partial}{\partial\vtheta}r_n(\vtheta,\vz) + \frac{\partial}{\partial\vtheta}d_n(\vtheta,\vz) \cdot[\hat\eta_{\vtheta,n}(\vz)-\eta_{\vtheta,n}(\vz)]+ d_n(\vtheta,\vz) \cdot\left[\frac{\partial}{\partial\vtheta}\hat\eta_{\vtheta,n}(\vz)-\frac{\partial}{\partial\vtheta}\eta_{\vtheta,n}(\vz)\right].
\end{equation}
It followed from Corollary \ref{corollary:rate of bias} that 
$$\sup_{\vtheta,\eta,\vz}\left|\frac{\partial^2}{\partial\eta^2\partial\vtheta}\E[\ell_n(\vtheta,\eta)|\vz]\right|=O(|A_n|).$$
Thus, 
\begin{equation}\label{eq:thm_kernel_rate_d_diff}
    \sup_{\vtheta,\vz}\left|\frac{\partial}{\partial\vtheta}d_n(\vtheta,\vz)\right|=O(|A_n|).
\end{equation}
Moreover, Lemma \ref{lemma:kernel objective function rate} shows that
\begin{equation}\label{eq:thm_kernel_rate_r_diff}
    \sup_{\vtheta,\vz}\left|\frac{\partial}{\partial\vtheta}r_n(\vtheta,\vz)\right| =o_p\left(|A_n|^{1-\frac{m}{2(m+k+q+1)}\cdot \frac{l}{l+q+1}}\right).
\end{equation}
Substituting \eqref{eq:thm_kernel_eta_consistency}, \eqref{eq:thm_kernel_rate_d_diff}, \eqref{eq:thm_kernel_rate_r_diff} into  \eqref{eq:thm_kernel_taylor_diff} gives us
$$ \sup_{\vtheta,\vz}\left|\frac{\partial}{\partial\vtheta}\hat\eta_{\vtheta,n}(\vz)-\frac{\partial}{\partial\vtheta}\eta_{\vtheta,n}(\vz) \right| = o_p\left(|A_n|^{-\frac{m}{2(m+k+q+1)}\cdot \frac{l}{l+q+1}}\right).$$
If we differentiate \eqref{eq:kernel taylor} twice with respect to $\vtheta$ and use a similar approach, we would obtain 
$$ \sup_{\vtheta,\vz}\left|\frac{\partial^2}{\partial\vtheta^2}\hat\eta_{\vtheta,n}(\vz)-\frac{\partial^2}{\partial\vtheta^2}\eta_{\vtheta,n}(\vz)\right| = o_p\left(|A_n|^{-\frac{m}{2(m+k+q+1)}\cdot \frac{l}{l+q+1}}\right).$$    

\subsection{Proof of Semiparametric Efficiency (Corollary \ref{corallary:Efficiency} in main manuscript)}

    The PCF of a Poisson spatial point process is always zero. Thus, when $X$ is a Poisson spatial point process, its covariance matrix $\Sigma(\vtheta^*,\eta^*,\nu^*,\psi^*)$ collapses to the sensitivity matrix $S(\vtheta^*,\eta^*,\nu^*)$. Then, by Theorem \ref{thm:normality}, the asymptotic covariance of $\hat\vtheta_n$ is $S_n^{-1}(\vtheta^*,\eta^*,\nu^*)$, which is the semiparametric lower bound in Theorem \ref{thm:least_favorable_direction}. Thus, $\hat\vtheta_n$ is semiparametrically efficient when $X$ is Poisson.

\newpage

\section{Proof of Auxiliary Lemmas}

\renewcommand{\theequation}{B.\arabic{equation}}
\setcounter{equation}{0}

\subsection{Multivariate Pythagorean Theorem}
\begin{lemma}[Multivariate Pythagorean theorem]\label{thm:pythagorean}
Let $L_2(A)$ be space of square-integrable functions on $A\in\R^2$, $\mathcal{U}\subset L_2(A)$ be a linear subspace, $L_2^{(k)}(A)$ be the $k$-product space of $L_2(A)$, $\mathcal{U}^k$ be the $k$-product space of $\mathcal{U}$. For any $f\in \mathcal{U}^k$ and $g\in L_2^{(k)}(A)$ that is orthogonal to $\mathcal{U}^k$,
$$\left|f+g\right\rangle \left\langle f+g\right|_A = \left|f\right\rangle \left\langle f\right|_A +  \left|g\right\rangle \left\langle g\right|_A.$$
\end{lemma}

\begin{proof}:

Denote the k-dimensional element $g = (g_1,\ldots,g_k)$, $f = (f_1,\ldots,f_k)$. $g$ is orthogonal to $\mathcal{U}^k$ if and only if for every $j=1,\ldots,k$, $g_j$ is orthogonal to $\mathcal{U}$. Consequently, for any $i,j\in[k]$,
    $$\int_A \lambda(\vu;\vtheta^*,\eta^*)f_i(\vu) g_j(\vu) \mathrm{d}\vu=0.$$
Thus, we have
    \begin{align*}
        \left|f\right\rangle \left\langle g\right|_A & = \int_A \lambda(\vu;\vtheta^*,\eta^*)f(\vu) g(\vu)^\top \mathrm{d}\vu=0^{k\times k},\\
        \left|f+g\right\rangle \left\langle f+g\right| & = \left|f\right\rangle \left\langle f\right|_A +  \left|g\right\rangle \left\langle g\right|_A + 2\left|f\right\rangle \left\langle g\right|_A,\\
        & =  \left|f\right\rangle \left\langle f\right|_A +  \left|g\right\rangle \left\langle g\right|_A.
    \end{align*}
\end{proof}

\subsection{Rate of Spatial Summation}
\begin{lemma}[Spatial Summation]\label{lemma:bias rate}
Let $X$ be a spatial point process with intensity function $\lambda(\vu;\vtheta,\eta)$ for some $\vtheta\in\Theta,\eta\in\mathcal{H}$. 
Let $\{A_n\}_{n=1}^\infty$ be an expanding sequence of region in $\R^2$ that $|A_n|\rightarrow \infty$. Let $f(\vu)$ be a bounded function defined on $\R^2$ such that $|f(\vu)|<C$ on $\cup_{n=1}^\infty A_n$ for some positive constant $C$. Under condition \ref{condition:smoothness}, \ref{condition:boundedness}, \ref{condition:pair_correlation}, we have
$$\E\left[\sum_{\vu\in X\cap A_n}f(\vu)\right]=O(|A_n|),\quad{}\sum_{\vu\in X\cap A_n}f(\vu)-\E\left[\sum_{\vu\in X\cap A_n}f(\vu)\right]=O_p(|A_n|^{\frac{1}{2}}).$$
Furthermore, if there exists a positive constant $c$ such that $|f(\vu)|>c$ on $\cup_{n=1}^\infty A_n$,
$$\E\left[\sum_{\vu\in X\cap A_n}f(\vu)\right]=\Theta(|A_n|).$$
\end{lemma}

\begin{proof}:
We proceed with the proof in two steps. Step 1 shows the rate of the expectation. Step 2 shows the rate of the deviation.

\textit{Step 1:} By Campbell's theorem, we have
$$\sum_{\vu\in X\cap A_n}f(\vu) = \int_{A_n}f(\vu)\lambda(\vu;\vtheta,\eta)\mathrm{d}\vu.$$
 Denote $B:=\{\vu:\lambda(\vu;\vtheta,\eta)\leq c_2\}$ as the set defined in condition \ref{condition:boundedness}. Then, we have
\begin{equation}\label{eq:rate_bias_exp}
     \E\left[\sum_{\vu\in X\cap A_n}f(\vu)\right]  = \int_{A_n\cap B^c} f(\vu)\lambda(\vu;\vtheta,\eta)\mathrm{d}\vu +  \int_{A_n\cap B} f(\vu)\lambda(\vu;\vtheta,\eta)\mathrm{d}\vu.
\end{equation}
By condition \ref{condition:smoothness} and the compactness of the target parameter and nuisance parameter space, there exists a positive constant $C_0$ such that $$\sup_{\vu\in\R^2,\vtheta\in\Theta,\eta\in\mathcal{H}}\lambda(\vu;\vtheta,\eta)<C_0.$$
If we further have $c<f(\vu)<C$ on $\cup_{n=1}^\infty A_n$, we have 
\begin{gather}
     c\cdot c_2(|A_n|-|B|)\leq \int_{A_n\cap B^c} f(\vu)\lambda(\vu;\vtheta,\eta)\mathrm{d}\vu\leq C\cdot C_0|A_n|\label{eq:rate_bias_exp_bound_1},\\
      0\leq \int_{A_n\cap B} f(\vu)\lambda(\vu;\vtheta,\eta)\mathrm{d}\vu\leq C\cdot c_2|B| \label{eq:rate_bias_exp_bound_2}.
\end{gather}
By condition \ref{condition:boundedness}, $B$ is bounded. Then, substituting \eqref{eq:rate_bias_exp_bound_1} and \eqref{eq:rate_bias_exp_bound_2} into \eqref{eq:rate_bias_exp} gives us
 $$\E\left[\sum_{\vu\in X\cap A_n}f(\vu)\right] = \Theta(|A_n|).$$
 When $f(\vu)<C$ on $\cup_{n=1}^\infty A_n$, only the upper bounds in \eqref{eq:rate_bias_exp_bound_1} and \eqref{eq:rate_bias_exp_bound_2} hold. Then, substituting \eqref{eq:rate_bias_exp_bound_1} and \eqref{eq:rate_bias_exp_bound_2} into \eqref{eq:rate_bias_exp} gives us
 $$\E\left[\sum_{\vu\in X\cap A_n}f(\vu)\right] = O(|A_n|).$$

 \textit{Step 2:} Denote
 \begin{gather*}
     R_1:=\int_{A_n}\int_{A_n} f(\vu)f(v)\lambda(\vu;\vtheta,\eta)\lambda(v;\vtheta,\eta)[g(\vu,v)-1]\mathrm{d}\vu \mathrm{d}v, \\
     R_2:=\int_{A_n}f(\vu)^2\lambda(\vu;\vtheta,\eta)\mathrm{d}\vu.
 \end{gather*}
 
 The variance of $\sum_{\vu\in X\cap A_n}f(\vu)$ satisfies
 $$\text{Var}\left[\sum_{\vu\in X\cap A_n}f(\vu)\right] = R_1+R_2.$$
By the boundedness of  $f(\vu)$ and $\lambda(\vu;\vtheta,\eta)$, 
\begin{align*}
    R_2\leq C_0 C^2|A_n| = O(|A_n|).
\end{align*}
Moreover, by condition \ref{condition:pair_correlation}, 
\begin{align*}
    R_1  &\leq C^2C_0^2 \int_{A_n}\int_{A_n} \left|g(\vu,v)-1\right|\mathrm{d}\vu \mathrm{d}v\\ 
   & = C^2C_0^2 \int_{A_n}\int_{\R^2} \left|g(\vu,v)-1\right|\mathrm{d}\vu \mathrm{d}v \\ 
    & = C^2C_0^2 C_2 \int_{A_n}\mathrm{d}v= O(|A_n|),
\end{align*}
where the constant $C_2$ is defined in condition \ref{condition:pair_correlation}. 
Therefore, 
$$\text{Var}\left[\sum_{\vu\in X\cap A_n}f(\vu)\right] = O(|A_n|).$$
Then, there exists a positive constant C and a positive integer $N$ such that for $n>N$, 
$$\text{Var}\left[\sum_{\vu\in X\cap A_n}f(\vu)\right] \leq C|A_n|.$$
By Markov inequality, for any $\epsilon>0$, let $\delta_\epsilon=(C/\epsilon)^{\frac{1}{2}}$. Then, for $n>N$, 
\begin{align*}
&\pr\left\{\sum_{\vu\in X\cap A_n}f(\vu)-\E\left[\sum_{\vu\in X\cap A_n}f(\vu)\right]\geq \delta_\epsilon|A_n|^{\frac{1}{2}}\right\}\\
\leq &\frac{\text{Var}\left[\sum_{\vu\in X\cap A_n}f(\vu)\right]}{\delta_{\epsilon}^2|A_n|}\\
=& \frac{C}{\delta_{\epsilon}^2}=\epsilon.
\end{align*}
Therefore, 
$$\sum_{\vu\in X\cap A_n}f(\vu)-\E\left[\sum_{\vu\in X\cap A_n}f(\vu)\right]=O_p(|A_n|^{\frac{1}{2}}).$$

\end{proof}

\begin{corollary}[Spatial Summation]\label{corollary:rate of bias}
Let $X$ be a spatial point process with intensity function $\lambda(\vu;\vtheta,\eta)$. Let $\{A_n\}_{n=1}^\infty$ be a sequence of expanding regions in $\R^2$. Under condition \ref{condition:smoothness}, \ref{condition:boundedness}, \ref{condition:pair_correlation}, we have  
\begin{equation}\label{eq:rate of points num}
    \E\left[|X\cap A_n|\right] = \Theta(|A_n|),\quad{}\left||X\cap A_n|-\E\left[|X\cap A_n|\right]\right|=O_p(|A_n|^{\frac{1}{2}}).
\end{equation}
 For $i,j\in\{0,1,2\}$, denote $$\ell_n^{(i,j)}(\vtheta,\eta)=\frac{\partial^{i+j}}{\partial\vtheta^i\partial\eta^j}\ell_n(\vtheta,\eta),$$ 
 we have
\begin{equation}\label{eq:rate of likelihood bias}
\sup_{\vtheta,\eta\in\mathcal{H}}\E\left[\ell_n^{(i,j)}(\vtheta,\eta)\right] = O(|A_n|),\quad{}\sup_{\vtheta\in\vtheta,\eta\in\mathcal{H}}\left|\ell_n^{(i,j)}(\vtheta,\eta)-\E\left[\ell_n^{(i,j)}(\vtheta,\eta)\right]\right|=O_p(|A_n|^{\frac{1}{2}}).
\end{equation}
\end{corollary}

\begin{proof}:
If we let $f(\vu)=1$ in Lemma \ref{lemma:bias rate}, equation \eqref{eq:rate of points num} follows. Note that 
$$\frac{\partial^{i+j}}{\partial\vtheta^i\partial\eta^j}\ell_n(\vtheta,\eta)=\sum_{\vu\in X\cap A_n}\frac{\partial^{i+j}}{\partial\vtheta^i\partial\eta^j}\log\lambda(\vu;\vtheta,\eta)-\int_{A_n}\frac{\partial^{i+j}}{\partial\vtheta^i\partial\eta^j}\lambda(\vu;\vtheta,\eta)\mathrm{d}\vu.$$
Therefore, if we apply Lemma \ref{lemma:bias rate} to
$$f(\vu)=\frac{\partial^{i+j}}{\partial\vtheta^i\partial\eta^j}\log\lambda(\vu;\vtheta,\eta),$$
and 
$$f(\vu)=\frac{\partial^{i+j}}{\partial\vtheta^i\partial\eta^j}\lambda(\vu;\vtheta,\eta),$$ 
We will have equation \eqref{eq:rate of likelihood bias}.
\end{proof}

\subsection{Key Lemma for Proof of Theorem \ref{thm:consistency}}
\begin{lemma}[Sufficient Separation]\label{lemma:sufficient separation}
Let $X$ be the spatial point process with an intensity function $\lambda(\vu;\vtheta^*,\eta^*)$. Let $\{A_n\}_{n=1}^\infty$ be a sequence of expanding regions in $\R^2$. Under condition \ref{condition:Sufficient Separation}, \ref{condition:boundedness}, for any $(\vtheta,\eta)\neq (\vtheta^*,\eta^*)$, 
$$\E[\ell_n(\vtheta^*,\eta^*)]-\E[\ell_n(\vtheta,\eta)]=\Theta(|A_n|)\cdot\min\{|\vtheta-\vtheta^*|,1\}.$$
\end{lemma}

\begin{proof}:
    For any $(\vtheta,\eta)\neq(\vtheta^*,\eta^*)$, denote $$\Phi(\vu;\vtheta,\eta) :=\log\lambda(\vu;\vtheta^*,\eta^*)-\log\lambda(\vu;\vtheta,\eta).$$ 
    Then, the left-hand side of desired equation in Lemma \ref{lemma:sufficient separation} can be written as
    \begin{align}
    &\E[\ell_n(\vtheta^*,\eta^*)]-\E[\ell_n(\vtheta,\eta)]\nonumber\\ 
    =&\int_{A_n}[\log\lambda(\vu;\vtheta^*,\eta^*)-\log\lambda(\vu;\vtheta,\eta)]\lambda(\vu;\vtheta^*,\eta^*)\mathrm{d}\vu - \int_{A_n} [\lambda(\vu;\vtheta^*,\eta^*)-\lambda(\vu;\vtheta,\eta)]\mathrm{d}\vu\nonumber\\
    =& \int_{A_n}\lambda(\vu;\vtheta^*,\eta^*)\left[\log\lambda(\vu;\vtheta^*,\eta^*)-\log\lambda(\vu;\vtheta,\eta)-1+\frac{\lambda(\vu;\vtheta,\eta)}{\lambda(\vu;\vtheta^*,\eta^*)}\right]\mathrm{d}\vu \nonumber\\
    =& \int_{A_n}\lambda(\vu;\vtheta^*,\eta^*)\left\{\exp[\Phi(\vu;\vtheta,\eta)]-\Phi(\vu;\vtheta,\eta)-1\right\}\mathrm{d}\vu \label{eq:lemma_sufficient_separation_rewrite}
    \end{align}
    Since the inequality $\exp(x)\geq x +1$ has equality if and only if $x=0$, it follows from condition \ref{condition:Sufficient Separation} that there exist positive constants $c_0^\prime,c_1^\prime$ such that 
    \begin{equation}\label{eq:lemma_suff_lower_bound}
        \inf_{\vu\in C}|\exp[\Phi(\vu;\vtheta,\eta)]-\Phi(\vu;\vtheta,\eta)-1|\geq \min\{c_0^\prime,c_1^\prime|\vtheta-\vtheta^*|\},
    \end{equation}
    where the set $C$ is defined in \ref{condition:Sufficient Separation} and $|A_n\cap C|=\Theta(|A_n|)$. Additionally, denote 
    $B:=\{\vu:\inf_{\vtheta,\eta}\lambda(\vu;\vtheta,\eta)<c_2\}$, by condition \ref{condition:boundedness}, 
    \begin{equation}\label{eq:lemma_sufficient_separation_B}
        |B^c\cap A_n|=\Theta(|A_n|).
    \end{equation}
    Combine \eqref{eq:lemma_suff_lower_bound} and \eqref{eq:lemma_sufficient_separation_B} with  \eqref{eq:lemma_sufficient_separation_rewrite}, we have
\begin{align*}
    \E[\ell_n(\vtheta^*,\eta^*)]-\E[\ell_n(\vtheta,\eta)]
    \geq & c_2\int_{A_n\cap B^c} \left\{\exp[\Phi(\vu;\vtheta,\eta)]-\Phi(\vu;\vtheta,\eta)-1\right\}\mathrm{d}\vu\nonumber\\
    \geq & c_2\int_{A_n\cap B^c\cap C}\min\{c_0^\prime,c_1^\prime|\vtheta-\vtheta^*|\} \mathrm{d}\vu\nonumber\\
    \geq & c_2 \min\{c_0^\prime,c_1^\prime|\vtheta-\vtheta^*|\}|\cdot\Theta(|A_n|)\nonumber\\
    =& \Theta(|A_n|)\cdot\min\{|\vtheta-\vtheta^*|,1\}.
\end{align*}
\end{proof}

\subsection{Rate of Plug-in Error}

\begin{lemma}[First Order Plug-in Error]\label{lemma:first order plug-in error}
    Let $\hat{\eta}_{{\vtheta},n}$ be a sequence of estimators of $\eta_{{\vtheta},n}$. Let
    \begin{gather}
         r_n^{(1)}(\vtheta)=\ell_n(\vtheta,\hat{\eta}_{{\vtheta},n}) -\ell_n(\vtheta,{\eta}_{{\vtheta},n})\label{eq:r_1},\\
         e_n^{(1)} =  \sup_{j\in\{0,1,2\},\vtheta\in\Theta,\vz\in\mathcal{Z}}  \left|\frac{\partial^j}{\partial\vtheta^j}\left[\hat{\eta}_{{\vtheta},n}(\vz)-{\eta}_{{\vtheta},n}(\vz)\right]\right|\label{eq:e_1}.
    \end{gather}
    \noindent Then, under condition \ref{condition:smoothness}, \ref{condition:boundedness}, \ref{condition:pair_correlation}, 
    \begin{gather}
\sup_{\vtheta\in\vtheta}|r_n^{(1)}(\vtheta)|=O(|A_n|\cdot e_n^{(1)})\label{eq:plug_in_error_1},\\
\sup_{\vtheta\in\vtheta}\left|\frac{\partial^2}{\partial\vtheta^2}r_n^{(1)}(\vtheta)\right| = O(|A_n|\cdot e_n^{(1)}) \label{eq:plug_in_error_2}.
    \end{gather}
\end{lemma}
\begin{proof}:
    For fixed $\vtheta$, denote
\begin{gather*}
\hat{\eta}_{{\vtheta},n}(t):=
\eta_{\vtheta,n}(\vz_\vu)+t\left[\hat{\eta}_{{\vtheta},n}(\vz_\vu)-{\eta}_{{\vtheta},n}(\vz_\vu)\right],\quad{} t\in[0,1],\\
        Q_{\vtheta,n}^{(1)}(\vu)
    := \int_0^1 \left\{\frac{\partial}{\partial\eta}\Psi[\tau_{\vtheta}(\vy_\vu),\eta(\vz_\vu)]\bigg|_{\eta = \hat{\eta}_{{\vtheta},n}(t)} \right\}dt,\\
    Q_{\vtheta,n}^{(2)}(\vu) 
     := \int_0^1 \left\{\frac{\partial}{\partial\eta}\log\Psi[\tau_{\vtheta}(\vy_\vu),\eta(\vz_\vu)]\bigg|_{\eta = \hat{\eta}_{{\vtheta},n}(t)} \right\}dt.
\end{gather*}

Then, by Taylor's Theorem, $r_n^{(1)}(\vtheta)$ defined in \eqref{eq:r_1} can be expressed as follows:
\begin{align}
    r_n^{(1)}(\vtheta)=& \sum_{\vu\in X\cap A_n}\left[\log\lambda(\vu;\vtheta,\hat{\eta}_{\vtheta,n})-\log\lambda(\vu;\vtheta,\eta_{\vtheta,n})\right]-\int_{A_n}\left[\lambda(\vu;\vtheta,\hat{\eta}_{\vtheta,n})-\lambda(\vu;\vtheta,\eta_{{\vtheta},n})\right]\mathrm{d}\vu\nonumber\\
    =&  \sum_{\vu\in X\cap A_n} Q_{\vtheta,n}^{(2)}(\vu)\cdot \left[\hat{\eta}_{{\vtheta},n}(\vz_\vu)-{\eta}_{{\vtheta},n}(\vz_\vu)\right] - \int_{A_n}Q_{\vtheta,n}^{(1)}(\vu)\cdot \left[\hat{\eta}_{{\vtheta},n}(\vz_\vu)-{\eta}_{{\vtheta},n}(\vz_\vu)\right]\mathrm{d}\vu \label{eq:r_1_decompose}.
\end{align}
The second order derivative of $r_n^{(1)}(\vtheta)$ can be expressed as follows:
\begin{multline}\label{eq:r_1_decompose_derivative}
     \frac{\partial^2}{\partial\vtheta^2}r_n^{(1)}(\vtheta)=\sum_{i=0}^2\left\{ \sum_{\vu\in X\cap A_n} \frac{\partial^i}{\partial\vtheta^i}Q_{\vtheta,n}^{(2)}(\vu)\cdot \frac{\partial^{2-i}}{\partial\vtheta^{2-i}}\left[\hat{\eta}_{{\vtheta},n}(\vz_\vu)-{\eta}_{{\vtheta},n}(\vz_\vu)\right]\right\} - \\
\sum_{i=0}^2\left\{\int_{A_n}\frac{\partial^i}{\partial\vtheta^i}Q_{\vtheta,n}^{(1)}(\vu)\cdot \frac{\partial^{2-i}}{\partial\vtheta^{2-i}}\left[\hat{\eta}_{{\vtheta},n}(\vz_\vu)-{\eta}_{{\vtheta},n}(\vz_\vu)\right]\mathrm{d}\vu\right\}.
\end{multline}
By the smoothness condition \ref{condition:smoothness}, there exists a positive constant $C$ such that for $i=0,1,2$ and $j=1,2$,
\begin{equation}\label{eq:Q_bound_1}
     \sup_{\vu\in\R^2,\vtheta\in\Theta}\left|\frac{\partial^i}{\partial\vtheta^i}Q_{\vtheta,n}^{(j)}(\vu)\right|<C.
\end{equation}

\noindent Corollary \ref{corollary:rate of bias} shows that $|X\cap A_n|=O(|A_n|)$. Thus, plugging \eqref{eq:Q_bound_1} into \eqref{eq:r_1_decompose} and \eqref{eq:r_1_decompose_derivative} gives us
\begin{gather*}
\sup_{\vtheta\in\vtheta}|r_n^{(1)}(\vtheta)|=O(|A_n|\cdot e_n^{(1)}),\\
\sup_{\vtheta\in\vtheta}\left|\frac{\partial^2}{\partial\vtheta^2}r_n^{(1)}(\vtheta)\right| = O(|A_n|\cdot e_n^{(1)}).
    \end{gather*}
\end{proof}

\begin{lemma}[Second Order Plug-in Error]\label{lemma:second-order plug-in error}
    Let $\hat{\eta}_{{\vtheta},n}$ be a sequence of estimators of $\eta_{{\vtheta},n}$. Let
    \begin{gather}
        r_n^{(2)}(\vtheta)= \ell_n(\vtheta,\hat{\eta}_{{\vtheta},n}) -\ell_n(\vtheta,{\eta}_{{\vtheta},n})-\frac{\partial}{\partial\eta}\ell_n(\vtheta,{\eta}_{{\vtheta},n})\left[\hat{\eta}_{{\vtheta},n}(\vz)-{\eta}_{{\vtheta},n}(\vz)\right]\label{eq:r_2},\\
        e_n^{(2)}{(\vtheta)} =  \sup_{j\in\{0,1\},\vz\in\mathcal{Z}}  \left|\frac{\partial^j}{\partial\vtheta^j}\left[\hat{\eta}_{{\vtheta},n}(\vz)-{\eta}_{{\vtheta},n}(\vz)\right]\right|\label{eq:e_2}.
    \end{gather}
    \noindent Then, under condition \ref{condition:boundedness}, \ref{condition:pair_correlation}, we have 
   \begin{equation}\label{eq:plug_in_error_3}
    \left|\frac{\partial}{\partial\vtheta}r_n^{(2)}(\vtheta)\bigg|_{\vtheta=\vtheta^*}\right|=O_p(|A_n|\cdot e_n^{(2)}(\vtheta^*)).
\end{equation}
\end{lemma}

\begin{proof}:
    For fixed $\vtheta$, we denote
    $$\hat{\eta}_{{\vtheta},n}(t):=
\eta_{\vtheta,n}(\vz_\vu)+t\left[\hat{\eta}_{{\vtheta},n}(\vz_\vu)-{\eta}_{{\vtheta},n}(\vz_\vu)\right],\quad{} t\in[0,1],$$
$$Q_{\vtheta,n}^{(3)}(\vu):= \frac{1}{2}\int_0^1 \left\{\frac{\partial^2}{\partial\eta^2}\Psi[\tau_{\vtheta}(\vy_\vu),\eta(\vz_\vu)]\bigg|_{\eta = \hat{\eta}_{{\vtheta},n}(t)} \right\}dt,$$
$$ Q_{\vtheta,n}^{(4)}(\vu):=\frac{1}{2} \int_0^1 \left\{\frac{\partial^2}{\partial\eta^2}\log\Psi[\tau_{\vtheta}(\vy_\vu),\eta(\vz_\vu)]\bigg|_{\eta = \hat{\eta}_{{\vtheta},n}(t)} \right\}dt.$$
By Taylor's Theorem, $r_n^{(2)}(\vtheta)$ defined in \eqref{eq:r_2} can be expressed as follows:
\begin{align*}
    r_n^{(2)}(\vtheta)=& 
     \sum_{\vu\in X\cap A_n} Q_{\vtheta,n}^{(4)}(\vu)\cdot \left[\hat{\eta}_{{\vtheta},n}(\vz_\vu)-{\eta}_{{\vtheta},n}(\vz_\vu)\right]^2 - \int_{A_n}Q_{\vtheta,n}^{(3)}(\vu)\cdot \left[\hat{\eta}_{{\vtheta},n}(\vz_\vu)-{\eta}_{{\vtheta},n}(\vz_\vu)\right]^2\mathrm{d}\vu. 
\end{align*}
Moreover, the first-order derivative of $r_n^{(2)}(\vtheta)$ can be expressed as follows:
\begin{multline}\label{eq:r_2_decompose}
     \frac{\partial}{\partial\vtheta}r_n^{(2)}(\vtheta)
    =\sum_{i=0}^1\left\{\sum_{\vu\in X\cap A_n}  \frac{\partial^i}{\partial\vtheta^i}Q_{\vtheta,n}^{(4)}(\vu)\cdot \frac{\partial^{1-i}}{\partial\vtheta^{1-i}}\left[\hat{\eta}_{{\vtheta},n}(\vz_\vu)-{\eta}_{{\vtheta},n}(\vz_\vu)\right]^2 \right\}\\
    - \left\{\int_{A_n}\frac{\partial^i}{\partial\vtheta^i}Q_{\vtheta,n}^{(3)}(\vu)\cdot \frac{\partial^{1-i}}{\partial\vtheta^{1-i}}\left[\hat{\eta}_{{\vtheta},n}(\vz_\vu)-{\eta}_{{\vtheta},n}(\vz_\vu)\right]^2 \mathrm{d}\vu\right\}.
\end{multline}

By the smoothness condition \ref{condition:smoothness}, there exists a positive constant $C$ such that for $i=0,1$ and $j=3,4$,
\begin{equation}\label{eq:Q_bound_2}
     \sup_{\vu\in\R^2,\vtheta\in\Theta}\left|\frac{\partial^i}{\partial\vtheta^i}Q_{\vtheta,n}^{(j)}(\vu)\right|<C.
\end{equation}
Corollary \ref{corollary:rate of bias} shows that $|X\cap A_n|=O(|A_n|)$. Thus, plugging \eqref{eq:Q_bound_2} into \eqref{eq:r_2_decompose} gives us
\begin{equation*}
    \left|\frac{\partial}{\partial\vtheta}r_n^{(2)}(\vtheta)\bigg|_{\vtheta=\vtheta^*}\right|=O_p(|A_n|\cdot e_n^{(2)}(\vtheta^*)).
\end{equation*}
\end{proof}

\subsection{Rate of Spatial Empirical Processes}

\begin{lemma}[Spatial Empirical Process]\label{lemma:spatial_ep}
    Let $X$ be a spatial point process with a log-linear intensity function and $\{A_n\}_{n=1}^\infty$ be a sequence of expanding windows.
    Also, let $\mathcal{F}$ be an arbitrary class of stochastic functions defined on $\R^2$ and $\mathcal{F}^\prime:=\left\{\partial/\partial\vtheta\left(\eta_{\vtheta,n}- \hat\eta_{\vtheta,n}\right)\right\}$. 
   For every $f\in\mathcal{F}\cup\mathcal{F}^\prime$, denote 
    \begin{gather*}
        V_n^{(1)}(f)(\vu) := \frac{\partial^2}{\partial\vtheta\partial\eta}\log\lambda(\vu;\vtheta,\eta_{\vtheta,n})\bigg|_{\vtheta=\vtheta^*}[f(\vu)], \quad{} 
    V_n^{(2)}(f)(\vu) := \frac{\partial}{\partial\eta}\log\lambda(\vu;\vtheta,\eta_{\vtheta,n})\bigg|_{\vtheta=\vtheta^*}[f(\vu)].
    \end{gather*}
If condition \ref{condition:pair_correlation} holds, we have  
\begin{gather}
    \sup_{f\in\mathcal{F}}\left|\mathbb{G}^{X}_n\left[V_n^{(1)}(f)\right]\right|=0\label{eq:lemma_spatial_ep_1},\\
    \sup_{f\in\mathcal{F}^\prime}\left|\mathbb{G}^{X}_n\left[V_n^{(2)}(f)\right]\right|=O_p\left(|\mathcal{F}^\prime|\right)\label{eq:lemma_spatial_ep_2}.
\end{gather}
\end{lemma}

\begin{proof}:
We establish \eqref{eq:lemma_spatial_ep_1} in step 1, and \eqref{eq:lemma_spatial_ep_2} in step 2.

\textit{Step 1:} When the intensity function of $X$ is log-linear, i.e. $\lambda(\vu;\vtheta,\eta)=\exp[\vtheta^\top\vy_\vu+\eta(\vz_\vu)])$, by the chain rule, 
 \begin{align*}
      & \frac{\partial^2}{\partial\vtheta\partial\eta}\log\lambda(\vu;\vtheta,\eta_{\vtheta,n})\bigg|_{\vtheta=\vtheta^*}\\
      =&\frac{\partial^2}{\partial\vtheta\partial\eta}\log\Psi[\tau_{\vtheta^*}(\vy),\eta^*(\vz_\vu)]+\frac{\partial^2}{\partial\eta^2}\log\Psi[\tau_{\vtheta^*}(\vy),\eta^*(\vz_\vu)]\nu^*_n(\vz_\vu)\\
     =& \frac{\partial^2}{\partial\vtheta\partial\eta}\left(\vtheta^{*\top}\vy_\vu+\eta^*(\vz_\vu)\right)
     +\frac{\partial^2}{\partial\eta^2}\left(\vtheta^{*\top}\vy_\vu+\eta^*(\vz_\vu)\right)\nu^*_n(\vz_\vu)\\
     =&0+ 0\cdot \nu^*_n(\vz_\vu)=0
 \end{align*}
Thus, \eqref{eq:lemma_spatial_ep_1} follows from the fact that for every $f\in\mathcal{F}$, $V_n^{(1)}(f)=0$. 

\textit{Step 2:} We establish \eqref{eq:lemma_spatial_ep_2} in this step. When the intensity function is log-linear,
\begin{align}
   &\frac{\partial}{\partial\eta}\log\lambda(\vu;\vtheta,\eta_{\vtheta,n})\bigg|_{\vtheta=\vtheta^*} \nonumber\\
   =& \frac{\partial}{\partial\eta}\left(\vtheta^{*\top}\vy_\vu+\eta^*(\vz_\vu)\right)=1 \label{eq:lemma_2_V_2}.
\end{align} 
Also, the kernel regression estimator $\widehat\E[\ell_n(\vtheta,\eta;X)|\vz]$ simplifies to
$$\sum_{\vu\in X\cap A_n}K_h\left({z}_\vu-{z}\right)\left(\vtheta^{\top}\vy_\vu+\eta(\vz_\vu)\right) -{\int_{A_n} K_h\left({z}_\vu-{z}\right)\exp\left(\vtheta^{\top}\vy_\vu+\eta(\vz_\vu)\right)\mathrm{d}\vu}.$$
Since the nuisance estimation  $\hat\eta_{\vtheta,n}(\vz)$ is the unique solution of 
\begin{align*}
    0=&\frac{\partial}{\partial\eta}\widehat\E[\ell_n(\vtheta,\eta;X)|\vz] \\
    =& \frac{\partial}{\partial\eta}\left\{\sum_{\vu\in X\cap A_n}K_h\left({z}_\vu-{z}\right)\left(\vtheta^{\top}\vy_\vu+\eta(\vz_\vu)\right) -{\int_{A_n} K_h\left({z}_\vu-{z}\right)\exp\left(\vtheta^{\top}\vy_\vu+\eta(\vz_\vu)\right)\mathrm{d}\vu}\right\}\\
    =& \sum_{\vu\in X\cap A_n}K_h\left({z}_\vu-{z}\right) -{\int_{A_n} K_h\left({z}_\vu-{z}\right)\exp\left(\vtheta^{\top}\vy_\vu+\eta(\vz_\vu)\right)\mathrm{d}\vu}.
\end{align*}
We thus have
$$\hat\eta_{\vtheta,n}(\vz) = \log\left\{\frac{\sum_{\vu\in X\cap A_n}K_h\left({z}_\vu-{z}\right)}{\int_{A_n} K_h\left({z}_\vu-{z}\right)\exp\left(\vtheta^{\top}\vy_\vu\right)\mathrm{d}\vu}\right\}.$$
The derivative of $\hat\eta_{\vtheta,n}(\vz)$ with respect to $\vtheta$ is
\begin{equation}\label{eq:lemma_spatial_ep_derivative}
    \frac{\partial}{\partial\vtheta}\hat\eta_{\vtheta,n}(\vz) = \frac{\int_{A_n} K_h\left({z}_\vu-{z}\right)\exp\left(\vtheta^{\top}\vy_\vu\right)\vy_\vu\mathrm{d}\vu}{\int_{A_n} K_h\left({z}_\vu-{z}\right)\exp\left(\vtheta^{\top}\vy_\vu\right)\mathrm{d}\vu}.
\end{equation}
The right-hand side of \eqref{eq:lemma_spatial_ep_derivative} is non-stochastic, so the function class $\mathcal{F}^\prime$ is independent with $X$. Condition \ref{condition:smoothness} and \ref{condition:boundedness} needed for Corollary \ref{corollary:rate of bias} automatically hold  when the intensity function is log-linear. Then, by Corollary $\ref{corollary:rate of bias}$ and equation \eqref{eq:lemma_2_V_2}, 
\begin{align*}
\sup_{f\in\mathcal{F}^\prime}\left|\mathbb{G}^{X}_n\left[V_n^{(2)}(f)\right]\right|&=|A_n|^{-\frac{1}{2}}\sup_{f\in\mathcal{F}^\prime}\left|\sum_{\vu\in X\cap A_n}f(\vu)-\int_{A_n}f(\vu)\lambda(\vu)\mathrm{d}\vu\right|\\
&\leq |A_n|^{-\frac{1}{2}}\sup_{f\in\mathcal{F}^\prime,\vu\in\R^2}|f(\vu)|\left||X\cap A_n|-\E[|X\cap A_n|]\right|\\
&= |A_n|^{-\frac{1}{2}}|\cdot |\mathcal{F}^\prime|\cdot O_p(|A_n|^{\frac{1}{2}})\\
&= O_p\left(|\mathcal{F}^\prime|\right).
\end{align*}
\end{proof}

\begin{lemma}\label{lemma:empirical process}
Let $X$ be a spatial point process with intensity function $\lambda(\vu;\vtheta^*,\eta^*)$ where $\vtheta^*,\eta^*$ are the true parameters. Let $\ell_n(\vtheta,\eta)$ be the pseudo-likelihood function of $X$. Let $\hat\eta_{\vtheta,n}$ be a estimator of $\eta_{\vtheta,n}$ satisfies Assumption \ref{assumption:np_condition_uniform}. Then under condition \ref{condition:smoothness}, \ref{condition:boundedness}, \ref{condition:pair_correlation}, if either $\hat\eta_{\vtheta,n}$ is independent with $X$ or the intensity function is log-linear, we have
\begin{gather}
    \left|\frac{\partial^2}{\partial\vtheta\partial\eta}  \ell_n(\vtheta,\eta_{\vtheta,n})\bigg|_{\vtheta=\vtheta^*}[\hat\eta_{\vtheta^*,n}-\eta_{\vtheta^*,n}^*]\right|=o_p(|A_n|^{\frac{1}{2}})\label{eq:empirical_bound_1},\\
    \left|\frac{\partial}{\partial\eta}\ell_n(\vtheta^*,\eta_{\vtheta^*,n})\left[\frac{\partial}{\partial\vtheta}\hat\eta_{\vtheta^*,n}-\frac{\partial}{\partial\vtheta}\eta_{\vtheta^*,n}^*\right]\right|=o_p(|A_n|^{\frac{1}{2}})\label{eq:empirical_bound_2}.
\end{gather}
\end{lemma}

\begin{proof}: The proof of the lemma consists of three steps. In step 1, we show that the left-hand side of \eqref{eq:empirical_bound_1} and \eqref{eq:empirical_bound_2} can be regarded as a generalization of the classic empirical processes in \textit{i.i.d.} setting to spatial point processes.
In step 2, we establish \eqref{eq:empirical_bound_1} and \eqref{eq:empirical_bound_2} when $\hat\eta_{\vtheta,n}$ is independent with $X$ or when the intensity function is log-linear. 

\textit{Step 1: } 
Let $\{Y_i\}_{i=1}^n$ be \textit{i.i.d.} samples drawn from a random variable $Y$, and $f$ be a function defined on the support of $Y$. 
The empirical process $\mathbb{G}_n[f]$ is defined as
$$\mathbb{G}_n[f]:=n^{-\frac{1}{2}}\sum_{i=1}^n\left(f(Y_i)-\E[f(Y_i)]\right).$$

Now, we consider the generalization of the classic empirical process to a spatial point process $X$ defined on $\R^2$ with intensity function $\lambda(\vu)$. Let $\{A_n\}_{n=1}^{\infty}$ be a sequence of expanding region, and $\mathcal{F}$ be function class defined on spatial region $\R^2$. For every $f\in\mathcal{F}$, the \textit{spatial empirical processes} $\mathbb{G}^{X}_n[f]$ is defined as
$$\mathbb{G}^{X}_n[f]:=|A_n|^{-\frac{1}{2}}\left\{\sum_{\vu\in X\cap A_n }f(\vu)-\int_{A_n}f(\vu)\lambda(\vu)\mathrm{d}\vu\right\}.$$
Note that $|A_n|^{\frac{1}{2}}\mathbb{G}_n^{s}$ is also referred to as innovation/residual measure in spatial point process literature (See \citet{baddeley2008properties}). 

If we further denote \begin{gather*}
    V_n^{(1)}(f)(\vu) := \frac{\partial^2}{\partial\vtheta\partial\eta}\log\lambda(\vu;\vtheta,\eta_{\vtheta,n})\bigg|_{\vtheta=\vtheta^*}[f(\vu)]\\
    V_n^{(2)}(f)(\vu) := \frac{\partial}{\partial\eta}\log\lambda(\vu;\vtheta,\eta_{\vtheta,n})\bigg|_{\vtheta=\vtheta^*}[f(\vu)]\\
    \mathcal{F}_n^{1}:=\left\{(\hat\eta_{\vtheta^*,n}-\eta_{\vtheta^*,n}^*\right)\circ \vz\}\\
    \mathcal{F}_n^{2}:=\left\{\frac{\partial}{\partial\vtheta}\left(\hat\eta_{\vtheta^*,n}-\eta_{\vtheta^*,n}^*\right)\circ \vz\right\}\\
    |\mathcal{F}|:=\sup_{f\in\mathcal{F},\vu\in\R^2}|f(\vu)|
\end{gather*}
Equations $\eqref{eq:empirical_bound_1}$ and $\eqref{eq:empirical_bound_2}$ can be written as
\begin{gather}
\sup_{f\in\mathcal{F}_n^{1}}\left|\mathbb{G}^{X}_n[V_n^{(1)}(f)]\right|=o_p(1) \label{eq:empirical_bound_1_reform},\\
\sup_{f\in\mathcal{F}_n^{2}}\left|\mathbb{G}^{X}_n[V_n^{(2)}(f)]\right|=o_p(1)\label{eq:empirical_bound_2_reform}.
\end{gather}
respectively.

\textit{Step 2:} When the intensity function of $X$ is log-linear, \eqref{eq:empirical_bound_1_reform} and \eqref{eq:empirical_bound_2_reform} directly follow from Lemma \ref{lemma:spatial_ep} and Assumption \ref{assumption:np_condition_pointwise}. Then, we consider the case when when $\hat\eta_{\vtheta,n}$ is independent with $X$. 

By condition \ref{condition:smoothness}, 
$$\sup_{\vu\in\R^2}\left|\frac{\partial^2}{\partial\vtheta\partial\eta}\log\lambda(\vu;\vtheta,\eta_{\vtheta,n})\bigg|_{\vtheta=\vtheta^*}\right|=O(1).$$
By Corollary \ref{corollary:rate of bias}, 
$$\left||X\cap A_n|-\E[|X\cap A_n|]\right| = O_p(|A_n|^{\frac{1}{2}}).$$
Thus, when $\hat\eta_{\vtheta,n}$ is independent with $X$, conditional on $\hat\eta_{\vtheta,n}$, we have 
\begin{align*}
&\sup_{f\in\mathcal{F}_n^{1}}\left|\mathbb{G}^{X}_n[V_n^{(1)}(f)]\right| \\
&=|A_n|^{-\frac{1}{2}}\sup_{f\in\mathcal{F}_n^{1}}\left|\sum_{\vu\in X\cap A_n}V_n^{(1)}(f)(\vu)-\int_{A_n}V_n^{(1)}(f)(\vu)\lambda(\vu)\mathrm{d}\vu\right|\\
& \leq |A_n|^{-\frac{1}{2}}\sup_{\vu\in\R^2}\left|\frac{\partial^2}{\partial\vtheta\partial\eta}\log\lambda(\vu;\vtheta,\eta_{\vtheta,n})\bigg|_{\vtheta=\vtheta^*}\right|\cdot  |\mathcal{F}_n^1|\cdot \left||X\cap A_n|-\E[|X\cap A_n|]\right|\\
&=O_p(|\mathcal{F}_n^1|).
\end{align*}
By Lemma 6.1 in \citet{chernozhukov2018double}, if the above equality hold conditionally on $\hat\eta_{\vtheta,n}$, it also holds unconditionally. We thus have $$\sup_{f\in\mathcal{F}_n^{1}}\left|\mathbb{G}^{X}_n[V_n^{(1)}(f)]\right|=O_p(|\mathcal{F}_n^1|)$$ Similarly, $$\sup_{f\in\mathcal{F}_n^{2}}\left|\mathbb{G}^{X}_n[V_n^{(2)}(f)]\right| = O_p(|\mathcal{F}_n^2|).$$ 
Under assumption \ref{assumption:np_condition_uniform},  $|\mathcal{F}_n^1|$ and $|\mathcal{F}_n^2|$ are both $o_p(1)$, we thus have \eqref{eq:empirical_bound_1_reform} and \eqref{eq:empirical_bound_2_reform}.

\end{proof}

\subsection{Random Field Central Limit Theorem}
\begin{lemma}[Central Limit Theorem]\label{lemma:central limit theorem}
Under condition \ref{condition:smoothness}, \ref{condition:nonsigularity}, \ref{condition:nonsingular_covariance}, \ref{condition:alpha-mixing}, 
\begin{equation}\label{eq:lemma_clt}
    |A_n|^{-\frac{1}{2}}\bar{\Sigma}^{-\frac{1}{2}}_n(\vtheta^*,\eta^*,\nu^*,\psi^*)\frac{\partial}{\partial\vtheta}\ell_n(\vtheta,\eta_{\vtheta,n})\bigg|_{\vtheta=\vtheta^*}\rightarrow_d N({0},{I}_k).
\end{equation}
\end{lemma}

\begin{proof}:
For every pair of integer $(i,j)\in \mathbb{Z}^2$, let $C(i,j)$ be the unit volume square centered at $(i,j)$, and let $\mathcal{D}_n = \left\{(i,j)\in\mathbb{Z}^2:C(i,j)\cap A_n\neq \varnothing\right\}$. We denote
$$Z_{i,j}=\sum_{\vu\in X\cap C(i,j)\cap A_n}\frac{\partial}{\partial\vtheta}\log\lambda(\vu;\vtheta,\eta_{\vtheta,n})\bigg|_{\vtheta=\vtheta^*}-\int_{ C(i,j)\cap A_n}\frac{\partial}{\partial\vtheta}\lambda(\vu;\vtheta,\eta_{\vtheta,n})\bigg|_{\vtheta=\vtheta^*}\mathrm{d}\vu.$$
Then, we have 
\begin{gather*}
    \frac{\partial}{\partial\vtheta}\ell_n(\vtheta,\eta_{\vtheta,n})\bigg|_{\vtheta=\vtheta^*} = \sum_{(i,j)\in\mathcal{D}_n} Z_{i,j}\\
    \E\left[\frac{\partial}{\partial\vtheta}\ell_n(\vtheta,\eta_{\vtheta,n})\right]\bigg|_{\vtheta=\vtheta^*}=\E\left[\sum_{(i,j)\in \mathcal{D}_n}Z_{i,j}\right]\\
|A_n|\bar{\Sigma}_n(\vtheta^*,\eta^*,\nu^*,\psi^*) =\text{Var}\left(\sum_{(i,j)\in \mathcal{D}_n}Z_{i,j}\right)
\end{gather*}
Then, the central limit theorem \eqref{eq:lemma_clt} can be written as
\begin{equation*}
    \text{Var}\left(\sum_{(i,j)\in \mathcal{D}_n}Z_{i,j}\right)^{-\frac{1}{2}}\left(\sum_{(i,j)\in \mathcal{D}_n}Z_{i,j}-\E\left[\sum_{(i,j)\in \mathcal{D}_n}Z_{i,j}\right]\right)\rightarrow_d N(0,I_{k}),
\end{equation*}
It is satisfied by Theorem 1 in \cite{biscio2019general}, if the following assumptions holds:
 \begin{enumerate}
     \item $A_1\subset A_2\subset \ldots$ and $|\bigcup_{n=1}^\infty A_n|=\infty$ \label{clt:A_n}
     \item 
     The $\alpha$-mixing coefficient of $X$ satisfies $\alpha_{2,\infty}^X(r) = O(r^{-(2+\epsilon)})$ for some $\epsilon > 0$ \label{clt:alpha}
     \item There exists $\tau > \frac{4}{\epsilon}$ such that $\sup_{n\in \mathbb N}\sup_{(i,j)\in \mathcal{D}_n}\E\left[\left\|Z_{(i,j)}-\E\left[Z_{(i,j)}\right]\right\|^{2+\tau}\right]<\infty$ \label{clt:moment}
     \item The limit infimum of the smallest eigenvalue of $\Sigma_n(\vtheta^*,\eta^*,\nu^*,\psi^*)$ is larger  than zero. \label{clt:nonsingularity}
 \end{enumerate}
Condition \ref{clt:A_n} holds in our asymptotic framework. Condition \ref{clt:alpha} is identical to condition \ref{condition:alpha-mixing}. Condition \ref{clt:moment} is satisfied because $Z_{i,j}$ are uniformly bounded by condition \ref{condition:smoothness}. Condition \ref{clt:nonsingularity} is identical to condition \ref{condition:nonsingular_covariance}. 

\end{proof}

\subsection{Lemmas for Spatial Kernel Regression}
\begin{lemma}[Kernel Approximation Bias]\label{lemma:kernel_bias_rate}
 Let $h(\vy,\vz)$ be a bounded function defined on the covariance domain $\mathcal{Y}\times \mathcal{Z}$ that is $l$-th order continuously differentiable with respect to $\vz$.
    Let $K_h(\cdot)$ be an $l$-th order scaled kernel function (See Definition \ref{def:higher order kernel}) where $h$ is the bandwidth. Then we have
    \begin{equation}\label{eq:kernel_bias_rate}    \sup_{\vz^*\in\mathcal{Z}}\left|\int_{A_n}h(\vy_\vu,\vz_\vu)K_h\left({z}_\vu-{z}^*\right)\mathrm{d}\vu-\int_{\mathcal{Y}}h(\vy,\vz^*)f_n(\vy,\vz^*)\mathrm{d}\vy\right|=O(|A_n|h^l).
    \end{equation}
\end{lemma}
\begin{proof}: Let ${t}=(\vz-\vz^*)/h$, and let $f_n(\vy,\vz^*+h{t})=0$ for ${t}\notin (\mathcal{Z}-\vz^*)/h$. By the change of variable, for any $\vz^*\in\mathcal{Z}$,
    \begin{align}
        &\int_{A_n}h(\vy_\vu,\vz_\vu)K_h\left({z}_\vu-{z}^*\right)\mathrm{d}\vu\nonumber\\
        =& \int_{\mathcal{Z}}\int_{\mathcal{Y}}h(\vy,\vz)K_h\left(\vz-{z}^*\right)f_n(\vy,\vz)\mathrm{d}\vy\mathrm{d}\vz\nonumber\\
        =& \int_{\mathcal{Y}}\int_{\R^q}h(\vy,\vz^*+h{t})K\left({t}\right)f_n(\vy,\vz^*+h{t})\mathrm{d}{t}\mathrm{d}\vy.\label{eq:kernel_variable_change}
    \end{align}
Substituting the above the change of variable into the left-hand-side of \eqref{eq:kernel_bias_rate}, we have
\begin{align}
    &\sup_{\vz^*\in\mathcal{Z}}\left|\int_{A_n}h(\vy_\vu,\vz_\vu)K_h\left({z}_\vu-{z}^*\right)\mathrm{d}\vu-\int_{\mathcal{Y}}h(\vy,\vz^*)f_n(\vy,\vz^*)\mathrm{d}\vy\right|\nonumber\\
    =&  \sup_{\vz^*\in\mathcal{Z}}\left|\int_{\mathcal{Y}}\int_{\R^q}h(\vy,\vz^*+h{t})K\left({t}\right)f_n(\vy,\vz^*+h{t})\mathrm{d}{t}\mathrm{d}\vy-\int_{\mathcal{Y}}\int_{\R^q}h(\vy,\vz^*)K\left({t}\right)f_n(\vy,\vz^*)\mathrm{d}{t}\mathrm{d}\vy\right|\nonumber\\
    \leq&  \int_{\mathcal{Y}}\int_{\R^q}K\left({t}\right)\sup_{\vz^*\in\mathcal{Z}}\left|h(\vy,\vz^*+h{t})f_n(\vy,\vz^*+h{t})-h(\vy,\vz^*)f_n(\vy,\vz^*)\right|\mathrm{d}{t}\mathrm{d}\vy.\label{eq:lemma_kernel_bias_rate_change_var}
\end{align}
Thus, it suffices to show that \eqref{eq:lemma_kernel_bias_rate_change_var} is $O(|A_n|h^l)$.

By the definition of the Radon–Nikodym derivative $f_n(\vy,\vz)$,
$$\int_{\mathcal{Y}}\int_{\mathcal{Z}} f_n(\vy,\vz)\mathrm{d}\vy\mathrm{d}\vz=|A_n|.$$
Since $\mathcal{Y}$ and $\mathcal{Z}$ are compact, 
$f_n(\vy,\vz)$ is $O(|A_n|)$ almost surely for $\vy\in\mathcal{Y}$ and $\vz\in\mathcal{Z}$. Also, $f_n(\vy,\vz)$ and $h(\vy,\vz)$ are $l$-th order continuously differentiable with respect to $\vz$, and $h(\vy,\vz)$ is bounded. We thus have 
\begin{align}
&\sup_{\vz\in\mathcal{Z}}\left|h(\vy,\vz^*+h{t})f_n(\vy,\vz^*+h{t})-h(\vy,\vz^*)f_n(\vy,\vz^*)\right|\nonumber\\
\leq & \sup_{\vy\in\mathcal{Y},\vz\in\mathcal{Z}}\left|\frac{\partial^l}{\partial\vz}\left[h(\vy,\vz)f_n(\vy,\vz)\right]h^l{t}^l\right|\nonumber\\
= & O( |A_n|h^l{t}^l).\label{eq:lemma_kernel_bias_cont}
\end{align}
Recall that $\int_{\R^q}K\left({t}\right){t}^l\mathrm{d}{t}=\kappa_q<\infty$ in definition \ref{def:higher order kernel}. 
Next, we substitute \eqref{eq:lemma_kernel_bias_cont} into \eqref{eq:lemma_kernel_bias_rate_change_var} and yield 
\begin{align*}
    &\int_{\mathcal{Y}}\int_{\R^q}K\left({t}\right)\sup_{\vz^*\in\mathcal{Z}}\left|h(\vy,\vz^*+h{t})f_n(\vy,\vz^*+h{t})-h(\vy,\vz^*)f_n(\vy,\vz^*)\right|\mathrm{d}{t}\mathrm{d}\vy\\
     = &O(|A_n|h^l)\cdot  \int_{\R^q}K\left({t}\right){t}^l\mathrm{d}{t}\int_{\mathcal{Y}}f_n(\vy)\mathrm{d}\vy\\
    =&O(|A_n|h^l).
\end{align*}
\end{proof}

\begin{lemma}[Moment of Spatial Summation]\label{lemma:higher order deviation rate}
    Let $f(\vu)$ be a bounded function defined on $\R^2$. Under condition \ref{condition:cumulant}, the $m$-th order centered moment of $\sum_{\vu\in X\cap A_n}f(\vu)$
    is $O_p(|A_n|^{\frac{m}{2}})$
    
\end{lemma}
\begin{proof}:
Since $f(\vu)$ is bounded and
$$\left|\sum_{\vu\in X\cap A_n}f(\vu)\right|\leq\sup_{\vu\in\R^2}|f(\vu)|\cdot |X\cap A_n|,$$
it suffices to show that the $m$-th order centered moment of $|X\cap A_n|$ is $O(|A_n|^{\frac{m}{2}})$. Let $\kappa_{m^\prime}$ be the $m^\prime$-th order factorial cumulant defined as  
$$\kappa_{m^\prime} := \int_{A_n}\cdots \int_{A_n} Q_{m^\prime}(\vu_1,\ldots,\vu_{m^\prime})\mathrm{d}\vu_1 \ldots \mathrm{d}\vu_{m^\prime}.$$
Then, by condition \ref{condition:cumulant}, for $2\leq m^\prime\leq m$
\begin{align*}
    \kappa_{m^\prime} 
\leq & \int_{A_n}\left\{\int_{\R^2} \cdots\int_{\R^2} Q_{m^\prime}(\vu_1,\ldots,\vu_{m^\prime})\mathrm{d}\vu_1 \ldots \mathrm{d}\vu_{m^\prime-1}\right\}\mathrm{d}\vu_{m^\prime}\\
=&\int_{A_n} C \mathrm{d}\vu_{m^\prime}= O(|A_n|).
\end{align*}
Let $\kappa^\prime_{m^\prime}$ be the $m$-th order ordinary cumulant where the  ``prime'' distinguishes it from the factorial cumulant. Let $\Delta_{j,m^\prime}$ be the Stirling number of the second kind. Then, the ordinary cumulants are related to factorial cumulants by
$$\kappa^\prime_{m^\prime} = \sum_{j=1}^m \Delta_{j,m^\prime}\kappa_j.$$ 
Thus, for $2\leq m^\prime\leq m$, $\kappa^\prime_{m^\prime}=O(|A_n|)$. 

Let $\mu_{m^\prime}$ be the $m^\prime$-th order centered moment of $|X\cap A_n|$.  $\mu_{m^\prime}$ can be expressed in terms of the ordinary cumulants as

$$\mu_{m^\prime} = \sum_{k=1}^{m^\prime} B_{m^\prime,k}(0,\kappa^\prime_{2},\ldots,\kappa^\prime_{m^\prime-k+1})=B_{m^\prime}(0,\kappa^\prime_{2},\ldots,\kappa^\prime_{m^\prime}),$$ 
where $B_{m^\prime}$ is the Bell polynomials given by 
$$B_{m^\prime}(0,\kappa^\prime_{2},\ldots,\kappa^\prime_{m^\prime})=m^\prime !\sum_{2j_2+3j_3+\ldots m^\prime j_{m^\prime}=m^\prime}\prod_{i=2}^{m^\prime}\frac{\kappa^{\prime j_i}_i}{(i!)^{j_i}j_i!},$$
which can be bounded by
\begin{align*}
    B_{m^\prime}(0,\kappa^\prime_{2},\ldots,\kappa^\prime_{m^\prime})= & \sum_{2j_2+3j_3+\ldots m^\prime j_{m^\prime}\leq m^\prime}\prod_{i=2}^{m^\prime}O(|A_n|^{j_i})\\
    =&  \sum_{2j_2+3j_3+\ldots m^\prime j_{m^\prime}\leq m^\prime}O(|A_n|^{\sum_{i=2}^{m^\prime}j_i})\\
    \leq&\sup_{2j_2+3j_3+\ldots m^\prime j_{m^\prime}\leq m^\prime}O(|A_n|^{\sum_{i=2}^{m^\prime}j_i})\\
    \leq & O(|A_n|^{\frac{m^\prime}{2}}).
\end{align*}
Thus, $\mu_{m} = O(|A_n|^\frac{m}{2})$.

\end{proof}

\begin{lemma}[Kernel Approximation Deviation]\label{lemma:kernel_sup_deviation_rate}
 Let $\phi(\vy,\vz;\pi)$ be a bounded function defined on the covariance domain $\mathcal{Y}\times \mathcal{Z}$ parameterized by $\pi\in\Pi\subset\R^s$ and $\Pi$ is compact. We assume $\phi(\vy,\vz;\pi)$ to be continuously differentiable with respect to $\pi$. Let $X$ be a spatial point process with intensity function $\lambda(\vu;\vtheta,\eta)$, and $\{A_n\}_{n=1}^\infty$ be a sequence of expanding observational windows. Let $K_h(\cdot)$ be a scaled $l$-th order kernel function (See definition \ref{def:higher order kernel}). 
 Denote
    $$\widehat{F}_n(\vz,\pi;\phi) := \sum_{\vu\in X\cap A_n} \phi(\vy_\vu,\vz_\vu;\pi)K_h\left({z}_\vu-{z}\right).$$
    Under Assumption \ref{assumption:kernel_est}, for any $\gamma>0$,
    \begin{equation}
        \sup_{\vz\in\mathcal{Z},\pi\in\Pi}\left| \widehat{F}_n(\vz,\pi;\phi)  -\E\left[ \widehat{F}_n(\vz,\pi;\phi) \right]\right|
        = o_p(|A_n|^{\frac{q+s+m/2}{q+s+m}+\gamma}h^{-q-\frac{q+s}{q+s+m}}).
        \label{eq:kernel_sup_deviation_rate}
    \end{equation}
\end{lemma}
\begin{proof}:
To establish \eqref{eq:kernel_sup_deviation_rate}, we proceed in three steps. Step 1 shows the main argument. Steps 2 and 3 show the auxiliary calculations.

\textit{Step 1:} Let $\mathcal{Z}_\delta$ and $\Pi_{\delta}$ be the uniform grids on $\mathcal{Z}$ and $\Pi$ respectively where $\delta$ is the spacing. Denote
\begin{gather*}
    R_{n,1}:=\max_{\vz\in \mathcal{Z}_\delta,\pi\in\Pi_\delta}\left| \widehat{F}_n(\vz,\pi;\phi)  -\E\left[ \widehat{F}_n(\vz,\pi;\phi) \right]\right|,\\
    R_{n,2}:=\sup_{|\vz-\vz^\prime|\leq\delta,|\pi-\pi^\prime|\leq\delta} \left|\widehat{F}_n(\vz,\pi;\phi)-\widehat{F}_n(\vz^\prime,\pi^\prime;\phi)\right|.
\end{gather*}
Then, by the triangular inequality, 
\begin{align*}
\sup_{\vz\in\mathcal{Z},\pi\in\Pi}\left| \widehat{F}_n(\vz,\pi;\phi)  -\E\left[ \widehat{F}_n(\vz,\pi;\phi) \right]\right|
    \leq &R_{n,1}+R_{n,2}.
\end{align*}
In Step 2,3 respectively, we will show that for any $\gamma>0$
\begin{gather}
    R_{n,1} = o_p(\delta^{-\frac{q+s}{m}}|A_n|^{\frac{1}{2}+\gamma}h^{-q})\label{eq:kernel_sup_deviation_R_1},\\
    R_{n,2} =O_p(|A_n|h^{-(q+1)}\delta)\label{eq:kernel_sup_deviation_R_2}.
\end{gather}
If the above equalities hold and  
$\delta = \Theta(|A_n|^{-\frac{1}{2}\cdot\frac{m}{q+s+m}}h^{\frac{m}{q+s+m}})$, we have 
$$R_{n,1} = O_p(|A_n|^{\frac{q+s+m/2}{q+s+m}+\gamma}h^{-q-\frac{q+s}{q+s+m}}),\quad{}
R_{n,2} = O_p(|A_n|^{\frac{q+s+m/2}{q+s+m}}h^{-q-\frac{q+s}{q+s+m}}),$$
which directly gives us \eqref{eq:kernel_sup_deviation_rate}.

\textit{Step 2:} In this step, we establish \eqref{eq:kernel_sup_deviation_R_1}. First, note that
$$\widehat{F}_n(\vz,\pi;\phi) = h^{-q}\sum_{\vu\in X\cap A_n} \phi(\vy_\vu,\vz_\vu;\pi)K\left(\frac{{z}_\vu-{z}}{h}\right).$$
Since $\phi(\cdot)$ and the kernel function $K(\cdot)$ are uniformly bounded, it follows from Lemma \ref{lemma:higher order deviation rate} that the $m$-th order centered moment of $\widehat{F}_n(\vz,\pi;\phi)$ is $O(h^{-qm}|A_n|^{\frac{m}{2}})$. 
Then, by Markov inequality, for any $\epsilon>0$,
\begin{align*}
&\pr\left(\left|\widehat{F}_n(\vz,\pi;\phi)  -\E\left[ \widehat{F}_n(\vz,\pi;\phi) \right]\right|\geq \epsilon\right)\\
=&\pr\left(\left|\widehat{F}_n(\vz,\pi;\phi)  -\E\left[ \widehat{F}_n(\vz,\pi;\phi) \right]\right|^m\geq \epsilon^m\right)\\
=&O(|A_n|^{\frac{m}{2}}h^{-qm}\epsilon^{-m}).
\end{align*}
Then, by union bound, 
\begin{align*}
\pr\left(R_{n,1}\geq \epsilon\right)
    \leq &\sum_{\vz\in \mathcal{Z}_\delta,\pi\in\Pi_\delta}  \pr\left(\left|\widehat{F}_n(\vz,\pi;\phi)  -\E\left[ \widehat{F}_n(\vz,\pi;\phi) \right]\right|\geq \epsilon\right)\\
    =&O\left(\delta^{-(q+s)}|A_n|^{\frac{m}{2}}h^{-qm}\epsilon^{-m}\right).
\end{align*}
Thus, we have \eqref{eq:kernel_sup_deviation_R_1}

\textit{Step 3:} In this step, we establish \eqref{eq:kernel_sup_deviation_R_2}. First, by triangular inequality,
\begin{equation}\label{eq:kernel_sup_deviation_R_2_decomp}
    R_{n,2}\leq \sup_{|\vz-\vz^\prime|\leq\delta} \left|\widehat{F}_n(\vz,\pi;\phi)-\widehat{F}_n(\vz^\prime,\pi;\phi)\right|+ \sup_{|\pi-\pi^\prime|\leq\delta} \left|\widehat{F}_n(\vz^\prime,\pi;\phi)-\widehat{F}_n(\vz^\prime,\pi^\prime;\phi)\right|.
\end{equation}
By the smoothness of the kernel function $K(\cdot)$, the first term in the right-hand side of \eqref{eq:kernel_sup_deviation_R_2_decomp} satisfies
\begin{align}
    &\sup_{|\vz-\vz^\prime|\leq\delta} \left|\widehat{F}_n(\vz,\pi;\phi)-\widehat{F}_n(\vz^\prime,\pi;\phi)\right|\nonumber\\
    = &  \sup_{|\vz-\vz^\prime|\leq\delta}\left|\sum_{\vu\in X\cap A_n} \phi(\vy_\vu,\vz_\vu;\pi)\left[K_h\left({z}_\vu-{z}\right)-K_h\left({z}_\vu-{z}^\prime\right)\right]\right|\nonumber\\
    \leq & \sup_{\vy,\vz,\pi}|\phi(\vy,\vz;\pi)|h^{-q} \sup_{|\vz-\vz^\prime|\leq\delta}\left|\sum_{\vu\in X\cap A_n} K\left(\frac{{z}(\vu)-{z}}{h}\right)-K\left(\frac{{z}(\vu)-{z}^\prime}{h}\right)\right|\nonumber\\
    \leq & \sup_{\vy,\vz,\pi}|\phi(\vy,\vz;\pi)|\cdot\sup_{\vz}\left|\frac{\partial}{\partial\vz}K(\vz)\right|\cdot |X\cap A_n|\cdot h^{-(q+1)}\delta\nonumber\\
    =&O_p(|A_n|h^{-(q+1)}\delta)\label{eq:kernel_sup_R_2_1}.
\end{align}
By the smoothness of $\phi(\vy,\vz;\pi)$ with respect to $\pi$, the second term in the right-hand side of \eqref{eq:kernel_sup_deviation_R_2_decomp} satisfies
\begin{align}
    &\sup_{|\vz-\vz^\prime|\leq\delta} \left|\widehat{F}_n(\vz^\prime,\pi;\phi)-\widehat{F}_n(\vz^\prime,\pi^\prime;\phi)\right|\nonumber\\
    = &  \sup_{|\vz-\vz^\prime|\leq\delta}\left|\sum_{\vu\in X\cap A_n} \left[\phi(\vy_\vu,\vz_\vu;\pi)-\phi(\vy_\vu,\vz_\vu;\pi^\prime)\right]K_h\left({z}_\vu-{z}^\prime\right)\right|\nonumber\\
    \leq & \sup_{\vz}|K(\vz)|h^{-q} \sup_{|\vz-\vz^\prime|\leq\delta}\left|\sum_{\vu\in X\cap A_n} \left[\phi(\vy_\vu,\vz_\vu;\pi)-\phi(\vy_\vu,\vz_\vu;\pi^\prime)\right]\right|\nonumber\\
    \leq &  \sup_{\vz}|K(\vz)|\cdot\sup_{\vy,\vz,\pi}\left|\frac{\partial}{\partial\pi}\phi(\vy,\vz;\pi)\right|\cdot |X\cap A_n|\cdot h^{-q}\delta\nonumber\\
    =&O_p(|A_n|h^{-q}\delta)\label{eq:kernel_sup_R_2_2}.
\end{align}
If we substitute \eqref{eq:kernel_sup_R_2_1} and \eqref{eq:kernel_sup_R_2_2} into \eqref{eq:kernel_sup_deviation_R_2_decomp}, we will have \eqref{eq:kernel_sup_deviation_R_2}.

\end{proof}

\begin{lemma}[Spatial Kernel Regression] \label{lemma:kernel objective function rate}
Recall that $k$ is the dimension of target parameter, $q$ is the dimension of nuisance covariates, $m$ is the order of factorial cumulants defined in condition \ref{condition:cumulant}. Let $\alpha = \frac{m}{2(k+q+m+1)}$, $\beta = \frac{k+q+1}{k+q+m+1}$.
Under all the conditions in Assumption \ref{assumption:kernel_est}, for $i = 0,1$, $j = 0,1,2$, for any $\gamma>0$,
\begin{equation}\label{eq:kernel_obj_func_rate_1}
    \sup_{\vtheta\in\Theta, \eta\in\mathcal{H}, \vz\in\mathcal{Z}}\left|\frac{\partial^{i+j}}{\partial\eta^i\partial\vtheta^j}\widehat\E[\ell_n(\vtheta,\eta)|\vz]-\frac{\partial^{i+j}}{\partial\eta^i\partial\vtheta^j}\E[\ell_n(\vtheta,\eta)|\vz]\right|=o_p(|A_n|^{1-\alpha+\gamma}h^{-q-\beta}+|A_n|h^l)
\end{equation}
If we further let $h=\Theta(|A_n|^{-\frac{\alpha}{l+q+\beta}})$,  
\begin{equation}\label{eq:kernel_obj_func_rate_2}
   \sup_{\vtheta\in\Theta, \eta\in\mathcal{H}, \vz\in\mathcal{Z}}\left|\frac{\partial^{i+j}}{\partial\eta^i\partial\vtheta^j}\widehat\E[\ell_n(\vtheta,\eta)|\vz]-\frac{\partial^{i+j}}{\partial\eta^i\partial\vtheta^j}\E[\ell_n(\vtheta,\eta)|\vz]\right|=o_p(|A_n|^{1-\frac{\alpha l}{l+q+1}})
\end{equation}
\end{lemma}

\begin{proof}: Recall that
\begin{gather*}
    \E[\ell_n(\vtheta,\eta)|\vz]  = \int_{\mathcal{Y}} \log\left\{\Psi [\tau_{\vtheta}(\vy),\eta(\vz)]\right\}\Psi[\tau_{\vtheta^*}(\vy),\eta^*(\vz)]f_n(\vy,\vz)\mathrm{d}\vy -\int_{\mathcal{Y}}\Psi[\tau_{\vtheta}(\vy),\eta(\vz)]f_n(\vy,\vz)\mathrm{d}\vy,\\
\widehat\E[\ell_n(\vtheta,\eta)|\vz]=\sum_{\vu\in X_n\cap A}K_h\left({z}_\vu-{z}\right)\log\left\{\Psi [\tau_{\vtheta}(\vy_\vu),\eta(\vz_\vu)]\right\} -{\int_{A_n} K_h\left({z}_\vu-{z}\right)\Psi[\tau_{\vtheta}(\vy_\vu),\eta(\vz_\vu)]\mathrm{d}\vu}.
\end{gather*}
Denote 
\begin{gather*}
\phi_5(\vy;\vtheta,\eta):=\log\left\{\Psi [\tau_{\vtheta}(\vy),\eta(\vz)]\right\},\\
\phi_6(\vy;\vtheta,\eta):= \Psi [\tau_{\vtheta}(\vy),\eta(\vz)].\\
\end{gather*}
For any $\phi(\vy;\vtheta,\eta)$, we further denote
\begin{gather*}
    \phi^{i,j}(\vy;\vtheta,\eta):= \frac{\partial^{i+j}}{\partial\eta^i\partial\vtheta^j}\phi(\vy;\vtheta,\eta),\\
    F_n(\vz,\vtheta,\eta;\phi) := \int_{\mathcal{Y}}\phi(\vy;\vtheta,\eta)f_n(\vy,\vz)\mathrm{d}\vy,\\
    \widehat{F}_n(\vz,\vtheta,\eta;\phi):=\sum_{\vu\in X\cap A_n}K_h\left(\vz_\vu-\vz\right)\phi(\vy_\vu;\vtheta,\eta).
\end{gather*}
Then, by triangular inequality,
\begin{align}
    &\sup_{\vtheta, \eta, \vz}\left|\frac{\partial^{i+j}}{\partial\eta^i\partial\vtheta^j}\widehat\E[\ell_n(\vtheta,\eta)|\vz]-\frac{\partial^{i+j}}{\partial\eta^i\partial\vtheta^j}\E[\ell_n(\vtheta,\eta)|\vz]\right|\nonumber\\
    \leq & \sup_{\vtheta, \eta, \vz}\left|\E\left[\widehat{F}_n(\vz,\vtheta,\eta;\phi_5^{i,j})\right]-F_n(\vz,\vtheta,\eta;\phi_5^{i,j})\right|+\sup_{\vtheta, \eta, \vz}\left|\widehat{F}_n(\vz,\vtheta,\eta;\phi_5^{i,j})-\E\left[\widehat{F}_n(\vz,\vtheta,\eta;\phi_5^{i,j})\right]\right|\nonumber\\ 
    &+\sup_{\vtheta, \eta, \vz}\left|\int_{A_n} K_h\left({z}_\vu-{z}\right)\phi^{i,j}_6(\vy;\vtheta,\eta)\mathrm{d}\vu-\int_{\mathcal{Y}}\phi^{i,j}_6(\vy;\vtheta,\eta)f_n(\vy,\vz)\mathrm{d}\vy\right|.
    \label{eq:lemma_obj_decomp}
\end{align}
By the smoothness condition \ref{condition:smoothness}, $\phi_5^{i,j}$ and $\phi_6^{i,j}$ are uniformly bounded for every $i,j$. Thus, when condition \ref{condition:higher order smoothness} is satisfied, it followed from Lemma \ref{lemma:kernel_bias_rate} that
\begin{gather}
    \sup_{\vtheta, \eta, \vz}\left|\int_{A_n} K_h\left({z}_\vu-{z}\right)\phi^{i,j}_6(\vy;\vtheta,\eta)\mathrm{d}\vu-\int_{\mathcal{Y}}\phi^{i,j}_6(\vy;\vtheta,\eta)f_n(\vy,\vz)\mathrm{d}\vy\right|=O(|A_n|h^l)\label{eq:lemma_obj_bias_1},\\
    \sup_{\vtheta, \eta, \vz}\left|\E\left[\widehat{F}_n(\vz,\vtheta,\eta;\phi_5^{i,j})\right]-F_n(\vz,\vtheta,\eta;\phi_5^{i,j})\right|=O(|A_n|h^l)\label{eq:lemma_obj_bias_2}.
\end{gather}
Additionally, $\phi_5^{i,j}(\vy,\vtheta,\eta)$ is continuously differentiable with respect to $\vtheta$ and $\eta$. Thus, if we apply Lemma \ref{lemma:kernel_sup_deviation_rate} to $\phi_5^{i,j}(\vy;\vtheta,\eta)$ where we treat $\vtheta$ and $\eta$ as parameters, i.e., $s=k+1$ in Lemma \ref{lemma:kernel_sup_deviation_rate}, we have that for any $\gamma>0$,
\begin{equation}\label{eq:lemma_obj_deviation}
    \sup_{\vtheta, \gamma, \vz}\left|\widehat{F}_n(\vz,\vtheta,\eta;\phi_5^{i,j})-\E\left[\widehat{F}_n(\vz,\vtheta,\eta;\phi_5^{i,j})\right]\right|=o_p(|A_n|^{\frac{q+k+1+m/2}{q+k+1+m}+\gamma}h^{-q-\frac{q+k+1}{q+k+1+m}}).
\end{equation}
If we substitute \eqref{eq:lemma_obj_bias_1}, \eqref{eq:lemma_obj_bias_2} and \eqref{eq:lemma_obj_deviation} into \eqref{eq:lemma_obj_decomp}, equation \eqref{eq:kernel_obj_func_rate_1} holds for any $\gamma>0$.
If we further let $h=\Theta(|A_n|^{-\frac{\alpha}{l+q+\beta}})$, and choose 
$\gamma<\frac{\alpha l}{l+q+\beta}-\frac{\alpha l}{l+q+1}$, we can have equation \eqref{eq:kernel_obj_func_rate_2}.

\end{proof}

\subsection{Independence between nuisance derivative and observed point process}\label{sec:ind_nui_pp}

The independence between the nuisance derivative of the nuisance estimation $\frac{\partial}{\partial\vtheta}\hat\eta_{\vtheta}$ and the observed spatial point process $X$ is critical to control the first-order plug-in rate of the pseudo-log-likelihood. In this subsection, we elaborate on the occurrence of such independence when the intensity function of $X$ is log-linear.

For $A\in \R$, the nuisance estimand $\eta_{\vtheta}$ is defined via $\eta_{\vtheta} = \arg\max_{\eta\in\mathcal{H}}\E[\ell(\vtheta,\eta;X)]$. We consider a general nuisance estimator 
\begin{align*}
    \hat\eta_{\vtheta} &:= \arg\max_{\eta\in\mathcal{F}}\ell(\vtheta,\eta;X) \\
    & := \arg\max_{\eta\in \mathcal{F}} \left\{\sum_{\vu\in X\cap A_n}\log\lambda(\vu;\vtheta,\eta)-\int_{A_n}\lambda(\vu;\vtheta,\eta)\mathrm{d}\vu\right\}.
\end{align*}
$\mathcal{F}$ represents the function class defined by the nonparametric estimator. For instance, if a spline estimator is used,$\mathcal{F}$ is the linear space spanned by the spline basis. Similarly, with decision trees, the function class encompasses all finite rectangular partitions of the covariate space. 

Under Assumption \eqref{assumption:kernel_est}, $\hat\eta_{\vtheta} = \arg\max_{\eta\in\mathcal{F}}\ell(\vtheta,\eta)$ is well-defined and the Fréchet derivative satisfies

$$\frac{\partial}{\partial\eta}\ell(\theta,\eta;X)\big|_{\eta = \hat\eta_{\vtheta}}=0.$$

Differentiating both sides of the preceding equations with respect to $\vtheta$, and applying the chain rule for functional derivatives, yields:
\begin{equation}\label{eq:second order derivative of log-likelihood}
    \frac{\partial^2}{\partial\vtheta\partial\eta}\ell(\theta,\eta;X)\big|_{\eta = \hat\eta_{\vtheta}} + \frac{\partial^2}{\partial\eta^2}\ell(\theta,\eta;X)\big|_{\eta = \hat\eta_{\vtheta}}\circ \frac{\partial}{\partial\vtheta} \hat\eta_{\vtheta} =0.
\end{equation}
Unlike the Gateaux derivative of the intensity function in \eqref{eq:gateaux derivative}, the Gateaux derivative of the log-likelihood function does not reduce to multiplication. Therefore, we use the symbol $\circ$ to denote the composition in the above equation.

Denote 
\begin{gather*}
    \ell_1(\vtheta,\eta;X): = \sum_{\vu\in X\cap A}\log\lambda(\vu;\vtheta,\eta)\\
    \ell_2(\vtheta,\eta) := \int_{A}\lambda(\vu;\vtheta,\eta)\mathrm{d}\vu
\end{gather*}
$\ell_1(\vtheta,\eta;X)$ is the random part in the log-likelihood that depends on $X$. $\ell_2(\vtheta,\eta)$ is the non-random part that does not depends on $X$. The log-likelihood can be separated into two parts $$\ell(\vtheta,\eta;X) = \ell_1(\vtheta,\eta;X)-\ell_2(\vtheta,\eta)$$
When the intensity function is log-linear, implied by the Fréchet derivative in \eqref{eq:Frechet}, the second derivatives of $\ell_1(\vtheta,\eta;X)$ disappear as follows:
\begin{align}
    \frac{\partial^2}{\partial\eta^2}\ell_1(\vtheta,\eta;X):= & \sum_{\vu\in X\cap A}\frac{\partial^2}{\partial\eta^2}\left\{\vtheta \vy_\vu+\eta(\vz_\vu)\right\} = 0\label{eq:diminish_ell_1_1}\\
    \frac{\partial^2}{\partial\vtheta\partial\eta}\ell_1(\vtheta,\eta;X):= & \sum_{\vu\in X\cap A}\frac{\partial^2}{\partial\vtheta\partial\eta}]\left\{\vtheta \vy_\vu+\eta(\vz_\vu)\right\} = 0 \label{eq:diminish_ell_1_2}
\end{align}
Thus, the second-order derivatives of $\ell(\theta,\eta;X)$ in \eqref{eq:second order derivative of log-likelihood} only include the second-order derivative of $\ell_2$ and is thereafter reduced to:
\begin{equation*}
    \frac{\partial^2}{\partial\vtheta\partial\eta}\ell_2(\theta,\eta)\big|_{\eta = \hat\eta_{\vtheta}} + \frac{\partial^2}{\partial\eta^2}\ell_2(\theta,\eta)\big|_{\eta = \hat\eta_{\vtheta}}\circ \frac{\partial}{\partial\vtheta} \hat\eta_{\vtheta} =0.
\end{equation*}
The above equation no longer depends on $X$ directly but might do so indirectly through the dependence between nuisance estimation $\hat\eta_{\vtheta}$ and $X$. If we further expand it and plug in \eqref{eq:gateaux derivative}, the composition can be reduced to multiplication inside the integral as follows:
\begin{equation*}
\int_A \exp[\theta^\top \vy_\vu +\eta(\vz_\vu)]\big|_{\eta = \hat\eta_{\vtheta}}\vy_\vu\mathrm{d}\vu +
\int_A \exp[\theta^\top \vy_\vu +\eta(\vz_\vu)]\big|_{\eta = \hat\eta_{\vtheta}}\cdot \frac{\partial}{\partial\vtheta} \hat\eta_{\vtheta}(\vz_\vu)\mathrm{d}\vu = 0
\end{equation*}
By pushing forward induced
by $\vu\mapsto (\vy,\vz)$ from $A$ to $\mathcal{Y}\times \mathcal{Z}$ with Radon–Nikodym derivative $f(\vy,\vz)$, we have
\begin{equation*}
\int_{\mathcal{Z}}\int_{\mathcal{Y}} \exp[\theta^\top \vy +\eta(\vz)]\big|_{\eta = \hat\eta_{\vtheta}}\vy f(\vy,\vz)\mathrm{d}\vy \mathrm{d}\vz +
\int_{\mathcal{Z}}\int_{\mathcal{Y}} \exp[\theta^\top \vy +\eta(\vz)]\big|_{\eta = \hat\eta_{\vtheta}}\cdot \frac{\partial}{\partial\vtheta} \hat\eta_{\vtheta}(\vz)f(\vy,\vz)\mathrm{d}\vy \mathrm{d}\vz = 0
\end{equation*}
Then, we can obtain the explicit formula of $\frac{\partial}{\partial\vtheta} \hat\eta_{\vtheta}(\vz)$:
\begin{align}
    \frac{\partial}{\partial\vtheta} \hat\eta_{\vtheta}(\vz) = & \left\{\int_{\mathcal{Y}} \exp[\theta^\top \vy +\eta(\vz)]\big|_{\eta = \hat\eta_{\vtheta}}\vy f(\vy,\vz)\mathrm{d}\vy\right\}^{-1}\int_{\mathcal{Y}} \exp[\theta^\top \vy +\eta(\vz)]\big|_{\eta = \hat\eta_{\vtheta}}\mathrm{d}f(\vy,\vz)\vy\nonumber\\
    = & \left\{\int_{\mathcal{Y}} \exp(\theta^\top \vy)\vy f(\vy,\vz)\mathrm{d}\vy\right\}^{-1}\int_{\mathcal{Y}} \exp(\theta^\top \vy )f(\vy,\vz)\mathrm{d}\vy\label{eq:nuisance estimation derivative}
\end{align}
Then, the indirect dependence between nuisance estimation derivative and $X$ through $\hat\eta_{\vtheta}$ is further eliminated. In other words, the derivative of the nuisance estimation is fully independent of $X$. 


\begin{remark}
    Though $\hat\eta_{\vtheta}:= \arg\max_{\eta\in\mathcal{F}}\ell(\vtheta,\eta)$ in the above discussion is not a rigorous representation of specific nonparametric estimation method, for example, the objective function of kernel estimation is not exactly $\ell(\vtheta,\eta)$, and the functional class $\mathcal{F}$ has to be differentiable while the random forest is not, the above discussion heuristically shows the independence of nuisance estimation derivative and observed spatial point process. Additionally, deriving the formula of the nuisance estimation derivative does not require the knowledge of the nuisance derivative. 
\end{remark}

\begin{remark}
    The key conditions in deriving the independence between nuisance estimation derivative and the point process $X$ are (1) the disappearance of direct dependence term $\ell_1$ in equation \eqref{eq:diminish_ell_1_1} and \eqref{eq:diminish_ell_1_2}, (2) the disappearance of indirect dependence through $\hat\eta_{\vtheta}$ in equation \eqref{eq:nuisance estimation derivative}.
\end{remark}

In the following subsection, we will show that these two conditions are satisfied by kernel estimation and spline estimation. 

\subsubsection{Independence Between Nuisance Estimation Derivative and Spatial Point Process: Spline Estimation}

Let $B(\vz)\in\R^b$ be spline basis. We redefine $\eta$ as the coefficients of the basis function. The log-likelihood function for spatial point process with log-linear intensity becomes
$$\ell(\vtheta,\eta) = \sum_{\vu\in X\cap A}[\vtheta^\top \vy_\vu+\eta^\top B(\vz_\vu)] - \int_{A}\exp\left[\vtheta^\top \vy_\vu+\eta^\top B(\vz_\vu)\right]\mathrm{d}\vu.$$
In spline case, all the differential derivatives reduce to regular derivative in finite-dimensional real space. The disappearance \eqref{eq:diminish_ell_1_1} and \eqref{eq:diminish_ell_1_2} still holds for $$\ell_1(\vtheta,\eta;X): = \sum_{\vu\in X\cap A}[\vtheta^\top \vy_\vu+\eta^\top B(\vz_\vu)].$$
Then, when spline regression is used, equation \eqref{eq:second order derivative of log-likelihood} is reduced to
$$\int_{A}\exp\left[\vtheta^\top \vy_\vu+\hat\eta_{\vtheta}^\top B(\vz_\vu)\right]\left(\vy_\vu+\frac{\partial}{\partial\vtheta}\hat\eta_{\vtheta}^\top B(\vz_\vu)\right)B(\vz_\vu)\mathrm{d}\vu  = 0$$
By pushing forward induced
by $\vu\mapsto (\vy,\vz)$ from $A$ to $\mathcal{Y}\times \mathcal{Z}$ with Radon–Nikodym derivative $f(\vy,\vz)$,
$$\int_{\mathcal{Y}}\int_{\mathcal{Z}}\exp\left[\vtheta^\top \vy+\hat\eta_{\vtheta}^\top B(\vz)\right]\left(\vy+\frac{\partial}{\partial\vtheta}\hat\eta_{\vtheta}^\top B(\vz)\right)B(\vz)f(\vy,\vz)\mathrm{d}\vy\mathrm{d}\vz  = 0.$$
Then, the derivative of nuisance estimation is:

\begin{align*}
    \frac{\partial}{\partial\vtheta}\hat\eta_{\vtheta}^\top B(\vz)
    = \left\{\int_{\mathcal{Y}} \exp(\theta^\top \vy)\vy f(\vy,\vz)\mathrm{d}\vy\right\}^{-1}\int_{\mathcal{Y}} \exp(\theta^\top \vy )f(\vy,\vz)\mathrm{d}\vy \label{eq:nuisance estimation derivative approx}
\end{align*}

\subsubsection{Independence Between Nuisance Estimation Derivative and Spatial Point Process: Kernel Estimation}

When using kernel estimation to estimate $\eta_{\vtheta}$ and the spatial point process has log-linear intensity, the objective function is kernel regression estimation of the log-likelihood function:
\begin{align*}
    &\hat\eta_{\vtheta}(\vz)
    \\
    =&\arg\max_{\eta(\vz)\in M}\left[\sum_{\vu\in X\cap A}K_h\left(\vz_{\vu}-\vz\right)\log\left[\Psi\left\{\tau_{\vtheta}(\vy_{\vu}),\eta(\vz)\right\}\right] -\int_{A} K_h\left(z_{\vu}-z\right)\Psi\left\{\tau_{\vtheta}(\vy_{\vu}),\eta(\vz)\right\}\mathrm{d}\vu\right]\\
    =&\arg\max_{\eta(\vz)\in M}\left[\sum_{\vu\in X\cap A}K_h\left(\vz_{\vu}-\vz\right)[\vtheta^\top \vy_\vu +\eta(\vz)] -\int_{A} K_h\left(z_{\vu}-z\right)\exp[\vtheta^\top \vy_\vu +\eta(\vz)]\mathrm{d}\vu\right]
\end{align*}
The disappearance \eqref{eq:diminish_ell_1_1} and \eqref{eq:diminish_ell_1_2} still holds for $$\ell_1(\vtheta,\eta;X): = \sum_{\vu\in X\cap A}K_h\left(\vz_{\vu}-\vz\right)[\vtheta^\top \vy_\vu +\eta(\vz)].$$
Then, equation \eqref{eq:second order derivative of log-likelihood} in kernel estimation case is reduced to
\begin{equation*}
\int_A K_h\left(\vz_{\vu}-\vz\right) \exp[\theta^\top \vy_\vu +\hat\eta_{\vtheta}(\vz)]\vy_\vu\mathrm{d}\vu +
\int_A \exp[\theta^\top \vy_\vu +\hat\eta_{\vtheta}(\vz)]\cdot \frac{\partial}{\partial\vtheta} \hat\eta_{\vtheta}(\vz)\mathrm{d}\vu = 0
\end{equation*}
Different from the case of spline regression and the general case, the $\vz$ inside the nuisance parameter is a fixed $\vz\in\mathcal{Z}$ rather than $\vz_\vu$. This is because the kernel estimation we defined implicitly applies the push-forward measure. Then, the derivative of nuisance estimation is:
\begin{equation*}
\frac{\partial}{\partial\vtheta} \hat\eta_{\vtheta}(\vz) = \left\{\int_A K_h\left(\vz_{\vu}-\vz\right)\exp(\theta^\top \vy_\vu )\mathrm{d}\vu\right\}^{-1}
\int_A K_h\left(\vz_{\vu}-\vz\right) \exp(\theta^\top \vy_\vu)\vy_\vu\mathrm{d}\vu 
\end{equation*}



\newpage
\renewcommand{\theequation}{C.\arabic{equation}}
\setcounter{equation}{0}
\section{Additional Simulation Results}\label{sec:additional_sim}

\subsection{Logistic Approximation versus Quadrature Approximation}\label{sec:additional_sim_log}

First, we briefly introduce the logistic approximation based on \citet{baddeley2014logistic}. We generate a dummy spatial point process $D$ independent of $X$ and has intensity function $\delta(\vu)$. Then, conditional on the observed points in $X\cup D$, each point $\vu$ in $X\cup D$ belongs to $X$ with probability $\left[\lambda(\vu;\vtheta,\eta)+ \delta(\vu)\right]^{-1}\lambda(\vu;\vtheta,\eta)$ and belongs to $D$ with probability $\left[\lambda(\vu;\vtheta,\eta)+ \delta(\vu)\right]^{-1} \delta(\vu)$. Therefore, the estimation of $\vtheta$ and $\eta$ can be obtained by maximizing the  following logistic loglikelihood
\begin{equation} \label{eq:logistic_likelihood}
    \ell_{\rm logistic}(\vtheta,\eta)=\sum_{\vu\in X}\log\left\{\frac{\lambda(\vu;\vtheta,\eta)}{\lambda(\vu;\vtheta,\eta)+ \delta(\vu)}\right\} + \sum_{\vu\in D}\log\left\{\frac{\delta(\vu)}{\lambda(\vu;\vtheta,\eta)+ \delta(\vu)}\right\}.
\end{equation}
Note that the computation can also be implemented with \texttt{mgcv}.

Although the target estimation $\hat\vtheta$ obtained using logistic approximation is generally unbiased, it has inflated asymptotic variance due to the extra randomness from the dummy point process $D$. We generalized the adjusted variance estimator in \cite{baddeley2014logistic} to our semiparametric model for Poisson processes. For non-Poisson processes, we use the same variance estimator as described in the main manuscript. This because \cite{baddeley2014logistic} studied the Gibbs process, which is a different model for the interaction of points than that based on the pair correlation function. 

We implement both approximation methods for Poisson spatial point processes  and log-Gaussian Cox processes (LGCP) under the same simulation settings in the manuscript. For each scenario, we simulate the processes for 1000 times and report their summary. Table \ref{tab:MPL v.s. Logistic Poisson} and Table \ref{tab:MPL v.s. Logistic LGCP} show that the bias of logistic approximation is generally smaller than the bias of quadrature approximation. Moreover, in some cases (bold in Table \ref{tab:MPL v.s. Logistic Poisson} and Table \ref{tab:MPL v.s. Logistic LGCP}), the bias of quadrature approximation is not negligible and the confidence interval under-covers the true parameters. 

\begin{table}[htbp]
\centering
\begin{tabular}{lcccccccc}
\toprule
 &  & Nuisance & Method & $\text{Bias}_{\times100}$ & rMSE & meanSE & CP90 & CP95 \\
\midrule
\multirow{8}{*}{$W_1$} & \multirow{4}{*}{ind} & linear & mpl & -0.0359 & 0.0442 & 0.0456 & 91 & 95.2 \\
 &  & linear & logi & -0.2477 & 0.0467 & 0.0459 & 89.8 & 93.5 \\
 &  & poly & mpl & -0.1478 & 0.0469 & 0.0468 & 89.6 & 95.2 \\
 &  & poly & logi & -0.0438 & 0.0473 & 0.0476 & 91.1 & 94.7 \\\cmidrule(lr){2-9}
 & \multirow{4}{*}{dep} & linear & mpl & -0.1790 & 0.0443 & 0.0448 & 89.6 & 95.1 \\
 &  & linear & logi & -0.1318 & 0.0438 & 0.0449 & 91.2 & 95.2 \\
 &  & poly & mpl & -1.1371 & 0.0517 & 0.0520 & 90.2 & 95.3 \\
 &  & poly & logi & -0.7548 & 0.0531 & 0.0532 & 89.3 & 95.3 \\\cmidrule(lr){1-9}
 \multirow{8}{*}{$W_2$}& \multirow{4}{*}{ind} & linear & mpl & \textbf{-1.3329} & 0.0264 & 0.0231 & 85.6 & 91.1 \\
 &  & linear & logi & 0.0317 & 0.0239 & 0.0238 & 89.3 & 94.4 \\
 &  & poly & mpl & -0.1595 & 0.0236 & 0.0246 & 90.9 & 96.2 \\
 &  & poly & logi & -0.0158 & 0.0249 & 0.0254 & 91.2 & 95.7 \\\cmidrule(lr){2-9}
 & \multirow{4}{*}{dep} & linear & mpl & \textbf{-1.1509} & 0.0244 & 0.0228 & 86.9 & 93.2 \\
 & & linear & logi & 0.1089 & 0.0232 & 0.0236 & 90.2 & 95.8 \\
 & & poly & mpl & -0.4310 & 0.0260 & 0.0266 & 92.6 & 95.9 \\
 & & poly & logi & -0.0219 & 0.0266 & 0.0275 & 91.3 & 96.2 \\
\bottomrule
\end{tabular}
\caption{Quadrature v.s. Logistic for Poisson processes. \textit{mpl}: quadrature approximation. \textit{logi}: logistic approximation.}
\label{tab:MPL v.s. Logistic Poisson}
\end{table}

\begin{table}[htbp]
\centering
\begin{tabular}{ccccccccc}
\toprule
 & & Nuisance& Method &  {$\text{Bias}_{\times 100}$} & {rMSE} &{meanSE} & {CP90 (\%)} & {CP95 (\%)} \\
\midrule
\multirow{8}{*}{$W_1$} &\multirow{4}{*}{ind} &linear&mpl & -0.0473 & 0.0613 & 0.0572 & 87.4 & 92.7 \\
& &linear&logi & -0.3824 & 0.0628 & 0.0559 & 86.4 & 91.5 \\
& &poly&mpl & -0.0305 & 0.0633 & 0.0582 & 87.7 & 92.9 \\
& &poly&logi & 0.0534 & 0.0582 & 0.0573 & 88.9 & 95.0 \\ \cmidrule(lr){2-9}
&\multirow{4}{*}{dep} &linear&mpl & -0.2265 & 0.0604 & 0.0552 & 86.5 & 92.2 \\
& &linear&logi & -0.0493 & 0.0598 & 0.0547 & 86.5 & 91.7 \\
& &poly&mpl & \textbf{-0.7660} & 0.0658 & 0.0617 & 88.6 & 93.6 \\
& &poly&logi & -0.5999 & 0.0673 & 0.0611 & 86.1 & 91.9 \\ \cmidrule(lr){1-9}
\multirow{8}{*}{$W_2$} & \multirow{4}{*}{ind}&linear&mpl & \textbf{-0.8791} & 0.0323 & 0.0306 & 87.3 & 93.2 \\
& &linear&logi & 0.0645 & 0.0304 & 0.0306 & 89.8 & 95.0 \\
& &poly&mpl & -0.5021 & 0.0324 & 0.0311 & 89.0 & 94.3 \\
& &poly&logi & -0.1370 & 0.0325 & 0.0311 & 89.4 & 93.6 \\\cmidrule(lr){2-9}
&\multirow{4}{*}{dep} &linear&mpl & \textbf{-0.8908} & 0.0311 & 0.0295 & 87.2 & 93.1 \\ 
& &linear&logi & 0.1506 & 0.0312 & 0.0298 & 88.7 & 93.3 \\
& &poly&mpl & -0.7064 & 0.0332 & 0.0326 & 87.8 & 95.1 \\
& &poly&logi & -0.2485 & 0.0328 & 0.0327 & 89.8 & 94.3 \\
\bottomrule
\end{tabular}
\caption{Quadrature v.s. Logistic for log-Gaussian Cox processes (LGCP). \textit{mpl}: quadrature approximation. \textit{logi}: logistic approximation.}
\label{tab:MPL v.s. Logistic LGCP}
\end{table}

\subsection{Boundary Bias of Kernel Estimation}\label{sec:additional_sim_bound}
In our implementation, we use spline smoothing instead of kernel smoothing to estimate the nuisance parameters due to the boundary bias of kernel smoothing. As noted in existing works \cite{jones1993simple}, \cite{racine2001bias}, a well-known problem in the classic kernel regression is the boundary bias issue where the bias of the nonparametrically estimated regression function increases near the boundary of the covariate support. When we use spatial kernel regression to estimate the nuisance parameter, the nuisance estimation suffers from the same boundary bias so that the target estimation has inflated variance. 

In this section, we conduct a simulation study to illustrate the boundary bias of spatial kernel regression and its effect to target estimation through a simulation study. We consider a Poisson point process observed in $A=[0,1]\times[0,1]$ with a log-linear intensity function of the form
$$\lambda(\vu;\theta,\eta) = \exp[\theta y_\vu +\eta(z_\vu)]$$

Let $u_1,u_2$ be the two coordinates of a point $\vu\in\R^2$ (i.e., $\vu=(u_1,u_2)$). We let target covariate $y(\vu)$ and the nuisance covariates $z(\vu)$ be dependent (i.e., $y(\vu) = u_1+u_2$, $z(\vu) = u_1 - u_2$). The true target parameter $\theta^*$ is set to $4$. The true nuisance parameter varies among $\eta^*(z)=\alpha-4z^2$ (i.e. negative polynomial), $\eta^*(z)=\alpha+4z^2$ (i.e. positive polynomial) and $\eta^*(z)=\alpha+\exp(z)$ (i.e. exponential). We remark that the nuisance parameter $\eta^*$ has a tuning parameter $\alpha$ to control the number of points at each point process. For every nuisance parameter, we alter the value of $\alpha$ to generate two processess, a sparse point process with roughly $500$ points, and a dense point process with roughly $2000$ points. 

\begin{figure}[htbp]
    \centering
  \includegraphics[width=1.0\linewidth]{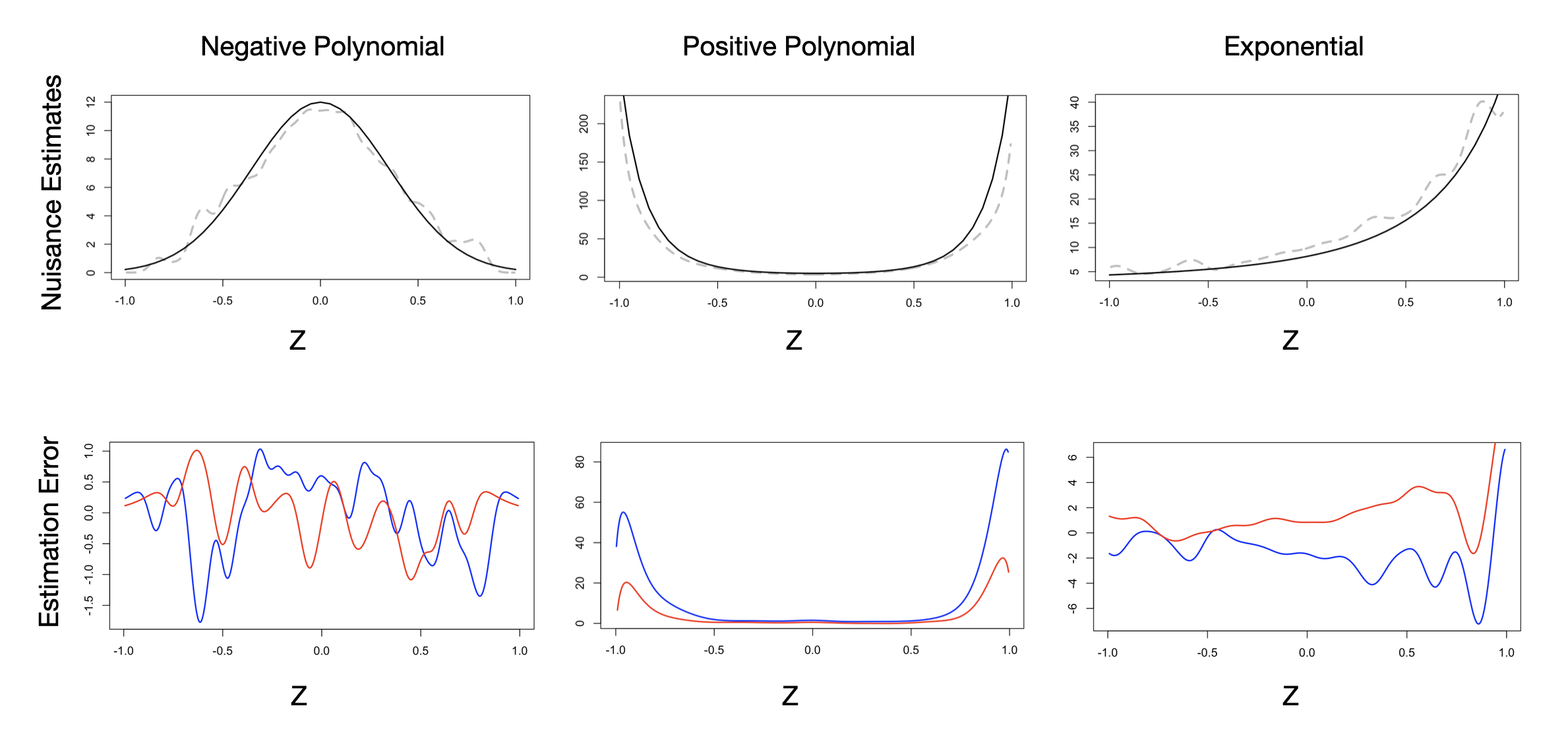}
    \caption{Upper: the true nuisance parameter (black line) and the estimated nuisance parameter (dashed line) with large $\alpha$. Bottom: estimation error of the nuisance parameter for small $\alpha$ (red line) and large $\alpha$ (blue line).}
    \label{fig:boundary}
\end{figure}

To illustrate the boundary bias of spatial kernel regression, we plot $\exp[\hat\eta(z_\vu)]$ and  $\exp[\eta^*(z_\vu)]$ on the top of Figure \ref{fig:boundary}, and the estimation error $\exp[\eta^*(z_\vu)]-\exp[\hat\eta(z_\vu)]$ on the bottom of Figure \ref{fig:boundary}. When $\exp[\eta^*(z_\vu)]$ is small near the boundary of support of $z$ (e.g. negative polynomial), the estimation error is fluctuating within a very small range. On the other hand, when $\exp[\eta^*(z_\vu)]$ is large near the boundary of the support of $z$ (e.g. positive polynomial and exponential), the estimation error greatly increases near the boundary. Specifically, if the intensity is concentrated near the boundary (i.e., positive polynomial nuisance function), the boundary bias is extremely large. Moreover, when the point process becomes denser (i.e., increase $\alpha$), the boundary bias becomes more severe. So, the boundary bias is not eased even if we observe more points in the same observational window.

 To see how the boundary bias affect target estimation, we simulate each scenario for 300 times and report the summary. For comparison, we fit an oracle model where the oracle models knows the true nuisance parameter $\eta^*$ a priori. Table \ref{table:simulation_dep_issue} shows the the empirical bias (BIAS), the root mean square error of the target estimation (RMSE), the mean of the estimated standard errors (SEM), and the 95\% coverage rate (CP95).

\begin{table}[h!]
    \begin{center}
\begin{tabular}{|l || l l l l l  l |} 
\hline
  Nuisance& a & exp($\alpha$) & $\text{BIAS}_{\times 100}$ & RMSE & SEM & CP95 \\  
 \hline 
 \hline
 Oracle & 1  & 6 & 0.651  & 0.135 & 0.140 &  0.960 \\  
 \hline
 \multirow{2}{5em}{Negative polynomial} & 1 & 6 & 0.059 & 0.144 & 0.144  & 0.950 \\
 & 1 & 12 & -0.231 & 0.105 & 0.101  & 0.960 \\ 
  \hline 
 \hline
 Oracle & 1  & 1.5 & -1.067  & 0.198 & 0.200  & 0.953 \\  
 \hline
 \multirow{2}{5em}{Positive polynomial} & 1 & 1.5 & 2.077 & 0.227 & \textbf{0.220}  & \textbf{0.940} \\
 & 1 & 3 & 0.231 & 0.174 &\textbf{0.155}   & \textbf{0.887} \\  
  \hline 
 \hline
 Oracle & 1  & 1.5 & 0.373  & 0.129 & 0.126  & 0.943 \\  
 \hline
 \multirow{2}{5em}{Exponential} & 1 & 1.5 & 1.309 & 0.149 & 0.153 & 0.953 \\  
  & 1 & 3 & 0.577 & 0.107 & 0.108  & 0.957 \\
  \hline 
 \hline
\end{tabular}
\end{center}
    \caption{Amplification of the boundary bias with increasing $\alpha$ when the target covariate and the nuisance covariate are dependent.}
    \label{table:simulation_dep_issue}
\end{table}

The estimation of the target parameter is consistent and asymptotically normal when the boundary bias is not severe (i.e., when the nuisance parameter is a negative polynomial or exponential). However, when the nuisance function is positive polynomial, the root mean square of the estimation of target parameter is greatly inflated due to the boundary bias of nuisance estimation, which is further amplified when $\alpha$ increases (bold in Table \ref{table:simulation_dep_issue}).

\newpage
\renewcommand{\theequation}{D.\arabic{equation}}
\setcounter{equation}{0}

\section{Preliminary}

\subsection{Higher Order Kernel}\label{sec:higher_kernel}
\begin{definition}[Higher Order Kernel]\label{def:higher order kernel} Let $l$ be a positive even integer. We say  $k(\cdot):\R\rightarrow\R$ to be a $l$-th order kernel function if it satisfies
\begin{gather*}
    \int_R k(z)\mathrm{d}z = 1,\\
    \int_R z^ik(z)\mathrm{d}z = 0,\quad{}(i=2,\ldots,l-1),\\
    \int_R z^l k(z)\mathrm{d}z = \kappa_l\neq 0.
\end{gather*}
Let $K(\vz)=\prod_{j=1}^q k(z_j)$ for $\vz=(z_1,\ldots,z_q)\in\R^q$. Then $K(\cdot)$ satisfies
\begin{gather*}
    \int_{R^q} K(\vz)\mathrm{d}\vz = 1,\\
    \int_{R^q} \vz^i K(\vz)\mathrm{d}\vz = 0,\quad{}(i=2,\ldots,l-1),\\
    \int_{R^q} \vz^l K(\vz)\mathrm{d}\vz = \kappa_l^q\neq 0.
\end{gather*}

\end{definition}

\subsection{Implicit Function Theorem}

\begin{theorem}[Implicit Function Theorem]\label{thm:implicit function theorem}
Let $f:\R^{n+m}\rightarrow\R^m$ be a $k$-times continuously differentiable function, and let $\R^{n+m}$ have coordinates $({x},{y})$. Fix a point $({a},{b})$ with $f({a},{b})={0}$, where ${0}\in\R^m$ is the zero vector. If the Jacobian matrix $$J_{f,{y}}({a},{b})=\left[\frac{\partial f_i}{\partial y_j}({a},{b})\right]$$
is invertible, then there exists an open set $U\in\R^n$ containing ${a}$ such that there exists a unique function $g:U\rightarrow\R^m$ such that $g({a})=b$, and $f[{x},g({x})]={0}$ for all ${x}\in U$. Moreover, $g$ is also $k$-times continuously differentiable and the Jacobian matrix of partial derivatives of $g$ in $U$ is given by
$$\left[\frac{\partial g_i}{\partial x_j}({x})\right]_{m\times n} = -\left\{J_{f,{y}}[{x},g({x})]\right\}^{-1}_{m\times m}\left\{J_{f,{x}}[{x},g({x})]\right\}^{-1}_{m\times n}.$$
\end{theorem}

\subsection{Mixing Coefficient of Spatial Point Process}\label{sec:mixing}

The $\alpha$-mixing coefficient is a widely used notion in the central limit theorem of spatial point processes (e.g.,  \citet{biscio2019general}). Let $\mathcal{F}$ and $\mathcal{G}$ be two $\sigma$-algebras defined on a common
probability space. The $\alpha$-mixing coefficient of  $\mathcal{F}$ and $\mathcal{G}$ is defined as 
$$\alpha(\mathcal{F}, \mathcal{G})=\sup \{|\mathrm{P}(F \cap G)-\mathrm{P}(F) \mathrm{P}(G)|: F \in \mathcal{F}, G \in \mathcal{G}\}.$$ 
Let $d(\vu,v) = \max\{|u_i-v_i|:1\leq i\leq 2\}$ be the distance between two points $\vu$ and $v$ in $\R^2$. Let $d(A_1,A_2) = \inf\{d(\vu,v);\vu\in A_1,v\in A_2\}$ be the distance between two subset $A_1$ and $A_2$ in $\R^2$. Given hyper-parameters $c_1,c_2,r$, the $\alpha$-mixing coefficient, $\alpha_{c_1,c_2}^X(r)$. of a spatial point process $X$ is defined as
\begin{align*}
    \sup\{\alpha(\sigma(X\cap A_1),\sigma(X\cap A_2)): A_1\in \R^2,A_2\in\R^2,|A_1|\leq c_1,|A_2|\leq c_2,d(A_1,A_2)\geq r\},
\end{align*}
$r$ is the minimum distance between two regions. $c_1,c_2$ are the upper bounds of $|A_1|$ and $|A_2|$ respectively.

\subsection{Factorial Cumulant Functions of Spatial Point Processes}\label{sec:factorial}
The $m$-th order factorial cumulant function describes the $m$-th order dependence of the spatial point process (See Definition 2.6 in \citet{biscio2016brillinger}). Given $m$ points $\{\vu_1, \vu_2, \ldots, \vu_m\}$ in $\R^2$, the $m$-th order factorial cumulant function $Q_m(\vu_1, \vu_2, \ldots, \vu_m)$ can be expressed as a function of the intensity functions up to the $m$th-order such that

$$Q_m(\vu_1, \vu_2, \ldots, \vu_m) = \sum_{G} (-1)^{|G| - 1} (|G| - 1)! \prod_{ g\in G} \lambda^{(|g|)}(\{\vu_i : i \in g\}),$$
where the sum is taken over all partitions $G$ of the set $\{1, 2, \ldots, m\}$, and $|G|$ denotes the number of elements in the partition.

For instance, the first-order factorial cumulant is the intensity function $\lambda(\vu)$:
\[
Q_1(\vu) = \lambda(\vu).
\]
The second-order factorial cumulant function that measures pairwise dependence of two points $\vu_1$ and $\vu_2$ is 
\[
Q_2(\vu_1, \vu_2) = \lambda^{(2)}(\vu_1, \vu_2) - \lambda(\vu_1) \lambda(\vu_2).
\]
The third-order factorial cumulant function that quantifies the joint interactions among three points is
\[
\begin{aligned}
Q_3(\vu_1, \vu_2, \vu_3) = & \ \lambda^{(3)}(\vu_1, \vu_2, \vu_3) - \lambda^{(2)}(\vu_1, \vu_2) \lambda(\vu_3) - \lambda^{(2)}(\vu_1, \vu_3) \lambda(\vu_2) \\
& - \lambda^{(2)}(\vu_2, \vu_3) \lambda(\vu_1) + 2 \lambda(\vu_1) \lambda(\vu_2) \lambda(\vu_3).
\end{aligned}
\]
Let $\kappa_m$ be the factorial cumulant of $|X\cap A_n|$, i.e., the number of the observed points. $\kappa_m$ satisfies 
$$\kappa_m=\int_{A^m}Q_m(\vu_1,\vu_2,\cdots,\vu_m)\mathrm{d}\vu_1\ldots \mathrm{d}\vu_m.$$
Let $\kappa_m^\prime$ be the ordinary cumulant of $|X\cap A_n|$. The ordinary cumulant can be express as a sum of factorial cumulants  via 
$$\kappa^\prime_{m} = \sum_{j=1}^m \Delta_{j,m}\kappa_j,$$
where $\Delta_{j,m}$ is Stirling number of the second kind.
Let $\mu_{m}$ be the centered moment of $|X\cap A_n|$ to the $m$-th order. $\mu_{m}$ can be expressed in terms of the ordinary cumulants via
$$\mu_{m} = \sum_{k=1}^{m} B_{m,k}(0,\kappa^\prime_{2},\ldots,\kappa^\prime_{m-k+1})=B_{m}(0,\kappa^\prime_{2},\ldots,\kappa^\prime_{m}),$$ 
where $B_{m^\prime}$ is the Bell polynomials.



\subsection{Gateaux Derivative}\label{sec:gateaux}

Gateaux derivative can be viewed as a generalization of the classical directional derivative. The Gateaux derivative of the intensity function is defined as follows:
\begin{definition}[Gateaux Derivative for Semiparametric Intensity Functions] \label{def:gateaux_derivative}
    The first-order Gateaux derivative of $\lambda(\vu;\vtheta,\eta) = \Psi\left\{\tau_{\vtheta}(\vy_{\vu}),\eta(\vz_\vu)\right\}$ with respect to $\eta$ along the direction $\bar\eta \in H$ is defined as 
\begin{align}
    \frac{\partial}{\partial\eta}\lambda(\vu;\vtheta,\eta)[\bar\eta] & =\lim_{h\rightarrow 0}\frac{\lambda(\vu;\vtheta,\eta+h\bar\eta)-\lambda(\vu;\vtheta,\eta)}{h}
\end{align}
provided the limit exists. If the Gateaux derivative is a linear operator of $\bar\eta$, $\lambda(\vu;\vtheta,\eta)$ is said to be Fréchet differentiable with respect to $\eta$ and such linear map is the Fréchet derivative.
\end{definition} 
When the intensity function satisfies $\lambda(\vu;\vtheta,\eta) = \Psi\left\{\tau_{\vtheta}(\vy_{\vu}),\eta(\vz_\vu)\right\}$, and $\Psi\{\cdot,\eta(z)\}$ is twice continuously differentiable with respect to $\eta(z)$, the $m$-order Gateaux derivative of $\lambda(\vu;\vtheta,\eta)$ along $\bar\eta_1,\ldots,\bar\eta_m$ equals:
\begin{equation}\label{eq:gateaux derivative}
   \frac{\partial}{\partial\eta^m}\lambda(\vu;\vtheta,\eta)[\bar\eta_1,\ldots,\bar\eta_m] = \frac{\partial^{m}}{\partial\gamma^m}\Psi\left\{\tau_{\vtheta}(\vy_{\vu}),\gamma\right\}\bigg|_{\gamma=\eta(\vz_{\vu})}\prod_{i=1}^m\bar{\eta}_i(\vz_{\vu}). 
\end{equation}
The intensity function is also Fréchet differentiable such that
\begin{equation}\label{eq:Frechet}
    \frac{\partial}{\partial\eta^m}\lambda(\vu;\vtheta,\eta) = \frac{\partial^{m}}{\partial\gamma^m}\Psi\left\{\tau_{\vtheta}(\vy_{\vu}),\gamma\right\}\bigg|_{\gamma=\eta(\vz_{\vu})}
\end{equation}
The variable $\gamma$ in the preceding equation denotes the value of the second argument of the bivariate function $\Psi(\cdot,\cdot)$. The derivative with respect to$\gamma$ is the standard partial derivative. For notational convenience and to enhance clarity, we employ the following definition for the $m-th$ order derivative of $\Psi$ with respect to its second argument, evaluated at $\eta(\vz)$
$$\frac{\partial^{m}}{\partial\eta^m}\Psi\left\{\tau_{\vtheta}(\vy_{\vu}),\eta(\vz_\vu)\right\} := \frac{\partial^{m}}{\partial\gamma^m}\Psi\left\{\tau_{\vtheta}(\vy_{\vu}),\gamma\right\}\bigg|_{\gamma=\eta(\vz_{\vu})}$$
In this context, $:=$ indicates a definitional equivalence.

\end{document}